\documentclass[12pt,a4j]{article}
\usepackage{amsmath,amsthm,amssymb,bm,graphicx,fancybox,multirow,hhline,psfrag}
\usepackage{comment}
\usepackage{ulem}
\usepackage{natbib}
\usepackage{mmacells}
\usepackage{algorithm}
\usepackage{algorithmic}

\def\hsymbu#1{\smash{\lower1.7ex\hbox{\huge$#1$}}}

\usepackage{indentfirst}
\usepackage{listings}
\usepackage{url}
\def\tr{\mathrm{Tr}}

\def\new{\mathrm{new}}

\def\l{{l}}

\newtheorem{prop}{Proposition}
\newtheorem{rem}{Remark}
\newtheorem{example}{Example}
\newtheorem{lem}{Lemma}
\newtheorem{thm}{Theorem}
\newcommand{\argmin}{\mathop{\rm arg~min}\limits}

  \makeatletter       
  \renewcommand{\@biblabel}[1]{\qquad}
  \makeatother

\begin{document}

{
\begin{center}
\textbf{\Large 
Clustering-based aggregate value regression
}
\end{center}







\begin{center}
\large {Kei Hirose$^{1}$, Hidetoshi Matsui$^{2}$ and Hiroki Masuda$^{3}$
}
\end{center}

\begin{flushleft}
{\footnotesize
$^1$ Institute of Mathematics for Industry, Kyushu University, 744 Motooka, Nishi-ku, Fukuoka 819-0395, Japan \\


$^2$ Faculty of Data Science, Shiga University, 1-1-1, Banba, Hikone, Shiga, 522-8522, Japan \\

\vspace{1.2mm}

$^3$ Graduate School of Mathematical Sciences, University of Tokyo, 3-8-1 Komaba Meguro-ku Tokyo 153-8914, Japan \\
}
{\it {\small E-mail: hirose@imi.kyushu-u.ac.jp, hmatsui@biwako.shiga-u.ac.jp, hmasuda@ms.u-tokyo.ac.jp}}	
\end{flushleft}

\vspace{1.5mm}

\begin{abstract}
In various practical situations, forecasting of aggregate values rather than individual ones is often our main focus. For instance, electricity companies are interested in forecasting the total electricity demand in a specific region to ensure reliable grid operation and resource allocation. However, to our knowledge, statistical learning specifically for forecasting aggregate values has not yet been well-established.  In particular, the relationship between forecast error and the number of clusters has not been well studied, as clustering is usually treated as unsupervised learning.  This study introduces a novel forecasting method specifically focused on the aggregate values in the linear regression model.  We call it the Aggregate Value Regression (AVR), and it is constructed by combining all regression models into a single model.  With the AVR, we must estimate a huge number of parameters when the number of regression models to be combined is large, resulting in overparameterization.  To address the overparameterization issue, we introduce a hierarchical clustering technique, referred to as AVR-C (C stands for clustering).  In this approach, several clusters of regression models are constructed, and the AVR is performed within each cluster.  The AVR-C introduces a novel bias-variance trade-off theory under the assumption of a misspecified model. In this framework, the number of clusters characterizes model complexity. Monte Carlo simulation is conducted to investigate the behavior of training and test errors of our proposed clustering technique.  The bias-variance trade-off theory is also demonstrated through the analysis of electricity demand forecasting.
 \end{abstract}

\section{Introduction}
With advances in censoring techniques, a large amount of micro-scale censoring data have been collected, and high-accuracy forecasting with censoring data has been essential in various fields of research and industry, such as energy, marketing, and traffic.  In many cases, we are interested in the forecast of aggregate value rather than individual values.  An example is the electricity demand forecast.  With the spread of smart meters, electricity companies have a large amount of household electricity demand data.  The electricity demand forecast is essential for trade in the electricity market, energy-saving interventions, and demand response (e.g., \citealp{guo2018residential,wang2018impact,ruiz2020integration}).  In reality, we are interested in the forecast of electricity demand aggregated by individual households, rather than the individual households themselves, to ensure reliable grid operation and resource allocation. For example, in the Japanese electricity market (JEPX: Japan Electric Power eXchange\footnote{\url{https://www.jepx.jp}}), the trade of electricity is made with not each household's demand but their aggregation in a specific area in Japan.  Another example would be purchase demand forecasting in marketing; a manager is often interested in the total sales from all stores rather than the sales in individual stores.

In many cases, the forecast is achieved with regression models, where predictor variables are previous demands and other factors 
(e.g., \citealp{Ramanathan.1997,Dudek:2016hy,Lusis:2017gd,Hirose.2021}).  The advantages of regression modeling include its computational efficiency and ease of model extension.  The aggregate value forecast can be achieved by directly applying regression analysis to individual-level data (e.g., individual households) and summing the individual forecast values.  This method is referred to as Individual Regression (IR).  However, we observe that the residuals of the IR models are often positively correlated, and such correlations cannot be incorporated with the individual forecast.  Moreover, the IR could occasionally be too simple and then the model could be misspecified; in this case, the predictors of the IR may not have sufficient information for accurate predictions.
Meanwhile, the aggregation with a large number of individuals may mitigate such an issue.  As a result, the aggregation would be essential to achieve good prediction accuracy.

In this study, we construct a regression model specifically for forecasting the aggregate values.  We directly forecast the aggregate values while making full use of all available individual information of predictors.  To achieve this, all responses are aggregated while the predictors are utilized at the individual level.  We call this model an Aggregate Value Regression (AVR). The AVR provides a significantly flexible model thanks to a variety of predictor variables. 
However, the AVR often results in overfitting; this is because the number of parameters is proportional to the number of individuals, resulting in a huge number of parameters. 
To estimate parameters applicable to high-dimensional data, we employ the ridgeless least squares \citep{Hastie.2022}, the ridge estimator with the regularization parameter converging to 0.  
Nevertheless, this approach does not always address the problem of overfitting.

To summarize what has been discussed so far, the AVR model is more flexible than IR, but it often suffers from overfitting. In contrast, the IR model may not suffer from overfitting but is less flexible due to the simplicity of the model.  
Therefore, a method that lies between IR and AVR is expected to improve the accuracy of both models.  To achieve this, we perform hierarchical clustering on regression models and estimate AVR models for each cluster separately. We call this method AVR-C (C stands for Clustering). The AVR-C bridges a gap between IR and AVR, allowing us to generate a model that is considerably more flexible than IR while preventing the overfitting of AVR.

The AVR-C introduces a novel bias-variance trade-off theory, contributing to the development of fundamental statistical theories in contemporary real-world problems. We will show that the training error is a monotone-decreasing with respect to its expectation as a function of the number of clusters under the assumption that the correlations of errors between two regression models are nonnegative. In other words, model complexity is characterized by the number of clusters. Although the assumption of non-negativity of the error correlation is not necessarily weak, it can often be satisfied in reality, such as in electricity demand forecasting. Meanwhile, the test error does not have the monotonicity property as in the training error; it often exhibits the U-shape function in our numerical experiments. 

It is noted that the bias-variance trade-off is particularly evident under the assumption that the model is misspecified.  In this case, the bias is expressed as the degrees of model misspecification, which quantifies the error arising from the difference between our assumed model and the true underlying model.  Meanwhile, the variance is characterized by the number of parameters within each cluster. The assumption of model misspecification plays an important role in providing a theoretical framework that explains how the trade-off arises in practical settings. See Section \ref{hm:sec_3} for details.

The AVR-C is dependent on the hierarchical clustering method.  We propose two clustering methods that are specialized for the AVR-C.  First, we minimize the training error for each step of hierarchical clustering. This method generally provides good performance in terms of forecast accuracy, but is computationally expensive because we must compute the training error for each step of hierarchical clustering.  The second clustering method focuses on the computational efficiency and interpretation, rather than prediction accuracy.  We employ standard hierarchical clustering, such as Ward's method, using ``$1 - \text{correlation coefficients}$" as a dissimilarity measure. The correlation matrix is constructed from the residuals generated by IR. With this procedure,  the error correlations within a cluster are made as positively correlated as possible.  
Therefore, regression models in the same cluster tends to have similar tendencies in residuals, resulting in an interpretable clustering results. 

Our bias-variance trade-off theory is investigated through the analyses of both synthetic and actual data.  We observe not only the conventional bias-variance trade off but also double descent phenomena (e.g., \citealp{Hastie.2022}), a contemporary bias-variance trade-off phenomenon, in both simulation and real data analyses. 


This paper is organized as follows: in Section 2, we introduce our proposed procedure, AVR-C.  The bias-variance trade-off theory of our proposed method is described in Section 3. Two clustering methods to construct the clusters in AVR-C are also presented.  In Section 4, we investigate the performance of our proposed method through  Monte Carlo simulations.  Section 5 illustrates the application of our proposed method to electricity demand data.  Conclusion and discussion are provided in Section 6.

\section{Proposed method}
\subsection{Aggregate value regression}
 Assume that $\{\bm{y}_m \mid m=1,\dots,M\}$ and $\{\bm{X}_m \mid m=1,\dots,M\}$ are $M$ sets of responses and predictors, respectively,  where $\bm{y}_m$ is an $n$-dimensional response vector and $\bm{X}_m$ is an $n\times p$ design matrix. Here, $n$ is the number of observations, $p$ is the number of variables, and $M$ is the number of regression models. For example, in household electricity demand forecasting, $M$ indicates the number of households. Specifically, we employ the following $M$ multiple linear regression models
\begin{equation}
\begin{aligned}
\bm{y}_1 &= \bm{X}_1\bm{\beta}_1 + \bm{\varepsilon}_1,\\
\bm{y}_2 &= \bm{X}_2\bm{\beta}_2 + \bm{\varepsilon}_2,\\
&\hspace{10mm}\vdots  \\
\bm{y}_M &= \bm{X}_M\bm{\beta}_M + \bm{\varepsilon}_M,
\end{aligned}
\label{eq:allmodels}
\end{equation}
where $\{\bm{\beta}_m \mid m=1,\dots, M\}$ and $\{\bm{\varepsilon}_m \mid m=1,\dots, M\}$ are sets of regression coefficients and error vectors with $E[\bm{\varepsilon}_m]=\bm{0}$ ($m=1,\dots,M$), respectively.   
Note that our methodology can be directly extended to any nonlinear statistical models and machine learning techniques; we use linear regression to introduce a novel bias-variance trade-off theory described in Section \ref{hm:sec_theory}.   

Let $\{(y_m^{\new},\bm{x}_m^{\new})\mid m=1,\dots,M\}$ be $M$ sets of responses and predictors independent from $\{(\bm{y}_m,\bm{X}_m)\mid m=1,\dots,M\}$.  Given the estimator of regression coefficients based on $\{(\bm{y}_m,\bm{X}_m)\mid m=1,\dots,M\}$, say $\hat{\bm{\beta}}_1,\dots,\hat{\bm{\beta}}_M$, we forecast aggregate value, $\hat{y}^{\new}:=\sum_{m=1}^M\hat{y}_m^{\new}$, where $\hat{y}_m^{\new}:=(\bm{x}_m^{\new})^T\hat{\bm{\beta}}_m$. Our aim is to generate an estimator $\hat{\bm{\beta}}_1,\dots,\hat{\bm{\beta}}_M$ that provides a good forecast accuracy of the aggregate values; that is, we estimate regression coefficients such that $E[(y^{\new} - \hat{y}^{\new})^2]$ is as small as possible, where $y^{\new}:=\sum_{m=1}^My_m^{\new}$.

One way to get the estimates of parameters, $\hat{\bm{\beta}}_1,\dots,\hat{\bm{\beta}}_M$, is to simply perform the regression model individually; that is, 
$$\hat{\bm{\beta}}_m=f_m(\bm{y}_m,\bm{X}_m),$$
where $f_m$ is a function of $(\bm{y}_m,\bm{X}_m)$. We call this method \textit{individual regression (IR)}. Typically, we employ the ordinary least squares (i.e., $f_m(\bm{y}_m,\bm{X}_m) = (\bm{X}_m^\top\bm{X}_m)^{-1}\bm{X}_m^T\bm{y}_m$) to get $\hat{\bm{\beta}}_1,\dots,\hat{\bm{\beta}}_M$.  

As an alternative to IR, we may consider the aggregate value forecasting method, in which the following regression model is derived by adding all equations of \eqref{eq:allmodels}; that is,
\begin{align}
\sum_{m=1}^M\bm{y}_m &= \sum_{m=1}^M\bm{X}_m\bm{\beta}_m + \sum_{m=1}^M\bm{\varepsilon}_m\nonumber\\
&= (\bm{X}_1 \cdots \bm{X}_M) 
\begin{pmatrix}
\bm{\beta}_1\\
\vdots\\
\bm{\beta}_M\\
\end{pmatrix}
+ \sum_{m=1}^M\bm{\varepsilon}_m.
\label{eq:model1}
\end{align}
Model \eqref{eq:model1} is considered as one regression model:
\begin{align}
\bm{y} &= \bm{X}\bm{\beta} +\bm{\varepsilon}, \label{eq:model1-2}
\end{align}
where
\begin{align}
\bm{y}=\sum_{m=1}^M\bm{y}_m, \quad
\bm{X}=(\bm{X}_1 \cdots \bm{X}_M),  \quad 
\bm{\beta} = \begin{pmatrix}
\bm{\beta}_1\\
\vdots\\
\bm{\beta}_M\\
\end{pmatrix}
, \quad 
\bm{\varepsilon} = \sum_{m=1}^M\bm{\varepsilon}_m.
\end{align}
We call the model \eqref{eq:model1} an \textit{aggregate value regression (AVR)}. We can estimate parameter  $\bm{\beta}$ with AVR model (e.g., $\hat{\bm{\beta}}=(\bm{X}^\top\bm{X})^{-1}\bm{X}^T\bm{y}$) to get $\hat{\bm{\beta}}_1,\dots,\hat{\bm{\beta}}_M$.

The AVR in \eqref{eq:model1-2} overcomes the limitations of IR by incorporating all predictor variables of individuals, resulting in much more flexible and enriched framework compared to IR. However, this flexibility also has its drawbacks. AVR involves a significantly larger number of parameters, $pM$, than that of IR, especially when $M$ is large. Therefore, AVR often leads to high variability due to overparameterization, resulting in low prediction accuracy in some cases.

\subsection{Aggregate value regression with clustering}
To address the simplicity of IR and the overparameterization issues in AVR, we perform clustering on regression models and estimate AVR models for each cluster separately. We call this method AVR-C (``C" stands for Clustering). Suppose that $\mathcal{G}$ is a set of $M$ regression models and each component consists of regression models; that is, $\mathcal{G} = \{ \mathcal{M}_m \mid m=1,\dots, M  \}$, where $\mathcal{M}_m$ indicates the $m$th regression model, $\bm{y}_m = \bm{X}_m\bm{\beta}_m + \bm{\varepsilon}_m.$ Assume that we have a set of $k$ clusters of regression models, $\{\mathcal{C}_1,\dots,\mathcal{C}_k\}$, where each $\mathcal{C}_j$ ($j=1,\dots,k$) is a subset of $\mathcal{G}$.  The clusters must satisfy $\mathcal{C}_j \subset \mathcal{G}$ $(j=1,\dots,k)$, $\mathcal{C}_j \cap \mathcal{C}_{j'} = \phi$ ($j \neq j'$) and $ \bigcup_{j=1}^k \mathcal{C}_j =  \mathcal{G}$.  Let $m_j = \#\mathcal{C}_j$, and 
$\mathcal{T}_j := \{l_1,\dots,l_{m_j}\} \subset \{1,\dots,M\}$ 
is a set of indices corresponding to $j$th cluster of regression models. In this case, $\mathcal{C}_j$ is expressed as 
$\mathcal{C}_j=\{\mathcal{M}_{l_1},\dots,\mathcal{M}_{l_{m_j}}\}$. 
We perform the AVR for the clustered regression models
\begin{align}
\sum_{q=1}^{m_j}\bm{y}_{l_q} &= (\bm{X}_{l_1} \cdots \bm{X}_{l_{m_j}}) 
\begin{pmatrix}
\bm{\beta}_{l_1}\\
\vdots\\
\bm{\beta}_{l_{m_j}}\\
\end{pmatrix}
+ \sum_{q=1}^{m_j}\bm{\varepsilon}_{l_q}
\quad (j=1,\dots,k),
\label{eq:model1_cl}
\end{align}
and get the estimates of regression coefficients, $\hat{\bm{\beta}}_{\mathcal{T}_1},\dots,\hat{\bm{\beta}}_{\mathcal{T}_k}$, where $\bm{\beta}_{\mathcal{T}_j}:=
\begin{pmatrix}
\bm{\beta}_{l_1}\\
\vdots\\
\bm{\beta}_{l_{m_j}}
\end{pmatrix}
$. 

A significant advantage of the AVR-C over AVR is that the number of parameters can be decreased, thanks to the clustering of the regression models.  Indeed, the number of parameters for the $j$th regression model is $p \times m_j$ ($j=1,\dots,k$), much smaller than $p \times M$ when $m_j \ll M$.  Another notable feature of the AVR-C is that it bridges a gap between IR and AVR.   Specifically, the AVR-C on $k=M$ and $k=1$ correspond to IR and AVR, respectively. 
With an appropriate number of clusters $k$, we obtain a model that is much more flexible than IR while preventing the overfitting of AVR.

\begin{rem}
The AVR-C achieves clustering of the regression models.  Another method of clustering in regression modeling is the application of sparse estimation to the regression coefficients \citep{Reich.2008,Zhang.2013,Tang.2016}.  
By clustering regression coefficients, some of the regression coefficients are identical, resulting in the increase in the sample size.  As a result, the estimate of the regression coefficient become more stable.  Our proposed AVR-C is fundamentally different from the clustering of regression coefficients since the regression coefficients are different across the model.  
\end{rem}

\begin{rem}
The problem of estimating a collection of regression coefficients can be seen as multi-task learning \citep{caruana1997multitask, zhang2021survey}, which considers multiple regression models whose variables are related to each other across models.
In particular, a method of clustering tasks (models) that are similar to each other in the context of multi-task learning is called task clustering \citep{thrun1996discovering, han2015learning,fiot2016electricity,yousefi2019multi,bonetti2024interpetable}. 
However, in general, task clustering is not intended to provide predictions for aggregate values and does not consider errors according to the number of clusters, which is discussed in the next section.  
\end{rem}

\subsection{Estimation}
The forecast accuracy depends on the estimation procedure of regression coefficients of $\bm{\beta}_{\mathcal{T}_j}$ ($j=1,\dots,k$). For simplicity, let us drop the suffix ${\mathcal{T}_j}$ and simply consider a linear regression model, $\bm{y}=\bm{X}\bm{\beta}+\bm{\varepsilon}$. The ordinary least squares (OLS), $(\bm{X}^T\bm{X})^{-1}\bm{X}^T\bm{y}$, is often used, but it cannot be applied when the number of parameters exceeds the number of observations. 

As an alternative to the OLS, we employ the ridgeless least squares \citep{Hastie.2022}:
\begin{equation}
	\hat{\bm{\beta}} = \min_{\bm{\beta} \in \mathcal{B}} \| \bm{\beta} \|_2, \quad \mathcal{B}=\{\argmin_{\bm{\beta}} \| \bm{y} - \bm{X}\bm{\beta}\|^2\}. 
 \label{eq:ridgeless}
  \end{equation}
 The ridgeless least squares estimator corresponds to the ridge estimator with the regularization parameter converging to 0.   When $\bm{X}^T\bm{X}$ is positive-definite, the ridgeless least squares estimator is equivalent to the OLS because $\argmin_{\bm{\beta}} \| \bm{y} - \bm{X}\bm{\beta}\|^2$ has a unique solution, $\hat{\bm{\beta}}=(\bm{X}^T\bm{X})^{-1}\bm{X}^T\bm{y}$.  Meanwhile, if $\bm{X}^T\bm{X}$ is rank-deficient, the ridgeless least squares estimate is computed with Moore-Penrose inverse that provides fast computation with both {\tt R} and {\tt Python}. 
In Section \ref{hm:sec_3} below, we assume that $\bm{X}^T\bm{X}$ is invertible to make the theoretical analysis of the proposed estimation procedure more tractable.  Meanwhile, in Sections 4 and 5, we will compute the estimator $\hat{\bm{\beta}}$ through \eqref{eq:ridgeless} in high-dimensional settings to investigate the numerical behavior of our proposed method. We remark that high-dimensional settings are not the primary focus of this work; rather, the main purpose of this research is to elucidate how clustering structure impacts the prediction error, particularly in terms of the bias-variance trade-off.

 \begin{rem}
 Instead of ridgeless least squares, one may use the regularization procedure; however, in our experience, the ridgeless least squares can often result in good performance. The high accuracy of the ridgeless least squares can be attributed to two reasons: first, in the AVR, we observe a sufficient number of observations thanks to the advancements in collecting sensor data; the second reason is that, unlike the regularization methods, the ridgeless least squares does not include a regularization parameter, which is usually selected by standard model selection criteria such as the cross-validation \citep{stone1974cross}.  
 With a large number of regression models, the model selection criterion occasionally fails to select appropriate regularization parameter values, resulting in unstable aggregate value forecasts.  The tuning-free ridgeless least squares yields better performance in many cases.
\end{rem}

\section{Relationship between training and test errors and the number of clusters}
\label{hm:sec_3}
Clustering is typically used to group observations, facilitating the interpretation of the underlying data structure.  With the AVR, for each step of the hierarchical clustering, two regression models are combined to create a single regression model, and the AVR is performed based on this new model.  A key observation is that the AVR-C has a significant relationship between the number of clusters and the training and test errors. The training error is almost a monotone-increasing function of the number of clusters (Figure \ref{fig:MSEresicor} below).  Meanwhile, the test error can often exhibit a U-shape function. Thanks to this property, the number of clusters is viewed as a model complexity, providing a novel bias-variance trade-off property specifically to the AVR-C.  


To clarify how the bias-variance trade-off arises in AVR-C, we consider the case where the model is misspecified; the assumed model does not include the true model. Suppose that the true model is given as follows:
\begin{align}
		\bm{y}_m &= \bm{X}_m\bm{\beta}_m + \bm{W}_m\bm{\theta}_m + \bm{\varepsilon}_m \quad (m=1,\dots,M), \label{eq:true_misspedified}
\end{align}
where $\bm{W}_m = (\bm{w}_{m,1},\dots,\bm{w}_{m,n})^\top$ ($m=1\dots,M$) are additional unobserved random predictors independent of $\bm{\varepsilon}_m$, and $\bm{\theta}_m$ are regression coefficient vectors that corresponds to $\bm{W}_m$.  We assume that the error variables between different regression models are allowed to be correlated, that is, ${\rm Cov}(\bm{\varepsilon}_m, \bm{\varepsilon}_{m'}) = \bm{\Sigma}_{mm'}$  ($m,m'=1\dots,M$). 
We also assume that $\bm{w}_{m,1},\dots,\bm{w}_{m,n}$ are mutually independent with $E[\bm{w}_{m,i}]=\bm{0}$ and $V[\bm{w}_{m,i}]=\bm{\Sigma}_{\bm{W}_m}$ ($m=1,\dots,M; i=1,\dots,n$), where $\bm{\Sigma}_{\bm{W}_m}$ 
is a positive-definite matrix. 

Our assumed model is
\begin{align}
		\bm{y}_m &= \bm{X}_m\bm{\beta}_m + \bm{\varepsilon}_m  \quad (m=1,\dots,M),\label{eq:assume_misspedified}
\end{align}
and the regression coefficients are estimated through the AVR-C. In this section, we assume that $n>Mp$ and design matrix based on the AVR-model, say $\bm{X}$, is full-rank; that is, $\bm{X}^T\bm{X}$ is invertible.  Under this assumption, the ridgeless least squares estimator is equivalent to the OLS estimator.  We remark that the AVR-C algorithm is applicable without such an assumption; it is only for the construction of bias-variance trade-off theory in the AVR-C context. 

\begin{rem}
It appears that the mean structure corresponding to the misspecified part, $\bm{W}_m\bm{\theta}_m$, can be simply represented by a single symbol $\bm{\xi}_m$. However, the form $\bm{W}_m\bm{\theta}_m$ is crucial for understanding the bias of AVR-C. We will illustrate how the bias is reduced through the aggregation of regression models in Example \ref{example:X1W2} presented in Section \ref{sec:Test error}.
\end{rem}

\subsection{Training error}

\subsubsection{Two regression models}
\label{hm:sec_theory}
To delve into the relationship between training error and the number of clusters, we first consider a simple case where the number of regression models is two, and investigate how the training error changes as two regression models are combined into one AVR model. 

\begin{prop}
We assume two regression models:
	\begin{align}
\bm{y}_1 &= \bm{X}_1\bm{\beta}_1 + \bm{\varepsilon}_1, \label{eq:lemreg-1}\\
\bm{y}_2 &= \bm{X}_2\bm{\beta}_2 + \bm{\varepsilon}_2.\label{eq:lemreg-2}
\end{align}
  Let $\bm{y} = \bm{y}_1+\bm{y}_2$.  The forecast values of $\bm{y}$ based on the AVR and IR are defined as $\hat{\bm{y}}$ and $\hat{\bm{y}}^*$, respectively.  We then obtain the following inequality:
\begin{equation*}
	\|\bm{y}-\hat{\bm{y}}\|_2^2 \leq \|\bm{y}-\hat{\bm{y}}^*\|_2^2.
\end{equation*}
\label{prop:biasvar}
\end{prop}
Proposition \ref{prop:biasvar} indicates training error decreases as the number of clusters decreases when two regression models are combined into one AVR model.  
We remark that whether the assumed models in \eqref{eq:lemreg-1} and \eqref{eq:lemreg-2} are misspecified or not is irrelevant to the proof of this proposition.

\begin{proof}
Let $\bm{X}=(\bm{X}_1,\bm{X}_2)$, and $\bm{\beta}=(\bm{\beta}_1^T,\bm{\beta}_2^T)^T$.  Let $\hat{\bm{\beta}}$ and $\hat{\bm{\beta}}^*$ be the least squares estimator of AVR and IR, respectively. Since $\hat{\bm{\beta}}$ is the minimizer of the least squares function, $\| \bm{y} - \bm{X}\bm{\beta}\|^2$, we have
\begin{eqnarray*}
	\|\bm{y}-\hat{\bm{y}}\|_2^2 &=& \|\bm{y}-\bm{X}\hat{\bm{\beta}}\|_2^2 \\
	&=& \|\bm{y}-\bm{X}_1\hat{\bm{\beta}}_1-\bm{X}_2\hat{\bm{\beta}}_2\|_2^2  \\
	&\leq& \|\bm{y}-\bm{X}_1\hat{\bm{\beta}}_1^*-\bm{X}_2\hat{\bm{\beta}}_2^*\|_2^2  \\
	&=& \|\bm{y}-\hat{\bm{y}}^*\|_2^2,
\end{eqnarray*}
which concludes the proof.
\end{proof}

\subsubsection{More than two regression models}
Now, we investigate the training error when the number of regression models is more than two. From the result of Proposition \ref{prop:biasvar}, it is expected that the monotonicity of the training error is shown for any $k>2$; if this is true, the training error is a monotone decreasing function as a function of $k$.   This conjecture is roughly correct in reality, but not always; we will need a condition that the errors are nonnegatively correlated. Our empirical studies of electricity demand forecasting show that the condition can hold in reality, suggesting that the nearly monotonicity of the training error with respect to $k$ holds in most cases.  

Before investigating a general case for arbitrary number of clusters, we first consider the case where there exist three regression models.  We compare the training errors of AVR and IR when the first two regression models are combined.  
\begin{lem}
	Consider three regression models:
	\begin{align}
\bm{y}_m &= \bm{X}_m\bm{\beta}_m + \bm{\varepsilon}_m \quad (m=1,2,3). \label{eq:lemreg2-1}
\end{align}
Let $\bm{y} = \bm{y}_1+\bm{y}_2$.  The forecast values of $\bm y$ based on the AVR and IR are defined as $\hat{\bm{y}}$ and $\hat{\bm{y}}^*$, respectively.  Also, the forecast value of $\bm{y}_3$ is denoted as  $\hat{\bm{y}}_3$.  Let $R$ be
\begin{align}
	R:=(\bm{y}_3 - \hat{\bm{y}}_3)^T\{(\bm{y} - \hat{\bm{y}}) - (\bm{y} - \hat{\bm{y}}^*)\},\label{eq:R}
\end{align}
and we assume that
\begin{equation}
	R \leq 0 \qquad\text{a.s.}
    \label{assumption:lem}
\end{equation}
Then we obtain the following inequality:
\begin{equation*}
	\|(\bm{y}+\bm{y}_3)-(\hat{\bm{y}}+\hat{\bm{y}}_3)\|_2^2 \leq \|(\bm{y}+\bm{y}_3)-(\hat{\bm{y}}^*+\hat{\bm{y}}_3)\|_2^2 \qquad\text{a.s.}
\end{equation*}
\label{lem:biasvar2}
\end{lem}
\begin{proof}
We can directly prove the inequality by making use of Proposition \ref{prop:biasvar} and an assumption \eqref{assumption:lem} as follows:
\begin{align}
		&\|(\bm{y}+\bm{y}_3)-(\hat{\bm{y}}+\hat{\bm{y}}_3)\|_2^2 - \|(\bm{y}+\bm{y}_3)-(\hat{\bm{y}}^*+\hat{\bm{y}}_3)\|_2^2\nonumber\\
		&=\|\bm{y}-\hat{\bm{y}}\|_2^2  - \|\bm{y}-\hat{\bm{y}}^*\|_2^2 + 2(\bm{y}_3 - \hat{\bm{y}}_3)^T\{(\bm{y} - \hat{\bm{y}}) - (\bm{y} - \hat{\bm{y}}^*)\} \label{eq:training3reg}\\ 
		&\leq 0\nonumber
		\end{align}
\end{proof}
Lemma \ref{lem:biasvar2} implies that under the assumption \eqref{assumption:lem}, the training error decreases when two regression models are combined into one regression model. However, it would not be easy to comprehend the meaning of the assumption \eqref{assumption:lem}.  To clarify the meaning of this assumption, we evaluate it with respect to its expectation.  The following lemma shows that the assumption \eqref{assumption:lem} holds with respect to its expectation when the errors are non-negatively correlated.

\begin{lem}
Assume that the three true models are given by
\begin{align}
		\bm{y}_m &= \bm{X}_m\bm{\beta}_m + \bm{W}_m\bm{\theta}_m + \bm{\varepsilon}_m \quad (m=1,2,3),
  \label{eq:latentmodel}
\end{align}
while our assumed models are \eqref{eq:lemreg2-1}.	Let $\bm{X} = (\bm{X}_1, \bm{X}_2)$,  $\bm{\varepsilon} = \bm{\varepsilon}_1+\bm{\varepsilon}_2$, $\bm{H}=\bm{X}(\bm{X}^T\bm{X})^{-1}\bm{X}^T$ and  $\bm{H}_m=\bm{X}_m(\bm{X}_m^T\bm{X}_m)^{-1}\bm{X}_m^T$ $(m=1,2,3)$.  Then, the expected value of $R$ defined by Eq. \eqref{eq:R} is expressed as follows:
	\begin{align}
			E[R \mid \bm{W}_1,\bm{W}_2,\bm{W}_3] &= \tr\{(\bm{I}_n - \bm{H}_3)(\bm{H}_1 - \bm{H})\bm{\Sigma}_{13}\} + \tr\{(\bm{I}_n - \bm{H}_3)(\bm{H}_2 - \bm{H})\bm{\Sigma}_{23}\} \nonumber \\
			 &\quad+(\bm{W}_3\bm{\theta}_3)^T(\bm{I}-\bm{H}_3)(\bm{H}_1 - \bm{H})(\bm{W}_1\bm{\theta}_1)   
             \nonumber\\
             &\quad +(\bm{W}_3\bm{\theta}_3)^T(\bm{I}-\bm{H}_3)(\bm{H}_2-\bm{H})(\bm{W}_2\bm{\theta}_2).  
			 \label{eq:proof_W3}
		\end{align}
Furthermore, taking the expectation with respect to $\bm{W}_1$, $\bm{W}_2$, and $\bm{W}_3$, we have   
	\begin{align*}
			E[R]&= \tr\{(\bm{I}_n - \bm{H}_3)(\bm{H}_1 - \bm{H})\bm{\Sigma}_{13}\} + \tr\{(\bm{I}_n - \bm{H}_3)(\bm{H}_2 - \bm{H})\bm{\Sigma}_{23}\}.		\end{align*}
In particular, we have $E[R] \leq 0$ if $\bm{\Sigma}_{13} \geq 0$ and $\bm{\Sigma}_{23} \geq 0$.
\label{lemma:training_3reg_misspecified}
\end{lem}
\begin{proof}
	The proof is provided in Appendix \ref{proof:lemma:training_3reg_misspecified}.
\end{proof}

By Lemma \ref{lemma:training_3reg_misspecified}, the training error is decreased with respect to its expectation when two regression models are combined into one regression model. 

We can immediately generalize the result of Lemma \ref{lemma:training_3reg_misspecified}.
\begin{thm}
Suppose that we have $k$ clusters of the following true regression models
\begin{align}
\bm{y}_{\mathcal{T}_j} &= \bm{X}_{\mathcal{T}_j}\bm{\beta}_{\mathcal{T}_j}+\bm{W}_{\mathcal{T}_j}\bm{\theta}_{\mathcal{T}_j}+ \bm{\varepsilon}_{\mathcal{T}_j}, \quad j=1,\dots,k,\label{eq:lemreg-clust}
\end{align}
and our assumed models
\begin{align}
\bm{y}_{\mathcal{T}_j} &= \bm{X}_{\mathcal{T}_j}\bm{\beta}_{\mathcal{T}_j}+ \bm{\varepsilon}_{\mathcal{T}_j}, \quad j=1,\dots,k.
\end{align}
The forecast values are denoted by $\hat{\bm{y}}_{\mathcal{T}_j}$ ($j=1,\dots,k$).  Without loss of generality, assume that the clusters of $\bm{y}_{\mathcal{T}_{1}}$ and $\bm{y}_{\mathcal{T}_2}$ are combined, and the regression coefficients are estimated through the AVR:
\begin{align}
\bm{y}_{\mathcal{T}_{1,2}} &= 
\begin{pmatrix}
\bm{X}_{\mathcal{T}_{1}} & \bm{X}_{\mathcal{T}_{2}}
\end{pmatrix}
\begin{pmatrix}
	\bm{\beta}_{\mathcal{T}_{1}}\\
	\bm{\beta}_{\mathcal{T}_{2}}
\end{pmatrix}
+ \bm{\varepsilon}_{\mathcal{T}_{1,2}},
\label{eq:lemreg-clust2}
\end{align}
where $\bm{y}_{\mathcal{T}_{1,2}}=\bm{y}_{\mathcal{T}_{1}} + \bm{y}_{\mathcal{T}_{2}}$ and $\bm{\varepsilon}_{\mathcal{T}_{1,2}}=\bm{\varepsilon}_{\mathcal{T}_{1}} + \bm{\varepsilon}_{\mathcal{T}_{2}}$.  Assume that ${\rm Cov}(\bm{\varepsilon}_{\mathcal{T}_{1}},\bm{\varepsilon}_{\mathcal{T}_{3}}) \geq 0$ and ${\rm Cov}(\bm{\varepsilon}_{\mathcal{T}_{2}},\bm{\varepsilon}_{\mathcal{T}_{3}})  \geq 0$. 
  Denote the corresponding forecast value by $\hat{\bm{y}}_{\mathcal{T}_{1,2}}$.  
Then, we obtain
\begin{align*}
	E\left[\left\|\sum_{j=1}^{k}\bm{y}_{\mathcal{T}_{j}}-\left(\hat{\bm{y}}_{\mathcal{T}_{1,2}} + \sum_{j=3}^{k}\hat{\bm{y}}_{\mathcal{T}_{j}} \right)\right\|^2\right]	\leq  
E\left[\left\|\sum_{j=1}^{k}\bm{y}_{\mathcal{T}_{j}} - \sum_{j=1}^{k}\hat{\bm{y}}_{\mathcal{T}_{j}} \right\|^2\right]. 
\end{align*}
\label{thm:bias-variance}
\end{thm}
\begin{proof}
Theorem \ref{thm:bias-variance} can be directly proved from Lemma \ref{lemma:training_3reg_misspecified} by setting $\bm{y}_1 = \bm{y}_{\mathcal{T}_{1}}$, $\bm{y}_2 = \bm{y}_{\mathcal{T}_{2}}$, and $\bm{y}_3 = \sum_{j=3}^{k}\bm{y}_{\mathcal{T}_{j}}$.  
\end{proof}
Theorem \ref{thm:bias-variance} indicates that when we perform hierarchical clustering on regression models, the training error is decreased when two regression models are combined into one AVR model.  More specifically, in the $(n-k+1)$th step of the hierarchical clustering algorithm ($k=n,n-1,\dots,2$), the expectation of the residual sum of squares is a monotone non-increasing function of $k$ under the assumption that the errors are nonnegatively correlated.  

\subsubsection{Numerical investigation}\label{sec:Numerical investigation}
Now, we consider a small numerical example to illustrate how the error covariance, $\bm{\Sigma}_{mm'}$, plays a role in the training error. We assume 50 regression models with $n=500$, and the predictor variables are constructed by random numbers from a $p=5$-dimensional i.i.d. uniform  distribution $U(0,3)$. The correlation pattern among regression models is $\bm{\Sigma}_{mm'} = \sigma_{mm'}\bm{I}_n$ for $m \neq m'$. Here, we set three types of correlation structures, that is, (i) $\sigma_{mm'} = (-0.8)^{|m-m'|}$, (ii) $\sigma_{mm'} =0.0$, and (iii) $\sigma_{mm'} = 0.6$.  For all settings, $\sigma_{mm}=1$.  We set $\bm{W}_m\bm{\theta}_m=\bm{0}$ in the true regression models for simplicity. 

Figure \ref{fig:MSEresicor} shows the MSE of training error along with the correlation heatmaps of residuals.  The residual is an estimate of errors; thus, the heatmap is an approximation of the correlation of errors. For each step of the hierarchical clustering algorithm, a new cluster is constructed such that the MSE is minimized.   On the top side of Figure \ref{fig:MSEresicor}, we observe that the training error is almost a monotone increasing function as a function of $k$ with correlation structures (ii) and (iii); meanwhile with (i), the MSE is not a monotone function as a function of $k$. 

The key point of the monotonicity of the training error is the correlation structure of error vectors.  As show in the Theorem \ref{thm:bias-variance}, the MSE of training error is a monotone increasing function as a function of the number of clusters when the error correlations are positive. With conditions (ii) and (iii), the error vectors appear to have almost no negative correlations, and then the condition of Theorem \ref{thm:bias-variance} may hold. On the other hand, the correlation structure of (i) is an AR(1) model with an AR coefficient being $-0.8$, implying that some of the errors are negatively correlated.   As a result, the training error may not always increase as the number of clusters increases. 

\begin{figure}[!t]
\centering
\includegraphics[width=\textwidth, bb=0 0 1031 651]{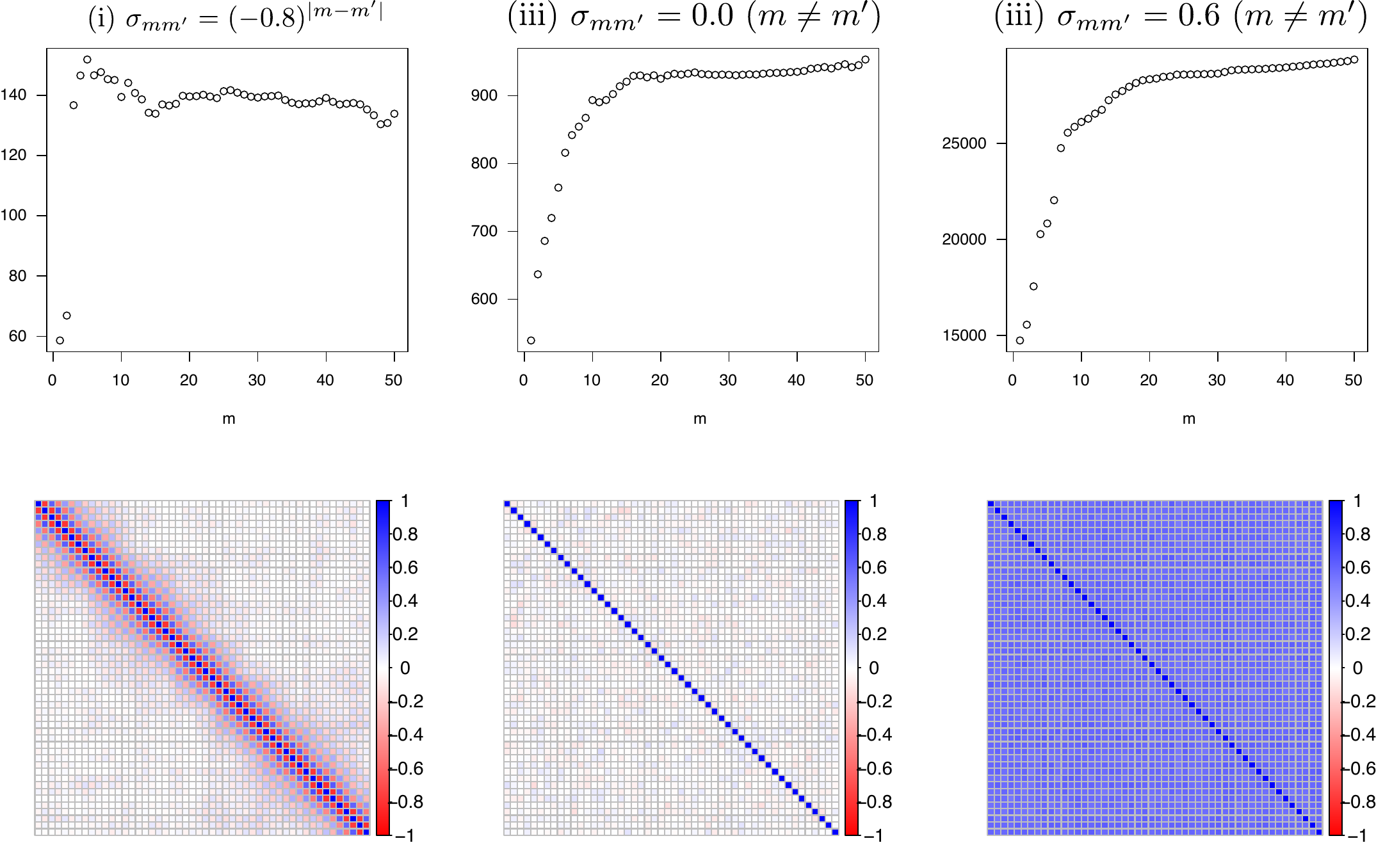}
\caption{MSE of training error (top side) along with correlation heatmap of residuals (bottom side).   We consider three types of correlation: (i) $\sigma_{mm'} = (-0.8)^{|m-m'|}$, (ii) $\sigma_{mm'} =0.0$ ($m \neq m'$), and (iii) $\sigma_{mm'} = 0.6$ ($m \neq m'$). For all settings, $\sigma_{mm}=1$.  Here, $\sigma_{mm'}$ is the covariance of errors; ${\rm Cov}(\bm{\varepsilon}_m, \bm{\varepsilon}_{m'}) = \sigma_{mm'}\bm{I}_n$.}\label{fig:MSEresicor}
\end{figure}

In summary, the non-negative correlations among errors play a key role in the monotonicity of the training error.  As mentioned before, in our empirical experiments of electricity demand forecast, the error covariances are largely positive, which is to be expected, and thus Theorem \ref{thm:bias-variance} might hold.  

\subsection{Test error}\label{sec:Test error}
We investigate the test error as a function of the number of clusters.  Specifically, it is sufficient to focus on the case where two regression models are combined when there are three regression models. Similar properties hold for cases with more than three regression models, as demonstrated in the proof of Theorem \ref{thm:bias-variance}.
\begin{thm}
	Consider a set of test data generated from the following regression models:
	\begin{align}
\bm{z}_m &= \bm{X}_m\bm{\beta}_m + \bm{W}_m\bm{\theta}_m+\bm{\eta}_m \quad (m=1,2,3). 
\end{align}
Let $\bm{z} = \bm{z}_1+\bm{z}_2$. We then evaluate the test errors as below:
\begin{align}
&E\left[\|(\bm{z}+\bm{z}_3)-(\hat{\bm{y}}+\hat{\bm{y}}_3)\|_2^2 \mid \bm{W}_1,\bm{W}_2,\bm{W}_3\right] - E\left[\|(\bm{z}+\bm{z}_3)-(\hat{\bm{y}}^*+\hat{\bm{y}}_3)\|_2^2\mid \bm{W}_1,\bm{W}_2,\bm{W}_3\right]\crcr
&= \tr\{(\bm{H}-\bm{H}_1)\bm{\Sigma}_{11}\} 
+ \tr\{(\bm{H}-\bm{H}_2)\bm{\Sigma}_{22}\}
+2\tr\{(\bm{H}-\bm{H}_1\bm{H}_2)\bm{\Sigma}_{12}\}\crcr
&\quad 
+2\tr\{(\bm{H}-\bm{H}_1)\bm{H}_3\bm{\Sigma}_{13}\}
+ 2\tr\{(\bm{H}-\bm{H}_2)\bm{H}_3\bm{\Sigma}_{23}\}\crcr
&\quad+ \|(\bm{I}-\bm{H})\bm{W}_1\bm{\theta}_1\|^2-\|(\bm{I}-\bm{H}_1)\bm{W}_1\bm{\theta}_1\|^2
+ \|(\bm{I}-\bm{H})\bm{W}_2\bm{\theta}_2\|^2-\|(\bm{I}-\bm{H}_2)\bm{W}_2\bm{\theta}_2\|^2\crcr
&\quad+2(\bm{W}_1\bm{\theta}_1)^T(\bm{I}-\bm{H})(\bm{W}_2\bm{\theta}_2)   - 2(\bm{W}_1\bm{\theta}_1)^T(\bm{I}-\bm{H}_1)(\bm{I}-\bm{H}_2)(\bm{W}_2\bm{\theta}_2)\crcr	 &\quad+2(\bm{W}_3\bm{\theta}_3)^T(\bm{I}-\bm{H}_3)(\bm{H}_1 - \bm{H})(\bm{W}_1\bm{\theta}_1)   +2(\bm{W}_3\bm{\theta}_3)^T(\bm{I}-\bm{H}_3)(\bm{H}_2-\bm{H})(\bm{W}_2\bm{\theta}_2).  \label{eq:testerror_noex}
\end{align}
Furthermore, taking the expectation with respect to $\bm{W}_1$, $\bm{W}_2$, and $\bm{W}_3$, we have   
\begin{align}
&E[\|(\bm{z}+\bm{z}_3)-(\hat{\bm{y}}+\hat{\bm{y}}_3)\|_2^2] - E[\|(\bm{z}+\bm{z}_3)-(\hat{\bm{y}}^*+\hat{\bm{y}}_3)\|_2^2]\crcr
&= \tr\{(\bm{H}-\bm{H}_1)\bm{\Sigma}_{11}\} 
+ \tr\{(\bm{H}-\bm{H}_2)\bm{\Sigma}_{22}\}
+2\tr\{(\bm{H}-\bm{H}_1\bm{H}_2)\bm{\Sigma}_{12}\}\crcr
&\quad 
+2\tr\{(\bm{H}-\bm{H}_1)\bm{H}_3\bm{\Sigma}_{13}\}
+ 2\tr\{(\bm{H}-\bm{H}_2)\bm{H}_3\bm{\Sigma}_{23}\}\crcr
   & \quad  -p\bm{\theta}_1^T\bm{\Sigma}_{\bm{W}_1}\bm{\theta}_1 -p\bm{\theta}_2^T\bm{\Sigma}_{\bm{W}_2}\bm{\theta}_2. \label{eq:difference_of_predictionerror}
\end{align}  
\label{thm:correlation_test_misspecified}
\end{thm}
\begin{proof}
	The proof is provided in Appendix \ref{proof:thm:correlation_test_misspecified}.
\end{proof}
Theorem \ref{thm:correlation_test_misspecified} implies that when we assume that $\bm{\Sigma}_{12} \geq 0$, $\bm{\Sigma}_{13} \geq 0$, and $\bm{\Sigma}_{23} \geq 0$, we obtain the following inequalities:
\begin{align*}
 \tr\{(\bm{H}-\bm{H}_1)\bm{\Sigma}_{11}\} 
+ \tr\{(\bm{H}-\bm{H}_2)\bm{\Sigma}_{22}\}
+2\tr\{(\bm{H}-\bm{H}_1\bm{H}_2)\bm{\Sigma}_{12}\}&\crcr
+2\tr\{(\bm{H}-\bm{H}_1)\bm{H}_3\bm{\Sigma}_{13}\}
+ 2\tr\{(\bm{H}-\bm{H}_2)\bm{H}_3\bm{\Sigma}_{23}\}
&\geq 0,\\
-p\bm{\theta}_1^T\bm{\Sigma}_{\bm{W}_1}\bm{\theta}_1 -p\bm{\theta}_2^T\bm{\Sigma}_{\bm{W}_2}\bm{\theta}_2 &\leq 0.
\end{align*}
Thus, test error in \eqref{eq:difference_of_predictionerror} is either increased or decreased with respect to its expectation when two regression models are combined into one regression model. 

To further understand the behavior of the test error, we examine the conditions under which the test error decreases when two regression models are combined into one regression model. Specifically, we have a look at the following term in \eqref{eq:testerror_noex}:
\begin{align*}
&\|(\bm{I}-\bm{H})\bm{W}_1\bm{\theta}_1\|^2-\|(\bm{I}-\bm{H}_1)\bm{W}_1\bm{\theta}_1\|^2
+ \|(\bm{I}-\bm{H})\bm{W}_2\bm{\theta}_2\|^2-\|(\bm{I}-\bm{H}_2)\bm{W}_2\bm{\theta}_2\|^2\crcr
&\quad+2(\bm{W}_1\bm{\theta}_1)^T(\bm{I}-\bm{H})(\bm{W}_2\bm{\theta}_2)   - 2(\bm{W}_1\bm{\theta}_1)^T(\bm{I}-\bm{H}_1)(\bm{I}-\bm{H}_2)(\bm{W}_2\bm{\theta}_2)\crcr	 &\quad+2(\bm{W}_3\bm{\theta}_3)^T(\bm{I}-\bm{H}_3)(\bm{H}_1 - \bm{H})(\bm{W}_1\bm{\theta}_1)   +2(\bm{W}_3\bm{\theta}_3)^T(\bm{I}-\bm{H}_3)(\bm{H}_2-\bm{H})(\bm{W}_2\bm{\theta}_2).  \end{align*}
For simplicity, we assume $\bm{X}_1$, $\bm{X}_2$, and $\bm{X}_3$ are uncorrelated; that is, $\bm{X}_1^\top\bm{X}_2=\bm{O}$, $\bm{X}_2^\top\bm{X}_3=\bm{O}$, and $\bm{X}_3^\top\bm{X}_1=\bm{O}$.  In  addition, we temporarily assume $\bm{W}_1^\top\bm{W}_3=\bm{O}$ and $\bm{W}_2^\top\bm{W}_3=\bm{O}$; although this is not consistent with our current assumption that $\bm{W}_j$ are random ($j=1,2,3$), it makes the exposition concise and will approximately hold if the covariances among $\bm{W}_1$, $\bm{W}_2$, and $\bm{W}_3$ are zero. Under these conditions, the last display becomes
\begin{align*}
&\|(\bm{I}-\bm{H})\bm{W}_1\bm{\theta}_1\|^2-\|(\bm{I}-\bm{H}_1)\bm{W}_1\bm{\theta}_1\|^2
+ \|(\bm{I}-\bm{H})\bm{W}_2\bm{\theta}_2\|^2-\|(\bm{I}-\bm{H}_2)\bm{W}_2\bm{\theta}_2\|^2.
\end{align*}
Let $D$ be the first two terms
\[
D:=\|(\bm{I}-\bm{H})\bm{W}_1\bm{\theta}_1\|^2-\|(\bm{I}-\bm{H}_1)\bm{W}_1\bm{\theta}_1\|^2.
\]
It should be noted that $D \leq 0$ holds because $(\bm{I}-\bm{H})$ projects onto a narrower space than $(\bm{I}-\bm{H}_1)$. 

For example, consider the case where $\bm{W}_1$ and $ \bm{X}_1$ are aligned in the same direction (e.g., $\bm{W}_1 = \bm{X}_1$). In such a scenario, $D=0-0=0$, implying that $D$ is maximized (i.e., test error may not be significantly reduced).  On the other hand, when $\bm{W}_1$ and $\bm{X}_2$ are aligned in the same direction (e.g., $\bm{W}_1 = \bm{X}_2$), we have
\[
D=-\|(\bm{I}-\bm{H}_1)\bm{W}_1\bm{\theta}_1\|^2 \leq 0.
\]
Thus, $D$ is always nonpositive, implying that the test error can be more frequently decreased compared to the case where $\bm{W}_1 = \bm{X}_1$. Similarly, the test error may be decreased when $\bm{W}_2$ and $\bm{X}_1$ are aligned in the same direction.

\begin{example}
We provide an intuitive explanation of why the prediction accuracy is improved when $\bm{W}_1$ and $\bm{X}_2$ are aligned in the same direction.  We assume the two true regression models as follows:
\begin{align*}
\bm{y}_1& = \bm{X}_1\bm{\beta}_1 + \bm{W}_1\bm{\theta}_1 + \bm{\varepsilon}_1,\\
\bm{y}_2 &= \bm{X}_2\bm{\beta}_2 + \bm{W}_2\bm{\theta}_2 + \bm{\varepsilon}_2.
\end{align*}
Suppose that $\bm{W}_1 = \bm{X}_2$. Additionally, for ease of presentation, we also assume that $\bm{W}_2 = \bm{X}_1$. Adding the above two regression models together, we have
\begin{align*}
\bm{y}_1 + \bm{y}_2 = \bm{X}_1(\bm{\beta}_1+\bm{\theta}_2)  + \bm{X}_2(\bm{\beta}_2 + \bm{\theta}_1) +\bm{\varepsilon}_1 + \bm{\varepsilon}_2
\end{align*}
The above AVR model turns out to be a correctly specified model, whereas the IR model is misspecified. Thus, the AVR model appropriately captures the effects of $\bm{W}_1\bm{\theta}_1$ and $\bm{W}_2\bm{\theta}_2$ through aggregation, allowing for more accurate estimation of the regression coefficients. As a result, the AVR model leads to improved prediction accuracy compared to the IR model.
\label{example:X1W2}
\end{example}
\begin{rem}
	When the model is correctly specified, that is, when $\bm{W}_m\bm{\theta}_m=\bm{0}$ for $m=1,2,3$, the difference of expected test error in \eqref{eq:difference_of_predictionerror} is 
	\begin{align*}
&E[\|(\bm{z}+\bm{z}_3)-(\hat{\bm{y}}+\hat{\bm{y}}_3)\|_2^2] - E[\|(\bm{z}+\bm{z}_3)-(\hat{\bm{y}}^*+\hat{\bm{y}}_3)\|_2^2]\\
&= \tr\{(\bm{H}-\bm{H}_1)\bm{\Sigma}_{11}\} 
+ \tr\{(\bm{H}-\bm{H}_2)\bm{\Sigma}_{22}\}
+2\tr\{(\bm{H}-\bm{H}_1\bm{H}_2)\bm{\Sigma}_{12}\}\crcr
&\quad 
+2\tr\{(\bm{H}-\bm{H}_1)\bm{H}_3\bm{\Sigma}_{13}\}
+ 2\tr\{(\bm{H}-\bm{H}_2)\bm{H}_3\bm{\Sigma}_{23}\}.
\end{align*}
The right-hand side is nonnegative as long as $\bm{\Sigma}_{mm'}\geq 0$, implying the test error is increased when two regression models are combined into one regression model.  More generally, in the AVR-C, the test error is a monotone-decreasing function as a function of the number of clusters.  Thus, performing the cluster analysis makes the prediction error worsen; we only need to perform the regression individually to get the best prediction error.  However, when the model is misspecified, which is often the case in reality, the test error may not be monotone-increasing function;  in practice, it can often be a U-shape function. 
\end{rem}
As mentioned at the beginning of this section, our AVR-C is viewed as the bias-variance trade-off, where the number of clusters represents the model complexity.  In the AVR-C framework, the bias-variance trade-off theory is constructed in the context of the misspecified model.  The bias of the OLS arises due to the assumption of a misspecified model. Although the OLS is biased under the IR framework, as demonstrated in example \ref{example:X1W2}, the bias can be reduced through the aggregation of regression models. On the other hand, when two regression models are combined, the parameters included in these two regression models must be estimated simultaneously, resulting in a larger number of parameters to be estimated.  As a result, the variance of the OLS estimator is increased.  Consequently, in the AVR-C framework, there exists a bias-variance trade-off, where the model complexity is determined by the number of clusters.

{
\subsection{Clustering method}
The result of AVR-C is highly dependent on the clustering method.  One can apply an ordinary hierarchical clustering method, such as Ward's method, to the original data points. However, the ordinary clustering is an unsupervised learning and does not take into account prediction accuracy. Since the purpose of clustering in our setting is to achieve high prediction accuracy of the regression model, it would be reasonable to perform clustering such that training or test error is minimized.  
Here, we propose two clustering methods to identify clusters of regression models.  

\subsubsection{Training Error Minimization (TEM)}
\label{sec:TEM}
We now describe the Training Error Minimization (TEM) procedure. Starting from $M$ regression models, we combine two regression models sequentially to create clusters based on the training or test errors.  Unfortunately, these errors are quite complicated as in Lemma \ref{lemma:training_3reg_misspecified} and Theorem \ref{thm:correlation_test_misspecified} and are unavailable in practice due to the dependence on unknown parameters. Thus, it would be difficult to analytically find a cluster such that training or test error is minimized.  One of the approaches to address this issue is to simply compute the training error numerically at each step of hierarchical clustering, and select a model that minimizes a training error. We call the algorithm training error minimization (TEM) algorithm. The details of the TEM algorithm are described in Algorithm \ref{alg:MSE}.
\begin{algorithm}[t]
\caption{Training Error Minimization (TEM) algorithm.}
\label{alg:MSE}
\begin{algorithmic}[1]
\STATE \textbf{Input:} $M$ regression models
\STATE Initialize $M$ clusters, each containing one regression model
\WHILE{number of clusters $> 1$}
    \STATE For each pair of clusters, compute the training error of the combined regression model
    \STATE Select a pair of clusters that minimizes the total training error
    \STATE Combine the selected pair into a new cluster
\ENDWHILE
\STATE \textbf{Output:} Clusters of regression models constructed by each step of the algorithm
\end{algorithmic}
\end{algorithm}

The TEM algorithm aims at minimizing the training error, allowing us to create a cluster that leads to a small prediction error.  However, Algorithm \ref{alg:MSE} is generally slow due to the computation of training error for each pair of regression models.  Indeed, for each step of the clustering algorithm in lines 4 of Algorithm \ref{alg:MSE}, we need to perform $\frac{m(m-1)}{2}$ regression analyses, resulting in a total of $\sum_{m=1}^M \frac{m(m-1)}{2} = \frac{M(M^2-1)}{6} = O(M^3)$ regression analyses to obtain the clustering result.  The computation is generally heavy when $M$ is large, but parallel computing would be helpful to alleviate the computational issue.

\subsubsection{Residual Correlation Matrix (RCM) algorithm}
\label{sec:RCM}
We introduce another clustering method that is more computationally efficient and also easy to interpret the clustering results.  Specifically, each cluster is constructed in such a way that the training error exhibits a similar distribution within the same cluster.  We first compute the correlation matrix of the residuals, say $\bm{R}=(r_{ij})$.  Here, $r_{ij}$ is the sample correlation coefficient between residuals of $i$th and $j$th regression models. Then, we perform the Ward's method using $\bm{R}^{-}$ as a distance matrix, where $\bm{R}^{-}$ is an $M \times M$ matrix whose $(i,j)$th elements is $1-r_{ij}$. Note that $\sqrt{1-r_{ij}}$ satisfies the axioms of distance.  
We call the above algorithm residual correlation matrix (RCM) algorithm. The details of the RCM algorithm are described in Algorithm \ref{alg:RCM}.

As $r_{ij}$ indicates the correlation coefficient between residuals of $i$th and $j$th regression models,  $1-r_{ij}$ indicates how far the correlation coefficient is from 1.  With the RCM algorithm, the residuals within the same cluster are expected to exhibit large positive correlations, reflecting similar residual patterns.  Thus, the RCM algorithm may lead to interpretable clusters in terms of similar residual structures. 
\begin{algorithm}[t]
\caption{Residual Correlation Matrix (RCM) algorithm.}
\label{alg:RCM}
\begin{algorithmic}[1]
\STATE \textbf{Input:} $M$ regression models
\STATE Compute the residuals for each regression model
\STATE Compute the correlation matrix of residuals, $\bm{R} =(r_{ij})$.  Here,  $r_{ij}$ is the correlation coefficient between the residuals of the $i$th and $j$th regression models.
\STATE Compute an $M \times M$ matrix $\bm{R}^{-}$.  Here, $(\bm{R}^{-})_{ij} =(1 - r_{ij})$.
\STATE Perform the standard clustering algorithm, such as Ward's method, on the matrix $\bm{R}^{-}$ to identify clusters of regression models.
\STATE \textbf{Output:} Clusters of regression models constructed by each step of the algorithm
\end{algorithmic}
\end{algorithm}

It would be worth investigating how the RCM algorithm affects the training error when two regression models are combined into one regression model.  To this end, we evaluate the reduction in training error by combining two regression models, specifically with respect to its expectation.  

\begin{prop}
We assume two regression models as in Proposition \ref{prop:biasvar}. The difference in training errors between AVR and IR with respect to its expectations is expressed as 
\begin{align*}
   E[\|\bm{y}-\hat{\bm{y}}\|_2^2] - E[\|\bm{y}-\hat{\bm{y}}^*\|_2^2] 
&= \tr\{(\bm{H}_1-\bm{H})\bm{\Sigma}_{11}\} +  \tr\{(\bm{H}_2-\bm{H})\bm{\Sigma}_{22}\} \\
& + 2  \tr\{(\bm{H}_1+\bm{H}_2 - \bm{H} - \bm{H}_1\bm{H}_2)\bm{\Sigma}_{12}\}
   -p\bm{\theta}_1^T\bm{\Sigma}_{\bm{W}_1}\bm{\theta}_1 -p\bm{\theta}_2^T\bm{\Sigma}_{\bm{W}_2}\bm{\theta}_2.
\end{align*}
In particular, when $\bm{\Sigma}_{mm'}=\sigma_{mm'}\bm{I}$, we obtain
\begin{align}
   E[\|\bm{y}-\hat{\bm{y}}\|_2^2] - E[\|\bm{y}-\hat{\bm{y}}^*\|_2^2] &=  -p\sigma_{11} - p\sigma_{22} -2\tr (\bm{H}_1\bm{H}_2)\sigma_{12} 
   -p\bm{\theta}_1^T\bm{\Sigma}_{\bm{W}_1}\bm{\theta}_1 -p\bm{\theta}_2^T\bm{\Sigma}_{\bm{W}_2}\bm{\theta}_2.\label{eq:Prop_trainingerrorExpectation}
\end{align}
\label{prop:residualclustering}
\end{prop}
\begin{proof}
   Proposition \ref{prop:residualclustering} can be proved in the same way as Theorem \ref{thm:correlation_test_misspecified}.
\end{proof}
   Eq. \eqref{eq:Prop_trainingerrorExpectation} of Proposition \ref{prop:residualclustering} indicates that the difference in training errors between AVR and IR increases as $\sigma_{12}$ increases. This observation implies that when two regression models are combined into one regression model, the training error decreases as $\sigma_{12}$ increases. Consequently, combining two regression models with a residual correlation close to 1 is expected to yield a lower training error.  The RCM algorithm may not always decrease the training error as effectively as the TEM algorithm, but our experiments show that the training error tends to decrease sufficiently as the number of clusters decreases.

}

{
\section{Simulation}
\label{sec:simulation}
We evaluate the performance of the proposed AVR-C through Monte-Carlo simulations. 
In particular, we generate datasets for predictors and responses and then investigate the training and test errors of the AVR-C model with varying numbers of clusters.  

We first generate the datasets according to the model \eqref{eq:latentmodel}, where the elements of $\bm X_m$ and $\bm W_m$ are obtained by the uniform distribution $U(0, 3)$.  
Values of the elements of parameter vectors $\bm\beta_m$ and $\bm\theta_m$ are generated from $U(0,1)$.  
Furthermore, the covariance matrix of the an error vector, $(\bm{\varepsilon}_1^\top,\dots,\bm{\varepsilon}_M^\top)^\top$, is set to be $\sigma^2 \bm R$, where $\sigma^2$ is a variance parameter and $\bm R$ is a correlation matrix given by
\begin{align}
    \bm R=\bm I_{10}\otimes \bm R_c, ~~~ \bm R_c = 0.9\times\bm 1_{M/10} \bm 1_{M/10}^\top + 0.1\bm I_{M/10}.
    \label{eq:sim_correlation}
\end{align}
With this covariance structure, the errors have a cluster structure (10 clusters) with positive correlations. 
For this dataset, we assume a regression model
\begin{align}
    \bm y_{\mathcal T_j} = \bm X_{\mathcal T_j} \bm\beta_{\mathcal T_j} + \bm\varepsilon_{\mathcal T_j}, ~~~j=1, \ldots, k,
    \label{eq:sim-model}
\end{align}
and then apply AVR-C, where we estimate the parameter vector $\bm\beta_{\mathcal T_j}$ by the ridgeless least squares \eqref{eq:ridgeless} and the TEM given in Section \ref{sec:TEM} is used for each step of hierarchical clustering.  
Results using the RCM in Section \ref{sec:RCM} are similar to those using TEM and then described in the Appendix \ref{sec:app:simulation}.  

We set the parameters as follows:
\begin{itemize}
    \item Training data size: $n = 500, 1000, 2000$
    \item Test data size: $n_{\text{test}} = 500$
    \item Number of variables: $p = 5$
    \item Variance parameter: $\sigma^2 = 0.25, 0.5, 1$
    \item Number of regression models: $M = 50, 500$
\end{itemize}
In this simulation, we set different predictors for training and test data, although we assume the same ones when introducing the bias-variance trade-off theory in Section 3.
The simulation setting is more realistic in practice,  and we will investigate whether similar tendencies described in Section 3 also appear in this setting.
For each setting, we repeat the analysis via the AVR-C 100 times and investigate the training and test errors.

Figure \ref{fig:sim_K50} shows training and test errors for $M=50$ as a function of the number of clusters.  
For all settings, the training errors increase as the number of clusters increases, which reflects the result of Theorem \ref{thm:bias-variance}.  
On the other hand, test errors show clear U-shaped functions, indicating that the results for the MSE on the test data reflect a bias-variance trade-off.  
This result is consistent with the implications of Theorem 2 and supports the effectiveness of the proposed AVR-C.  

\begin{figure}[!t]
	\begin{center}
		\includegraphics[width=0.3\hsize, bb= 0 0 450 300]{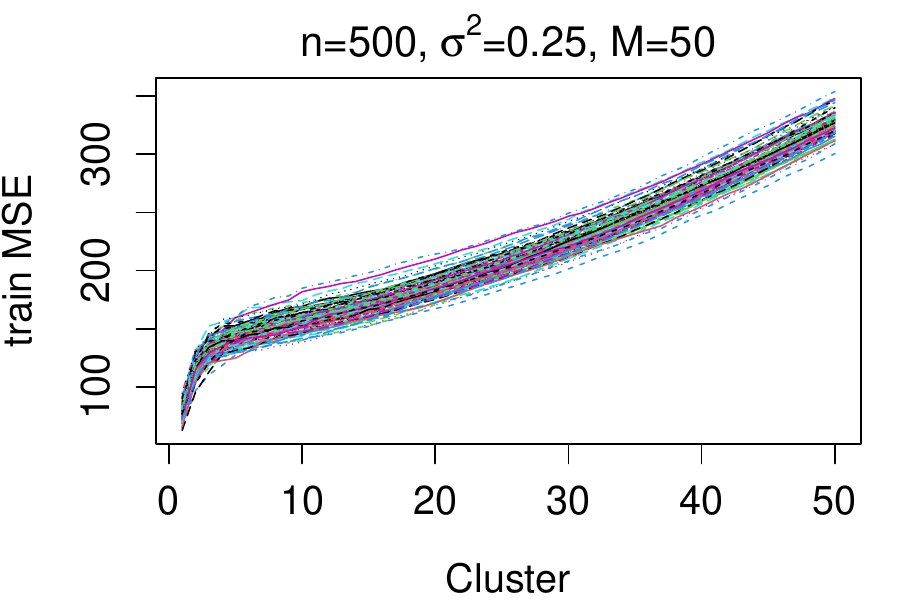}
		\includegraphics[width=0.3\hsize, bb= 0 0 450 300]{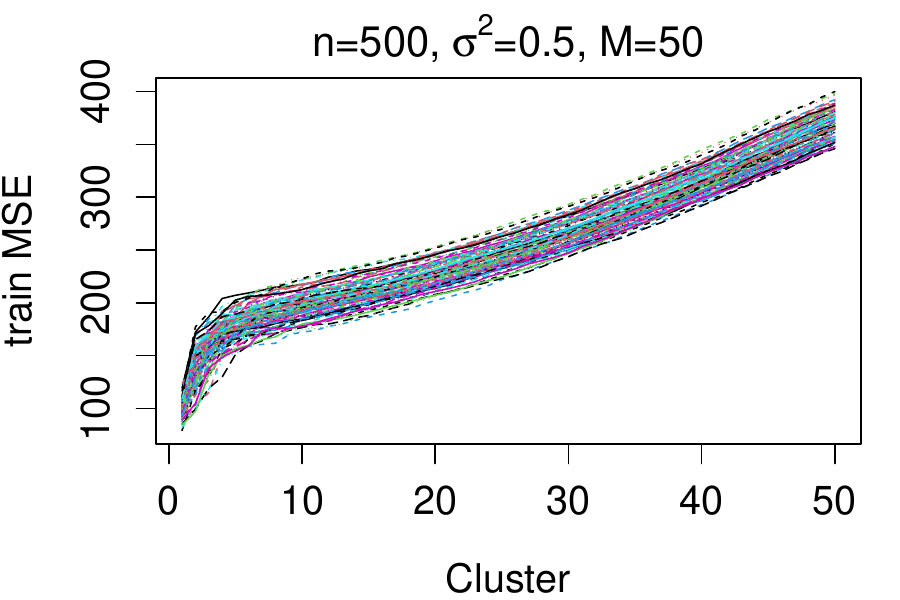}
		\includegraphics[width=0.3\hsize, bb= 0 0 450 300]{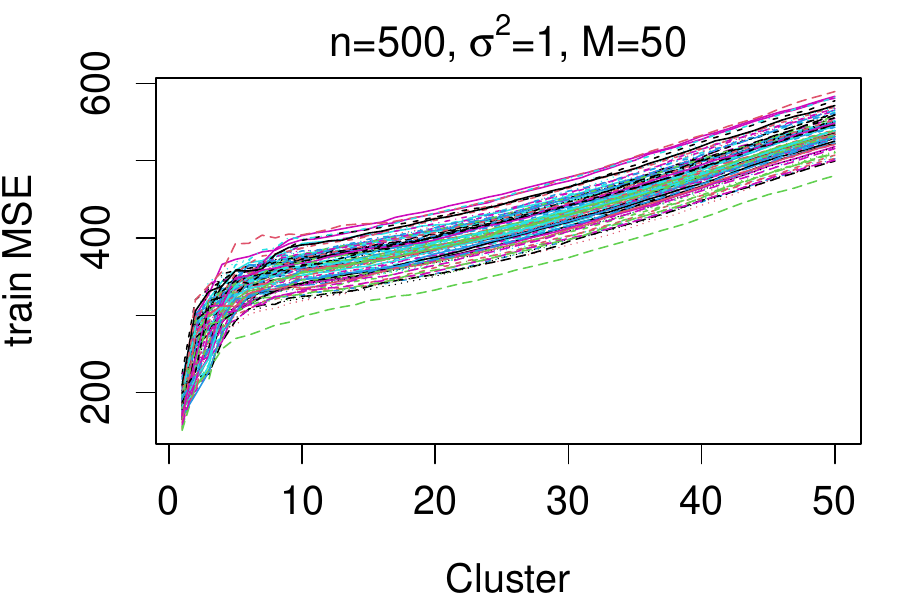} \\
		\includegraphics[width=0.3\hsize, bb= 0 0 450 300]{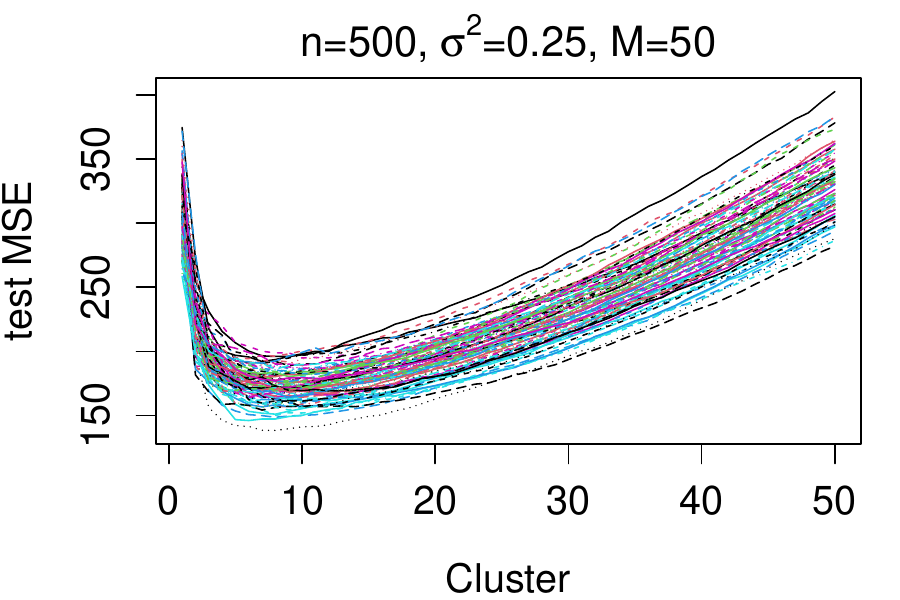}
		\includegraphics[width=0.3\hsize, bb= 0 0 450 300]{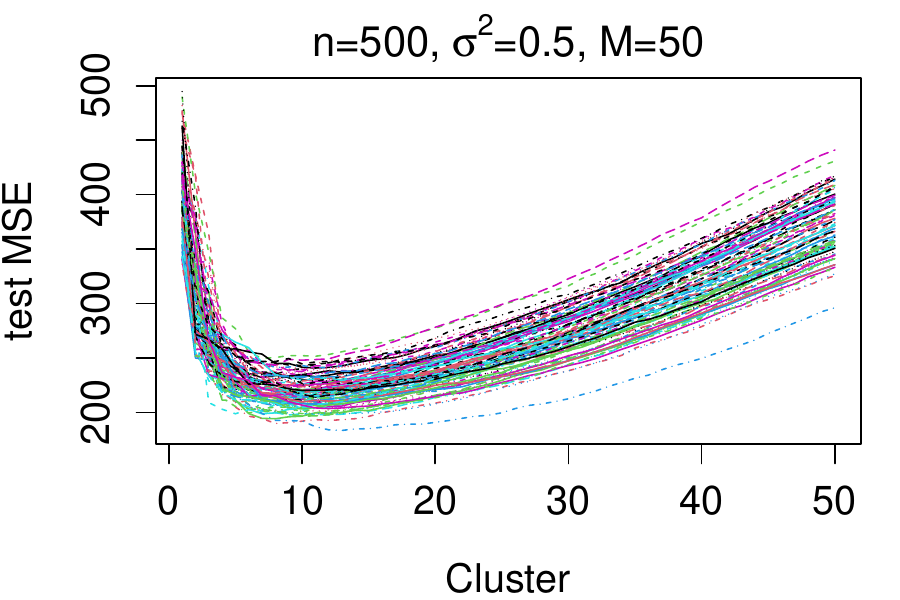}
		\includegraphics[width=0.3\hsize, bb= 0 0 450 300]{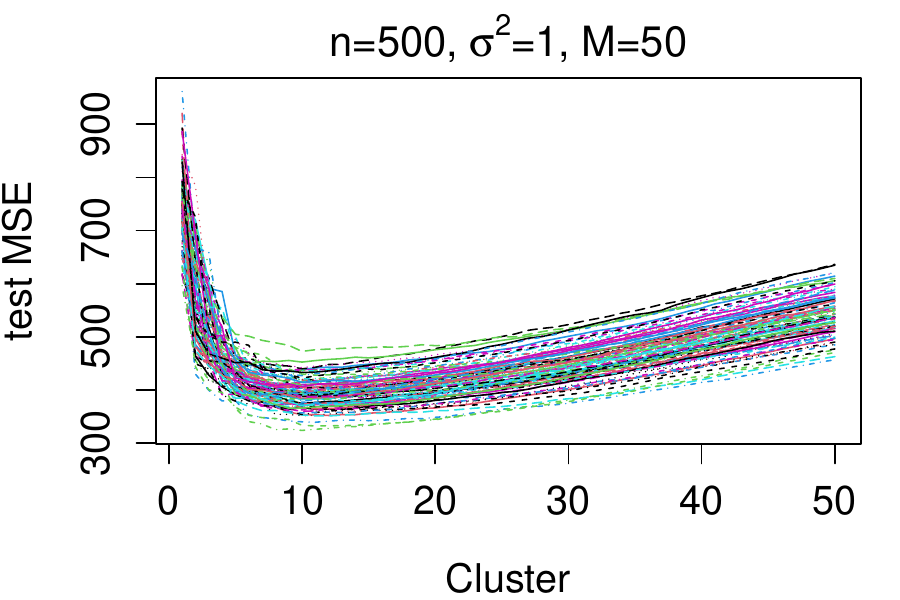}         
	\end{center}
         \vspace{-0.5cm}
	\begin{center}
		\includegraphics[width=0.3\hsize, bb= 0 0 450 300]{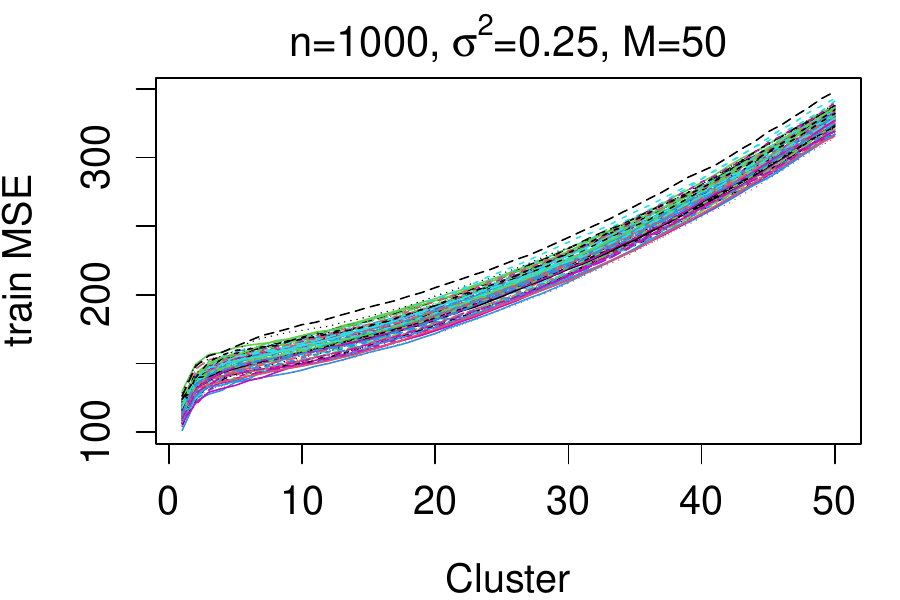}
		\includegraphics[width=0.3\hsize, bb= 0 0 450 300]{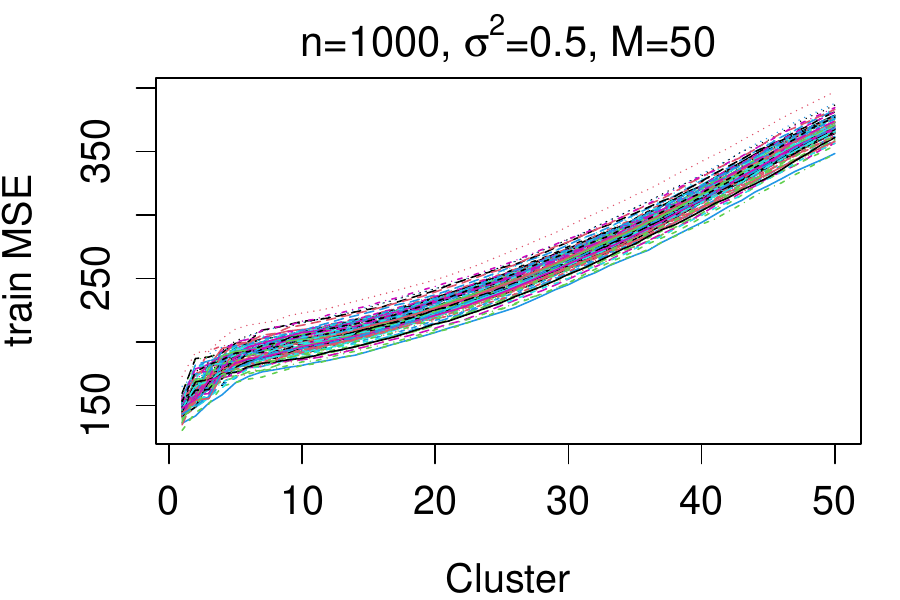}
		\includegraphics[width=0.3\hsize, bb= 0 0 450 300]{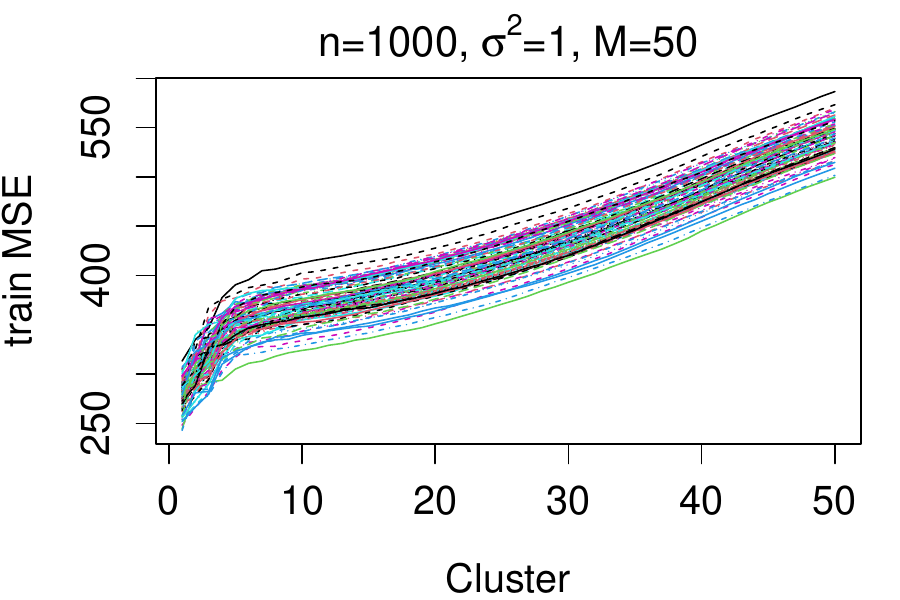} \\
		\includegraphics[width=0.3\hsize, bb= 0 0 450 300]{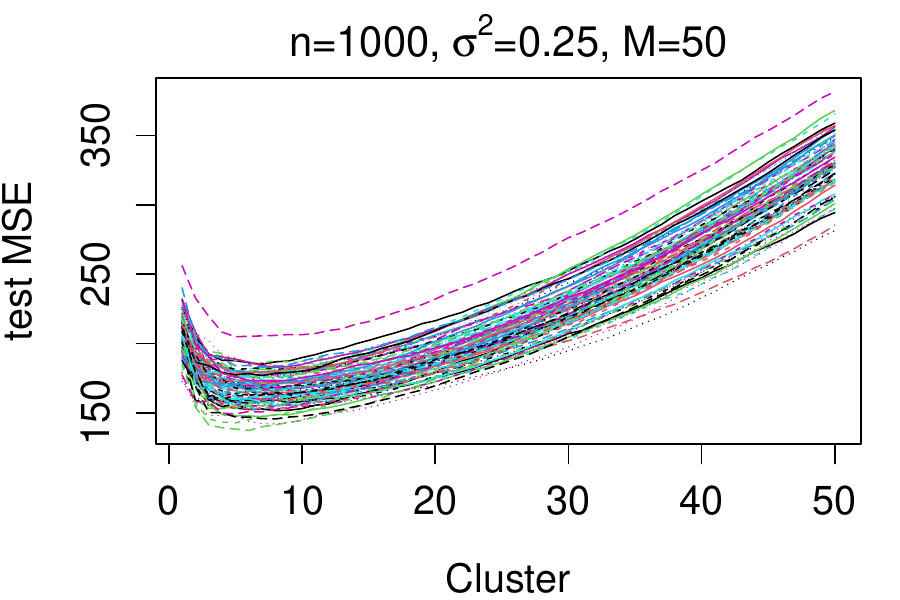}
		\includegraphics[width=0.3\hsize, bb= 0 0 450 300]{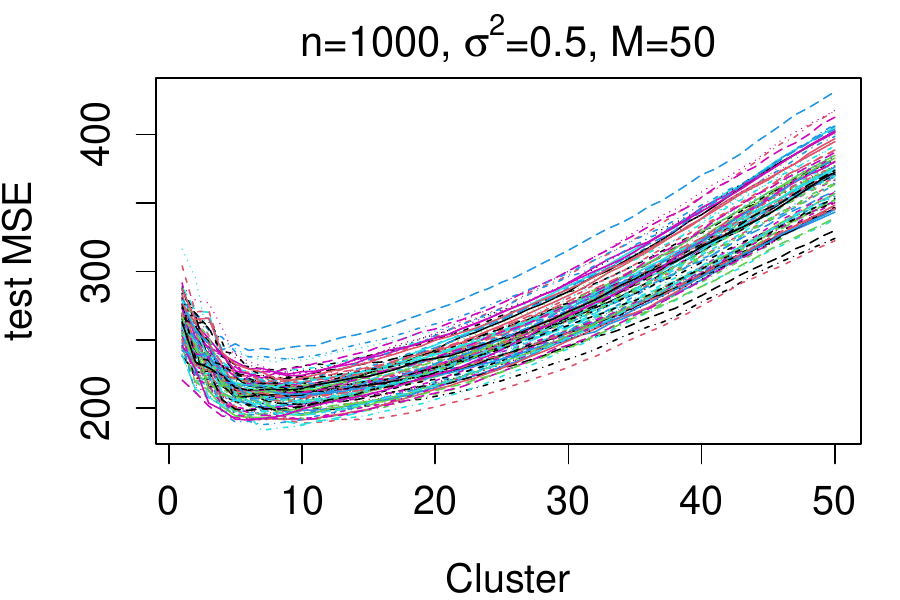}
		\includegraphics[width=0.3\hsize, bb= 0 0 450 300]{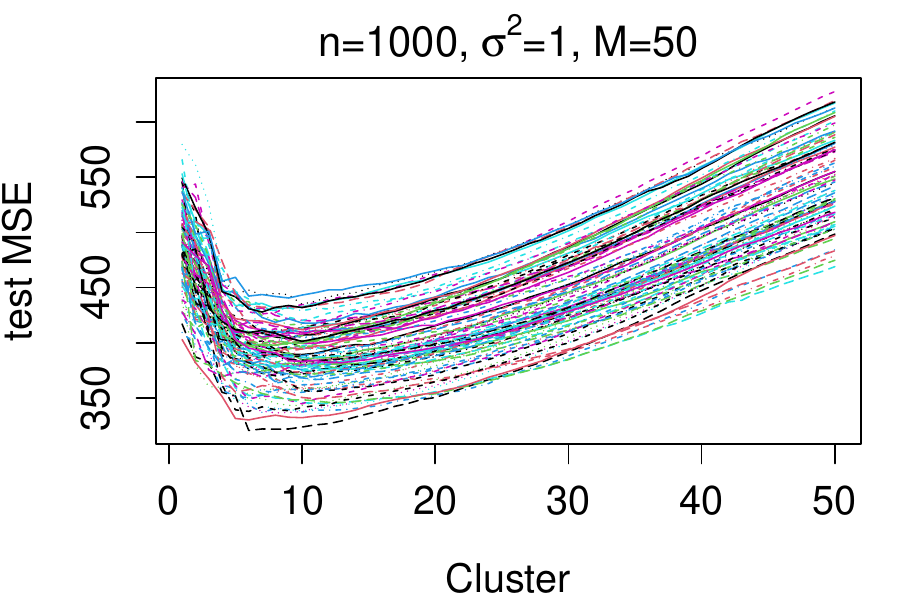}         
	\end{center}
         \vspace{-0.5cm}
 	\begin{center}
		\includegraphics[width=0.3\hsize, bb= 0 0 450 300]{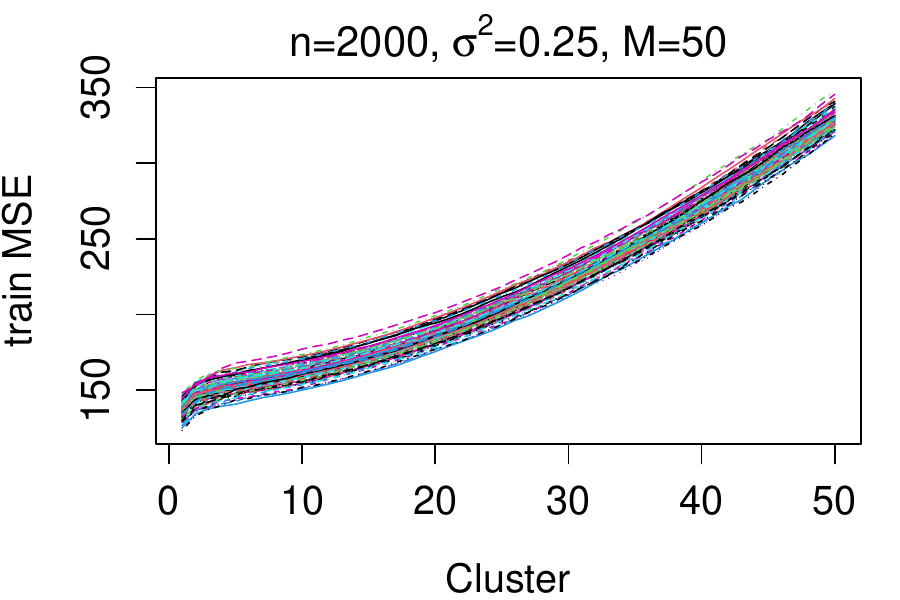}
		\includegraphics[width=0.3\hsize, bb= 0 0 450 300]{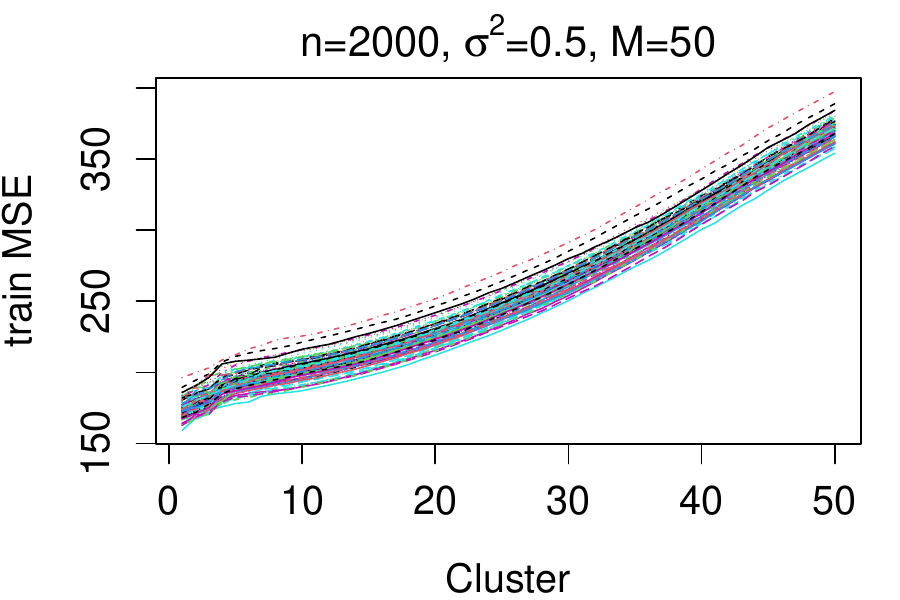}
		\includegraphics[width=0.3\hsize, bb= 0 0 450 300]{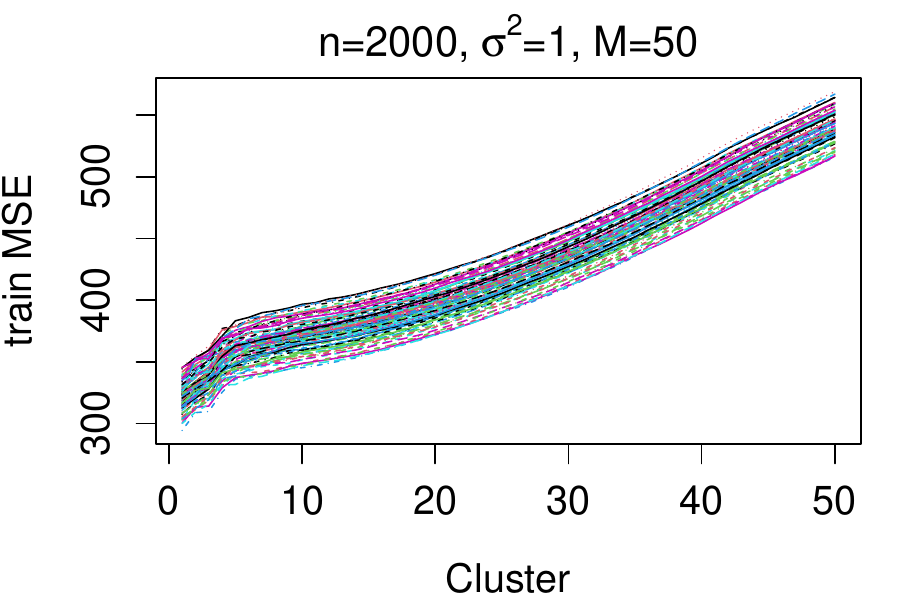} \\
		\includegraphics[width=0.3\hsize, bb= 0 0 450 300]{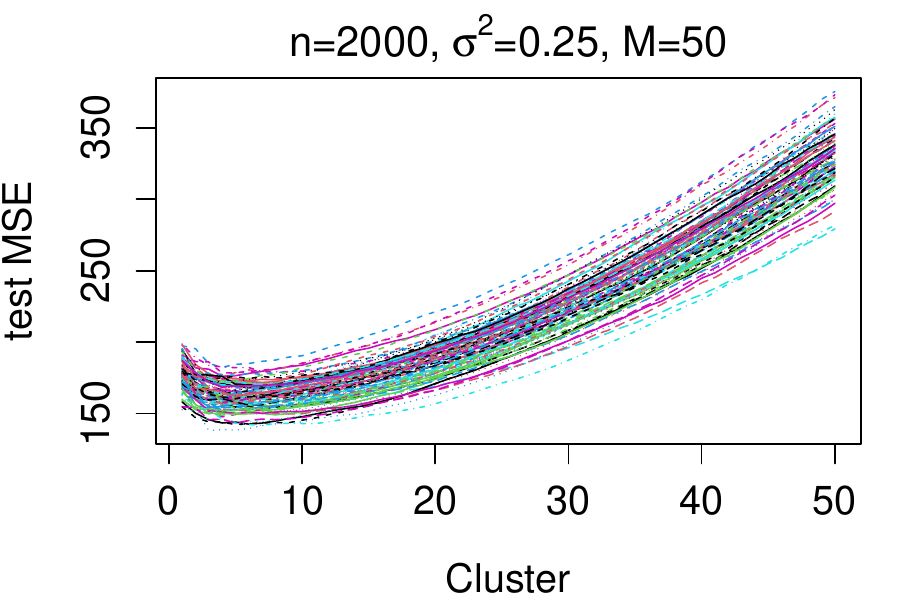}
		\includegraphics[width=0.3\hsize, bb= 0 0 450 300]{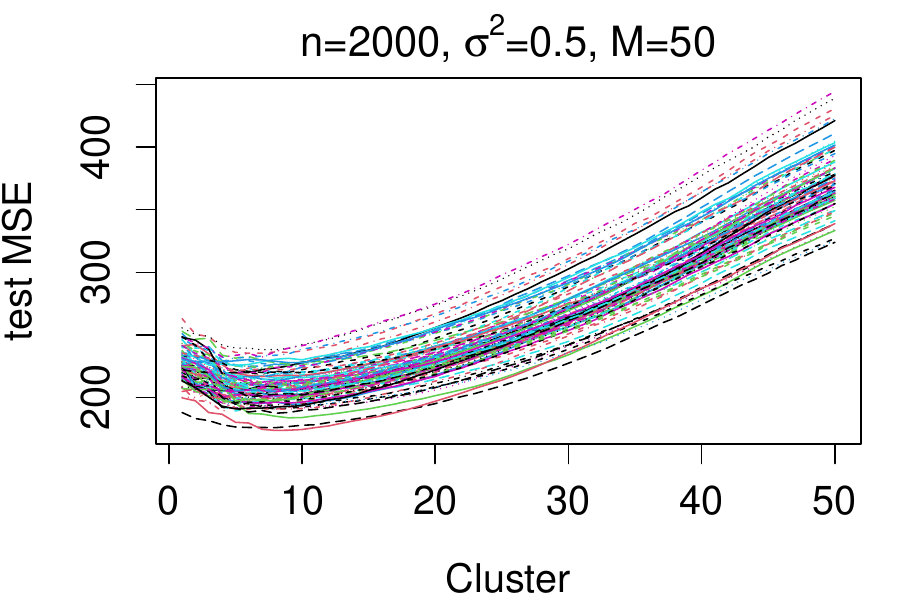}
		\includegraphics[width=0.3\hsize, bb= 0 0 450 300]{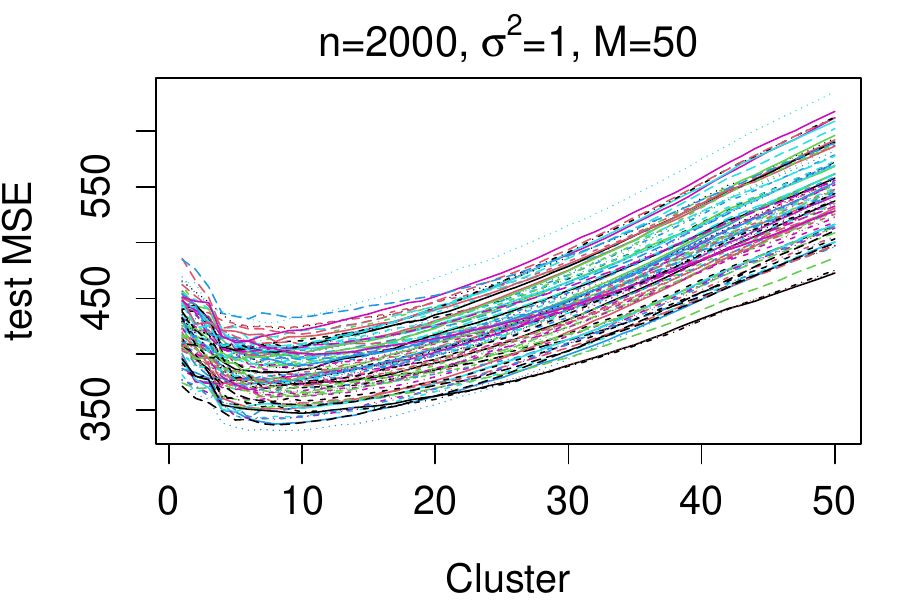}         
	\end{center}
	\caption{Simulation results for $M=50$. 
    } 
	\label{fig:sim_K50}

\end{figure}

The number of clusters that minimizes the test error is around 10 in many cases, suggesting that an appropriate number of clusters may be selected with test error minimization. We also observe that the minimizer for the number of clusters tends to be slightly smaller than the true number of clusters. This implies that AVR-C tends to select a more complex model than the true model when minimizing the test error (Note that the model becomes more complex as the number of clusters decreases). This result is reasonable because minimizing the test error does not always lead to the true model. 

The test error is also affected by the variance, $\sigma^2$. The result of Theorem 2 suggests that when $\sigma^2$ is small, the increase in test error in combining two regression models is not significant compared to large $\sigma^2$. Therefore, it would be reasonable to observe that as $\sigma^2$ becomes large, the decrease in test error for $k = 10, \ldots, 50$ is small, while the increase in test error for a smaller number of clusters, $k = 1, \ldots, 10$, is substantial.


 \begin{figure}[!t]
	\begin{center}
		\includegraphics[width=0.3\hsize, bb= 0 0 450 300]{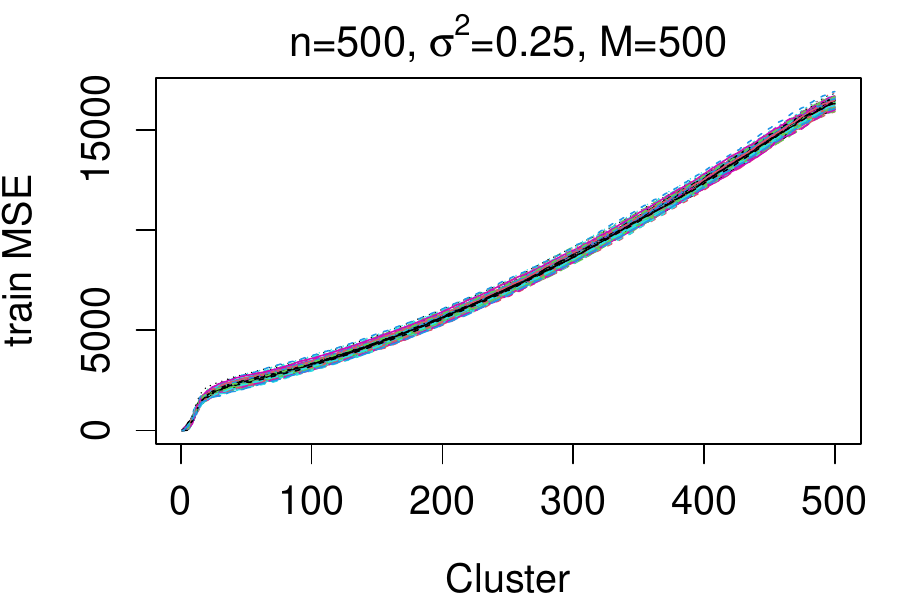}
		\includegraphics[width=0.3\hsize, bb= 0 0 450 300]{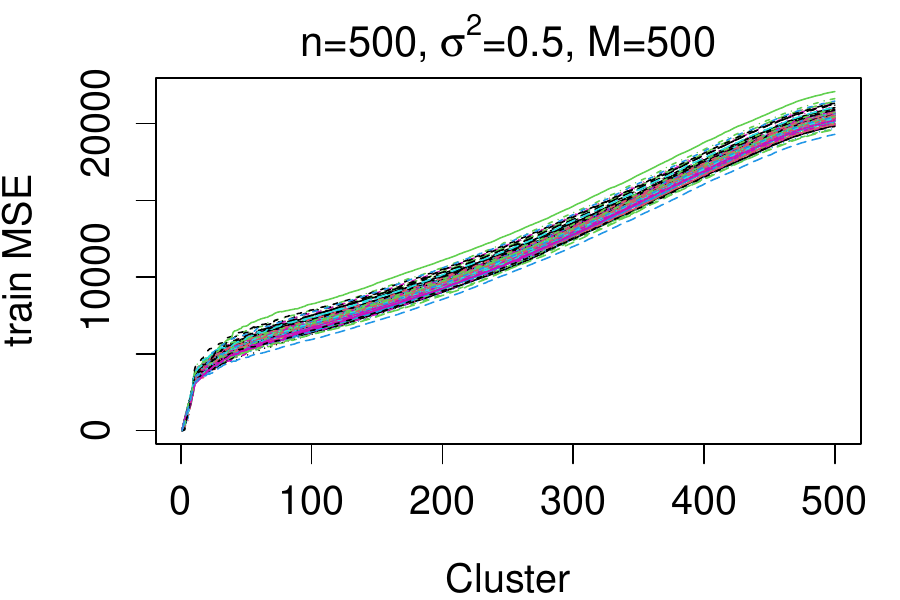}
		\includegraphics[width=0.3\hsize, bb= 0 0 450 300]{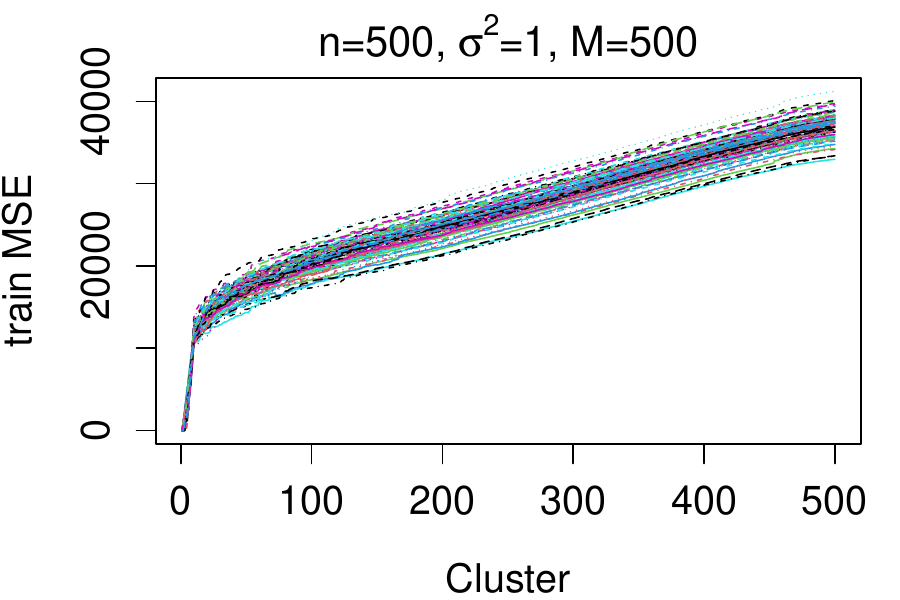} \\
		\includegraphics[width=0.3\hsize, bb= 0 0 450 300]{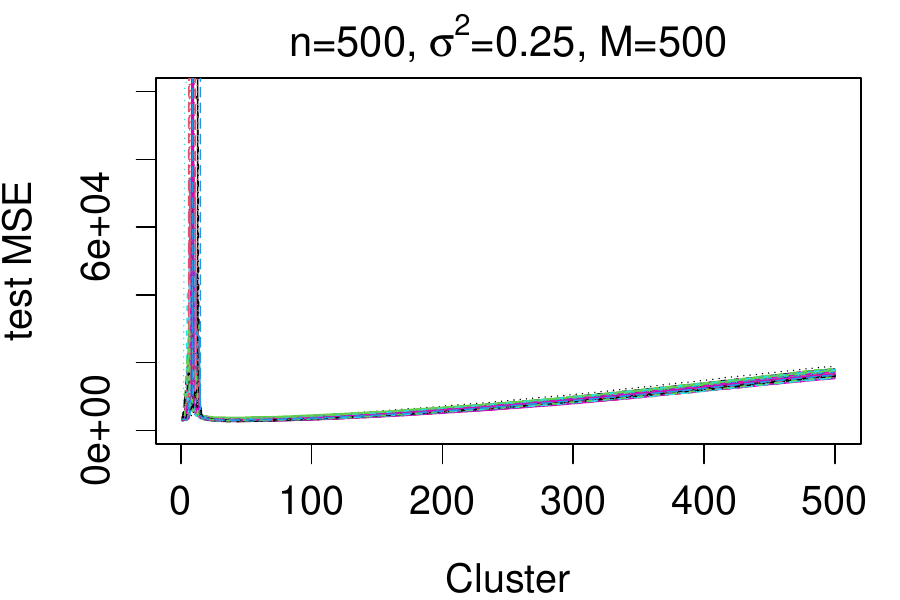}
		\includegraphics[width=0.3\hsize, bb= 0 0 450 300]{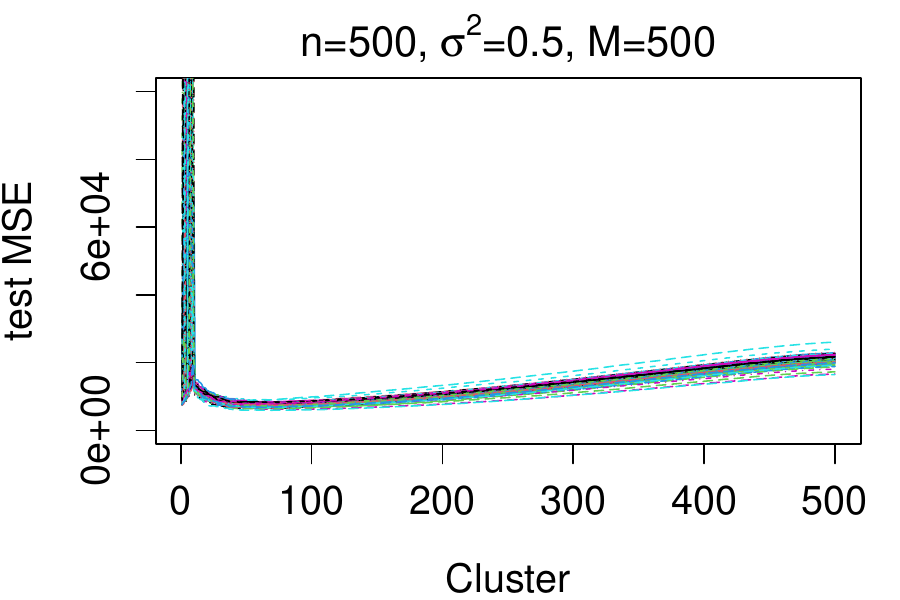}
		\includegraphics[width=0.3\hsize, bb= 0 0 450 300]{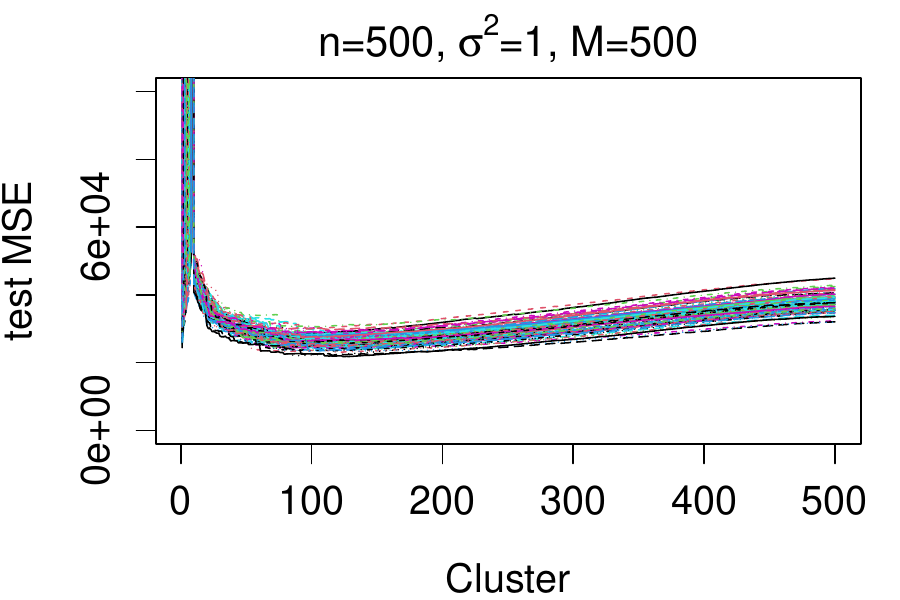}         
	\end{center}
         \vspace{-0.5cm}
	\begin{center}
		\includegraphics[width=0.3\hsize, bb= 0 0 450 300]{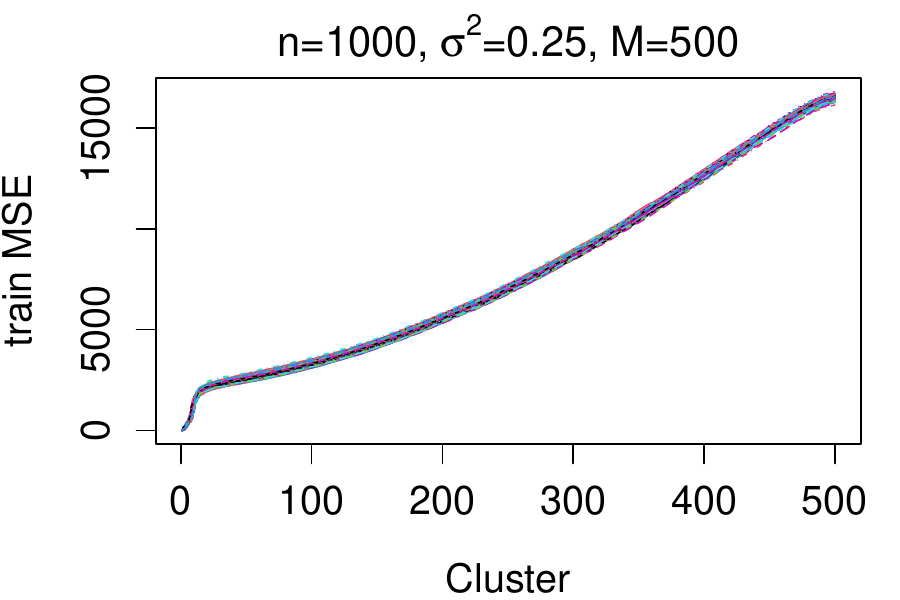}
		\includegraphics[width=0.3\hsize, bb= 0 0 450 300]{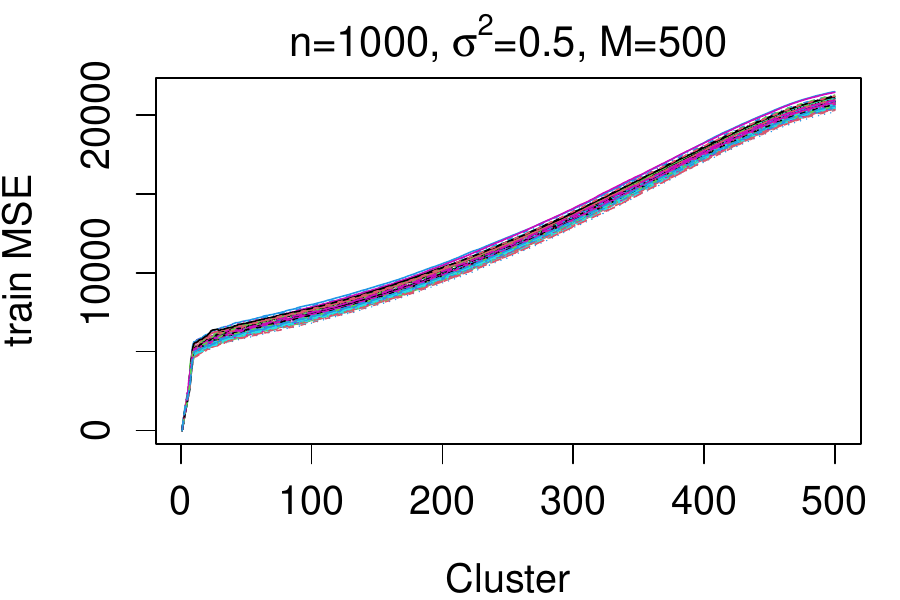}
		\includegraphics[width=0.3\hsize, bb= 0 0 450 300]{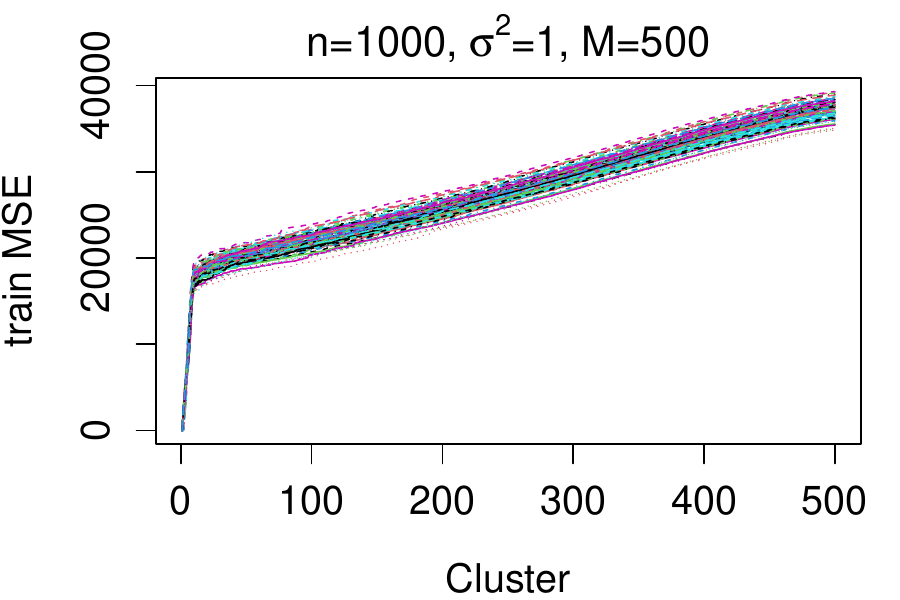} \\
		\includegraphics[width=0.3\hsize, bb= 0 0 450 300]{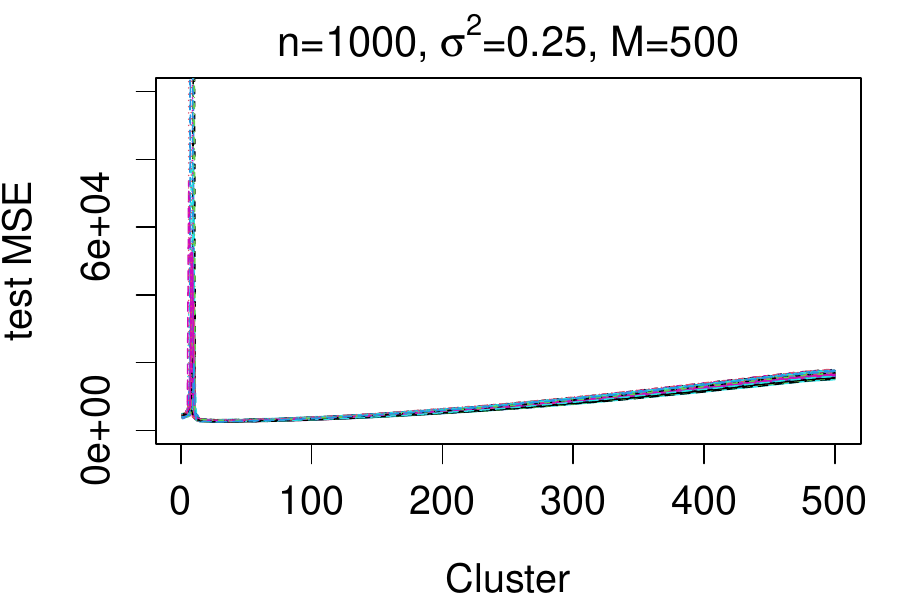}
		\includegraphics[width=0.3\hsize, bb= 0 0 450 300]{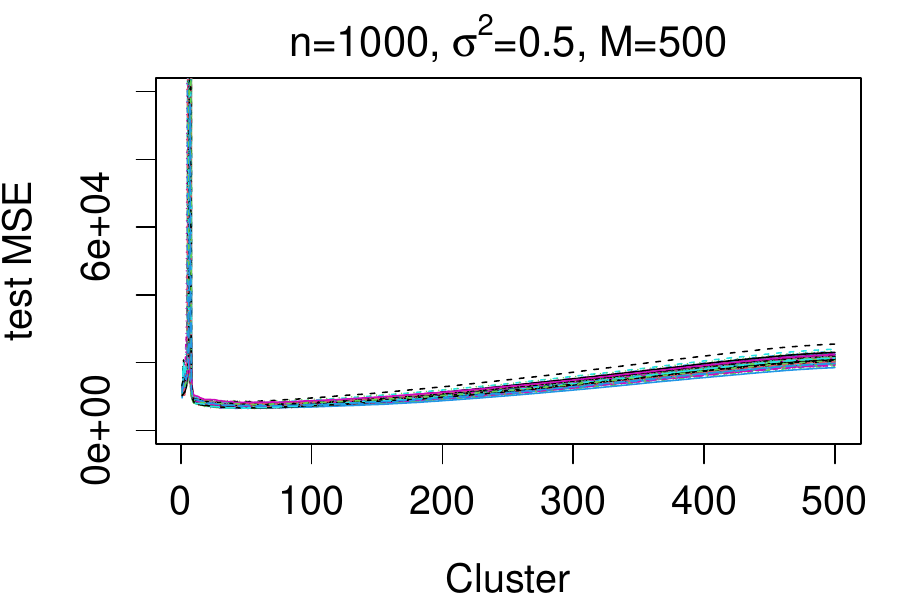}
		\includegraphics[width=0.3\hsize, bb= 0 0 450 300]{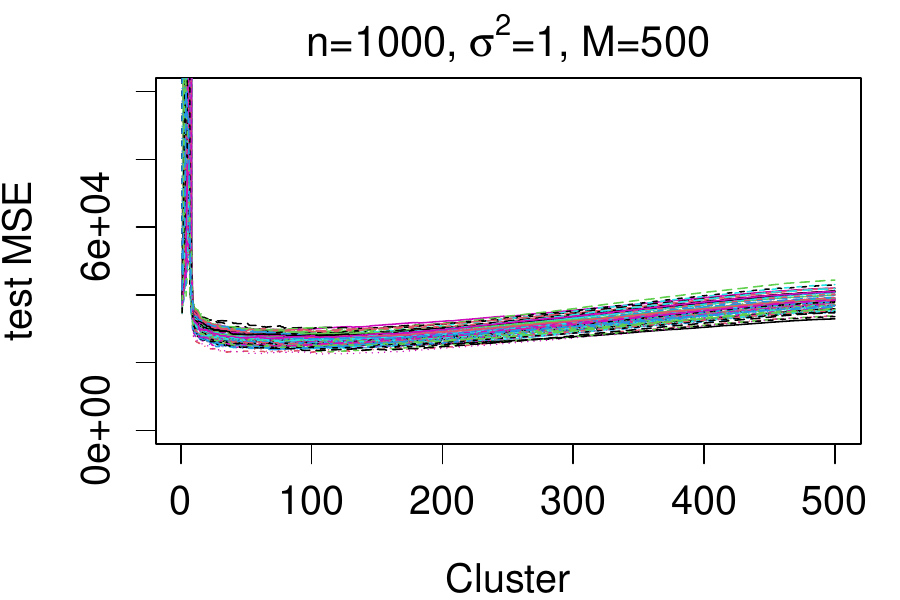}         
	\end{center}
         \vspace{-0.5cm}
 	\begin{center}
		\includegraphics[width=0.3\hsize, bb= 0 0 450 300]{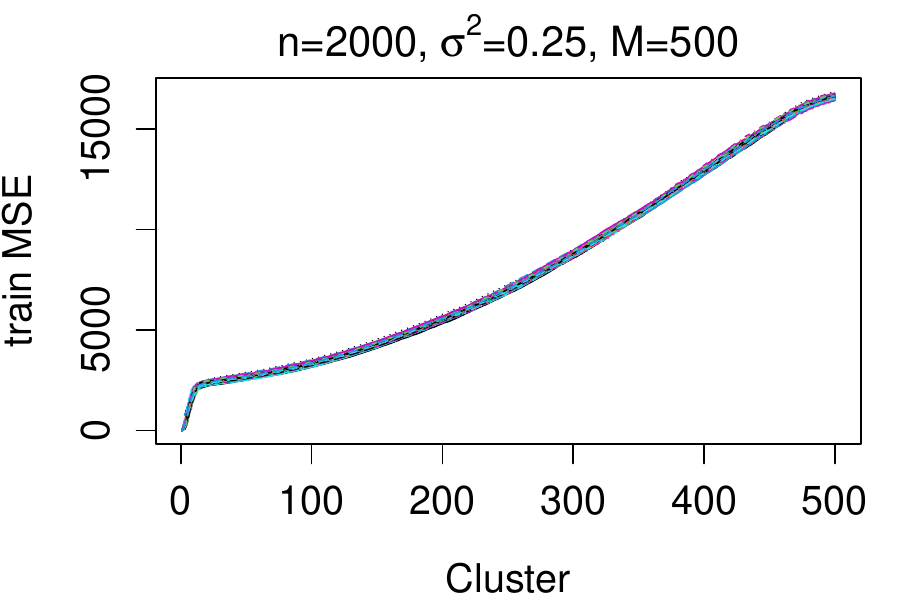}
		\includegraphics[width=0.3\hsize, bb= 0 0 450 300]{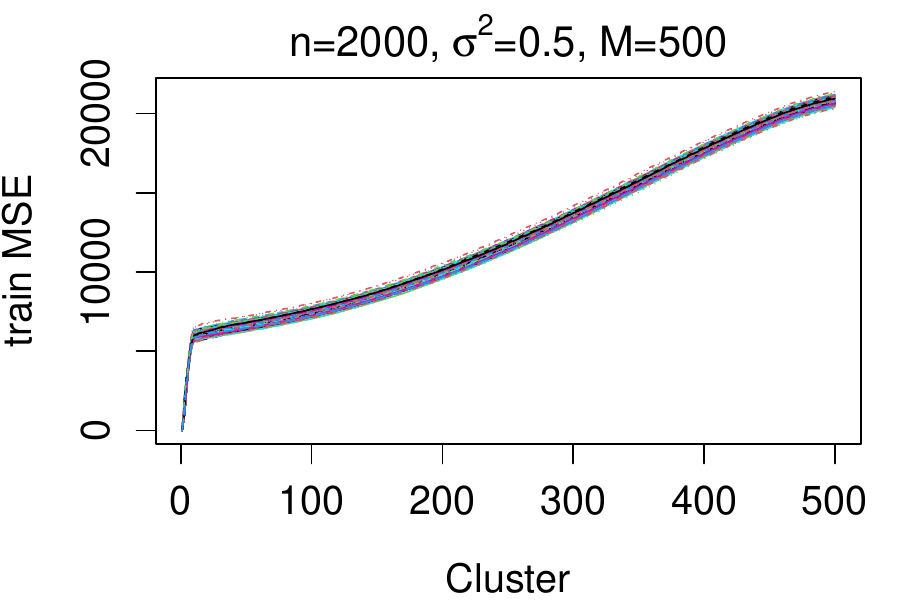}
		\includegraphics[width=0.3\hsize, bb= 0 0 450 300]{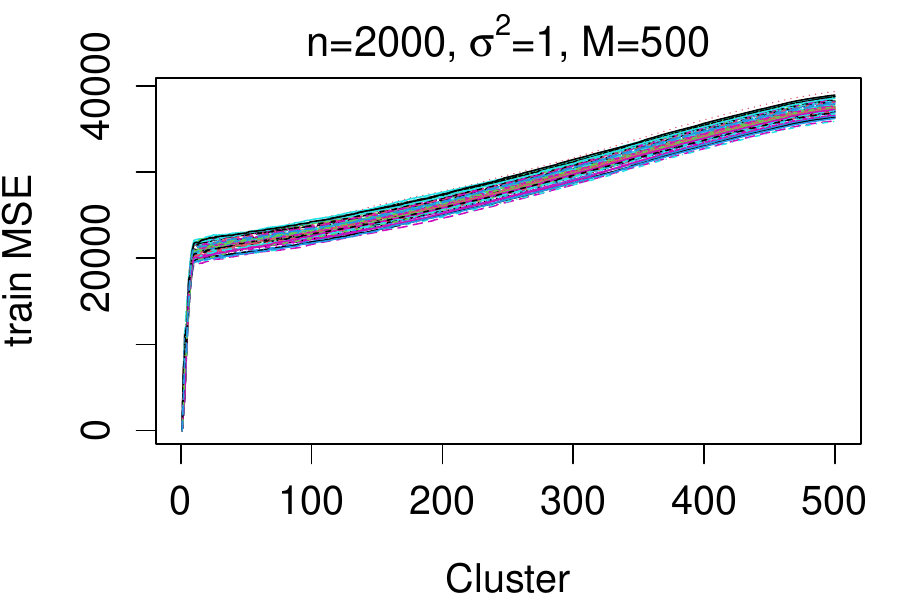} \\
		\includegraphics[width=0.3\hsize, bb= 0 0 450 300]{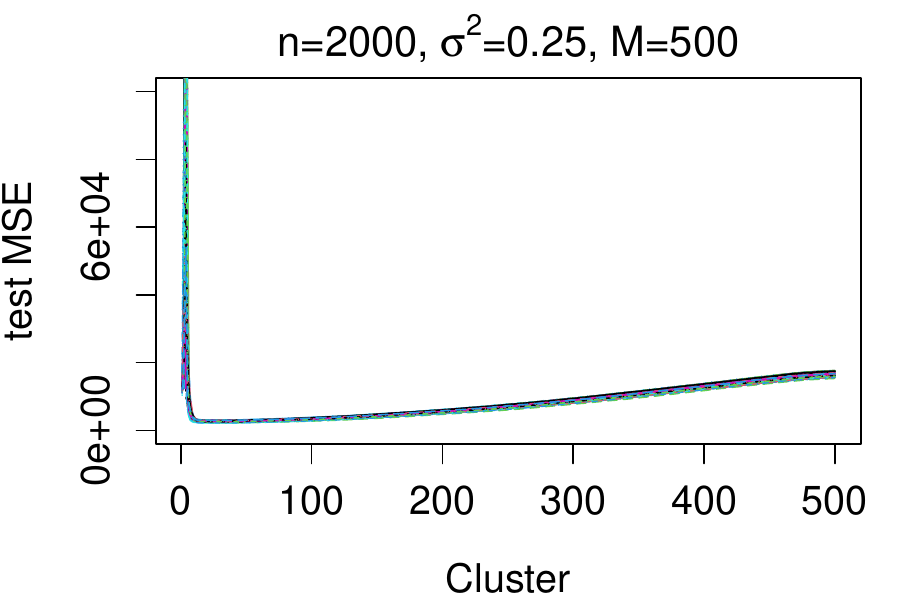}
		\includegraphics[width=0.3\hsize, bb= 0 0 450 300]{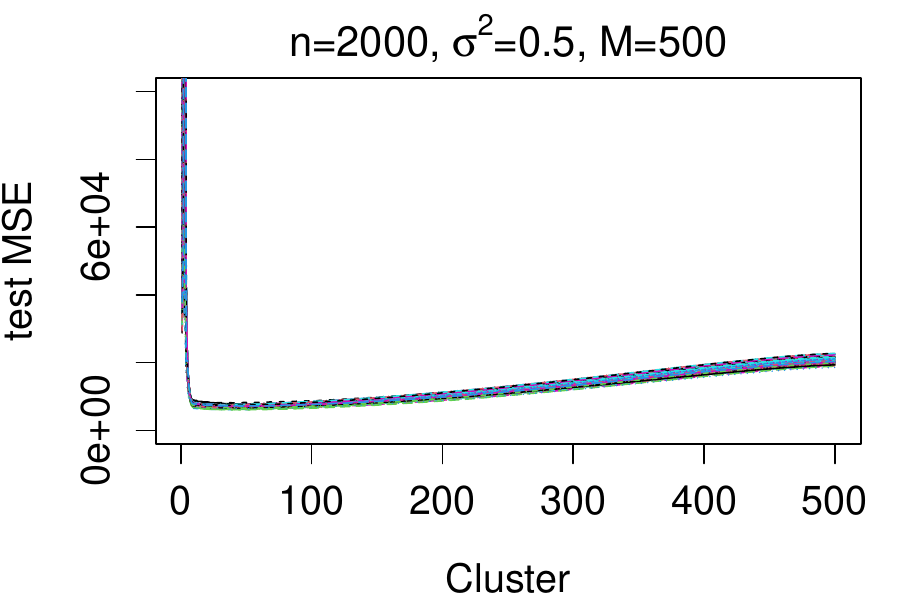}
		\includegraphics[width=0.3\hsize, bb= 0 0 450 300]{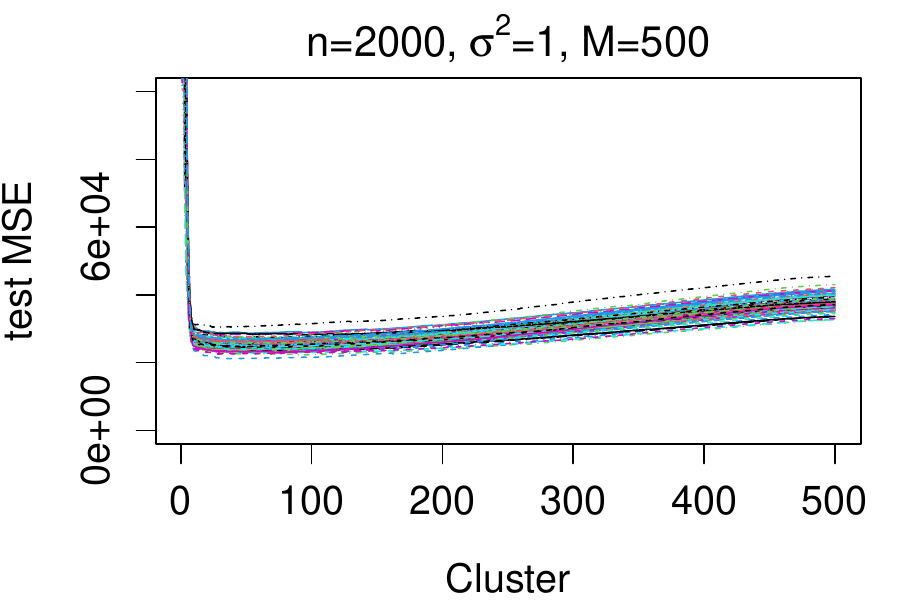}         
	\end{center}
	\caption{Simulation results for $M=500$.}
	\label{fig:sim_K500}
\end{figure}

Results for $M=500$ are similar to those for $M=50$, as shown in Figure \ref{fig:sim_K500}.  
However, for all $M=500$ settings, test errors diverge when the number of clusters is small.  At the divergence point, the training error for one of the clusters is close to 0, as there exists a situation where the sample size nearly equals the number of variables.
This phenomenon is referred to as double descent, and a number of researchers have studied it with the spread of deep learning theory (e.g., \citealp{belkin2019reconciling,Hastie.2022,schaeffer2023double}).
In particular, when $n=500$ and $\sigma^2=0.25$, we observe situations where the test error is minimized when the number of clusters equals 1.  This result indicates that the AVR-C under an interpolation regime, where the model achieves near-zero training error with high-dimensional predictors, can provide lower test errors than AVR-C with low-dimensional predictors.

}

\section{Real data analysis}
In this section, we illustrate the usefulness of our proposed AVR-C through the analysis of electricity demand data from various facilities\footnote{\url{https://www.ems-opendata.jp}}. We obtain the electricity consumption data of all available facilities from the website on 2022.10.30; at this time, the data consists of hourly electricity consumption data from $5727$ facilities.  The terms of the collected data differ among facilities; thus, we choose facilities that contain complete data from Jan. 5th, 2014 to Dec. 31st, 2014, resulting in $M=847$ facilities.   The demand is shown in megawatts (MW) at 1-hour intervals (i.e., $J=24$).  Training and test data consist of the first 100 days and the last 258 days, respectively.

We first construct a statistical model to forecast the electricity demand for each facility.  Let $y_{ijm}$ be the electricity demand on $i$th day ($i=1,\dots,I$), $j$th time interval ($j=1,\dots,J$), and $m$th facility ($m=1,\dots,M$).  In this analysis, we use the mean temperature as a daily weather information, and it is denoted as $u_{ir}$.  Here, $r$ is the index of the area ($r=1,\dots,R$). In this study, we use the temperature data from representative cities of $R=8$ areas in Japan as below:
\begin{center}
{\small
\begin{tabular}{ll}
\textbf{Area}:& Kanto, Chubu, Kansai, Chugoku, Shikoku, Tohoku, Kyushu-Okinawa, Hokkaido. \\
\textbf{City}:& Tokyo, Nagoya, Osaka, Hiroshima, Matsuyama, Sendai, Fukuoka, Sapporo. \\
\end{tabular}
}	
\end{center}
The temperature data are available at Japan Meteorological Agency\footnote{\url{https://www.jma.go.jp/jma/indexe.html}}. 

We employ the following statistical model \citep{Hirose.2021}, which was carefully developed through extensive trial and error and was inspired by similar sophisticated models proposed by several researchers \citep{Fan.2012,Hong.2016,Lusis:2017gd}:
\begin{equation}
y_{ijmr} = \sum_{t=1}^T\alpha_{tm}y_{i-t,j,m,r} + \sum_{h=1}^H \beta_{h}(j) g_{h}(u_{ir})  + \sum_{\ell=1}^L\gamma_{\ell m}z_{\ell} + \varepsilon_{ijmr}.\label{eq:forecastmodel}
\end{equation}
The first term is a linear combination of the past demand, $y_{i-t,j,m,r}$, to express the AR-type effect of electricity demand in the past $T$ days. Here, $\alpha_{tm}$ are regression coefficients related to the past electricity demand. We set $T=7$ to incorporate the effect of the electricity demand from the previous week. The second term corresponds to the daily weather effects. In particular, the terms $g_{h}({u}_{im}) $  ($h=1,...,H$) are B-spline functions to express the nonlinear relationship between daily temperature ${u}_{ir}$ and electricity demand $y_{ijmr}$. $\beta_h(j)$ are regression coefficients as a function of time interval $j$, and are expressed as a linear combination of $Q$ cyclic B-spline functions. The number of basis functions is $H=5$ and $Q=5$ to express the smooth nonlinearity with a relatively small number of parameters.  For details of the second term of \eqref{eq:forecastmodel}, please refer to \citet{Hirose.2021}. The third term corresponds to the weekday effects including weekends. The $z_{\ell}$ are dummy variables of the day of the week, and $\gamma_{\ell m}$ are the corresponding regression coefficients. Public holidays are allocated to Sundays, resulting in $L=6$ dummy variables.  Finally, the last term, $\varepsilon_{ijmr}$, corresponds to the error.   Note that when two facilities belong to the same area (i.e., they share the same index $r$), their predictor variables are also shared. Accordingly, the temperature and other inputs are treated as common predictor variables for all facilities in the same area.

 The numbers of parameters in the IR and AVR models are 38 $(=T + QH+L=7+5\times 5+6)$, and 6135 $(=MT + R QH+L=7\times 847+8\times 5 \times 5+6)$, respectively.  In this section, we first present the analysis in the Chugoku area to illustrate the usefulness of our proposed clustering method with a small number of facilities.  Then, we describe the analysis in all areas; in this case, the number of parameters of the AVR ($=6135$) exceeds the number of observations in the training data ($=100\times J=2400$); thus, the double descent phenomenon would occasionally be observed as seen in Figure \ref{fig:sim_K500} in the simulation study.

\subsection{Chugoku area}
In the Chugoku area, there are 63 facilities to be analyzed.  We first illustrate the usefulness of AVR-C with RCM.  The regression model in \eqref{eq:forecastmodel}  is applied to each facility separately (i.e., IR) and compute the residuals for all facilities.  Figure \ref{fig:dendrogram} shows the dendrogram obtained by the correlation matrix.  
\begin{figure}[!t]
\centering
\includegraphics[width=\textwidth, bb=0 0 865 422]{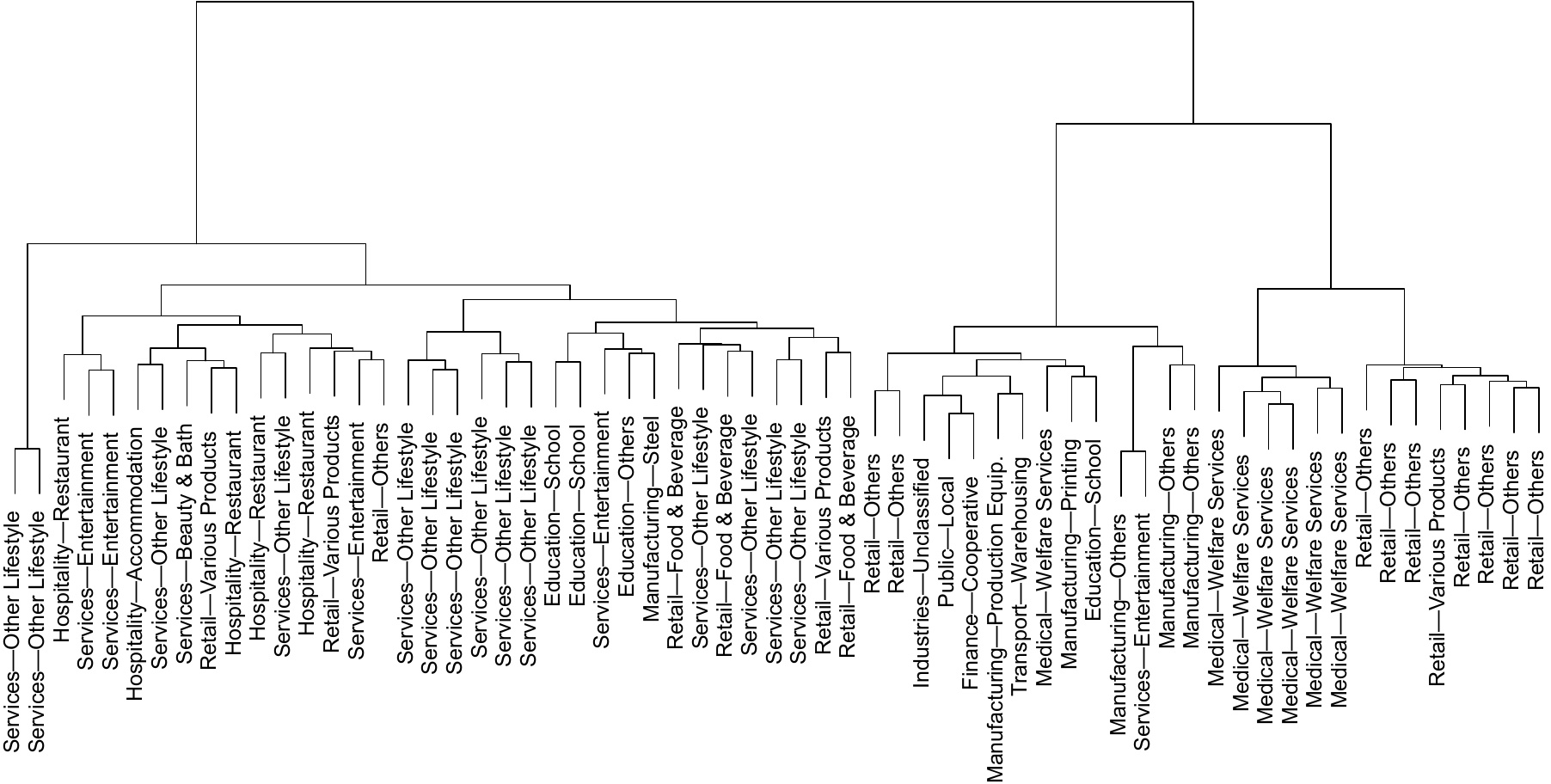}
\caption{Dendrogram obtained by correlation of residuals.  
For ease of comprehension, texts for each observation correspond to the facility's category and subcategory.  }
\label{fig:dendrogram}
\end{figure}
For ease of comprehension, texts for each observation correspond to the facility's category and subcategory.  The results show that several facilities that are likely to have a similar tendency of electricity usage belong to the same cluster. For example, it can be roughly seen that there are two main clusters. The facilities in the first cluster that corresponds to the left-hand side part of the dendrogram are related to lifestyle and entertainment (e.g., services of lifestyles, hospitality, retail--foods). Meanwhile, the second cluster, corresponding to the right-hand side part of the dendrogram, contains other various services, such as medical welfare, finance, manufacturing, and retail--others.  Furthermore, we observe three main sub-clusters in the second cluster; that is, Retail--Others, Medical--Welfare services, and other services.  These interpretable results are helpful in finding the tendency of electricity usage patterns, resulting in effective energy intervention and demand response.

The correlation plots of the residuals are depicted in Figure \ref{fig:corresrealdata}.  The left panel corresponds to the correlation plot for all facilities.  We observe that the residuals of many facilities are non-negatively correlated, while approximately $34\%$ of residuals of facilities are negatively correlated.  Thus, it would not be easy to interpret the whole correlation structure; the cluster analysis may be helpful in categorizing the electricity usage pattern.  The right panel corresponds to the correlation matrices of 12 clusters, determined by test error minimization based on RCM. Most of the correlations are non-negative; indeed, we observe that the negative correlations are less than $5\%$.   Thus, our clustering technique successfully constructed clusters such that the correlations of residuals within the cluster are mostly non-negative.  
\begin{figure}[!t]
\centering
\includegraphics[width=\textwidth, bb=0 0 701 436]{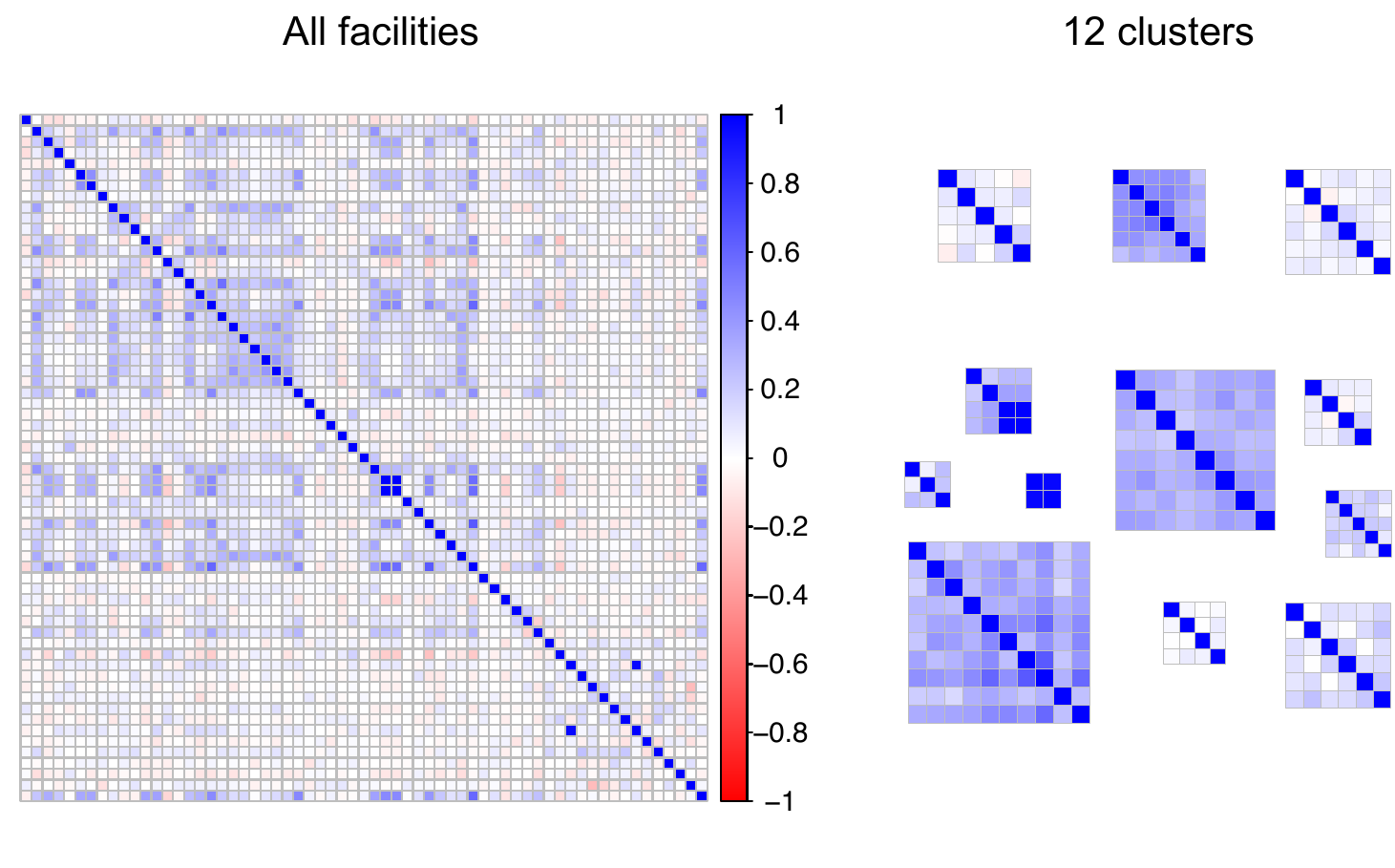}
\caption{Correlation of residuals in the Chugoku area: all facilities (left panel) and 12 clusters (right panel). The number of clusters ($k = 12$) is selected so that the test error based on RCM is minimized.}
\label{fig:corresrealdata}
\end{figure}

Figure \ref{fig:RMSE Chugoku} shows the RMSE for training and test datasets as a function of the number of clusters, $k$.  
\begin{figure}[!t]
\centering
\includegraphics[width=\textwidth, bb=0 0 653 275]{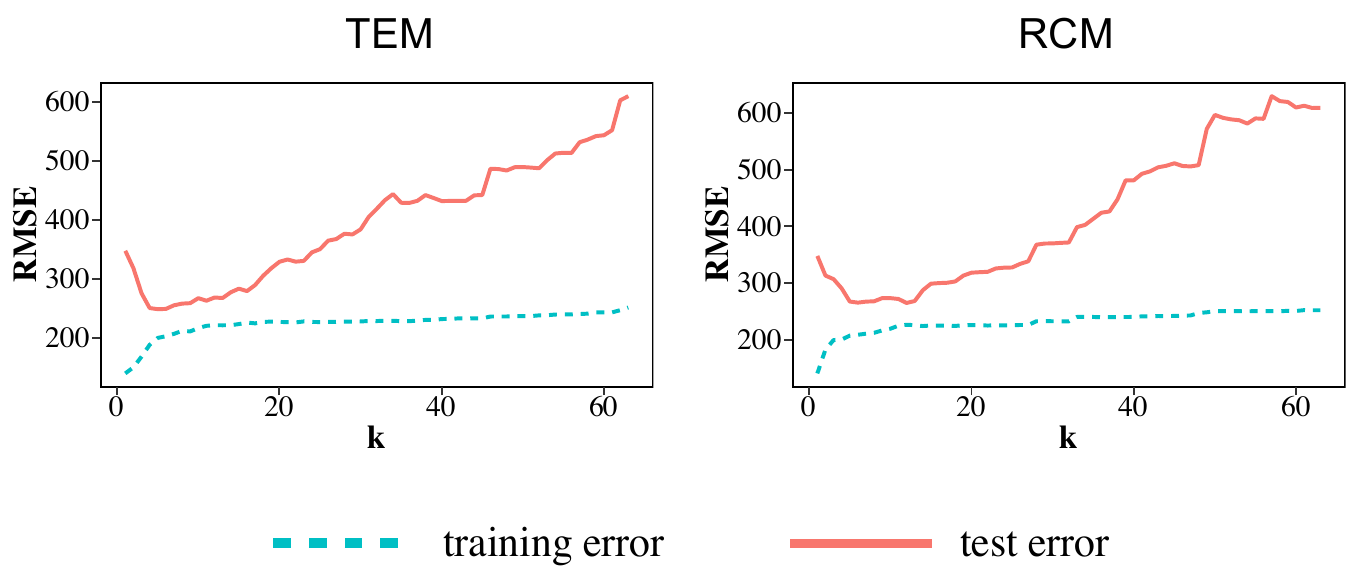}
\caption{RMSE for training and test datasets as a function of the number of clusters $k$ in the Chugoku area.}
\label{fig:RMSE Chugoku}
\end{figure}
Here we employ both TEM and RCM for the clustering method in AVR-C. 
In both clustering methods, the training error is almost (but not always) a monotone-increasing function of $k$.  Therefore, our proposed clustering technique successfully decreases the training error in most steps, and then the number of clusters is viewed as the model complexity.  The test error is roughly a U-shape function and is minimized at $k=4$ and $k=12$ for TEM and RCM, respectively.  Furthermore, although the TEM provides better performance in RMSE, the minimum value of RMSE obtained by these two clustering methods are quite similar. The TEM is slow because it performs the regression models every time when combining two regression models.  On the other hand, RCM is fast and interpretable in terms of correlations of residuals.  Considering that the difference of RMSEs between the two clustering methods is not significant, it would be reasonable to use RCM due to its fast computation and interpretability for this dataset.

\subsection{All areas}
To investigate the performance of AVR-C on a large-scale dataset, we apply it to all $M=847$ facilities.  In this case, the number of parameters can exceed the number of observations with a small number of clusters, suggesting that there is a possibility that the double descent phenomenon can occur.

Figure \ref{fig:RMSE ALL}  shows the RMSE for training and test datasets as a function of $k$ in all areas.  We apply both TER and RCM. 
\begin{figure}[!t]
\centering
\includegraphics[width=\textwidth, bb=0 0 654 278]{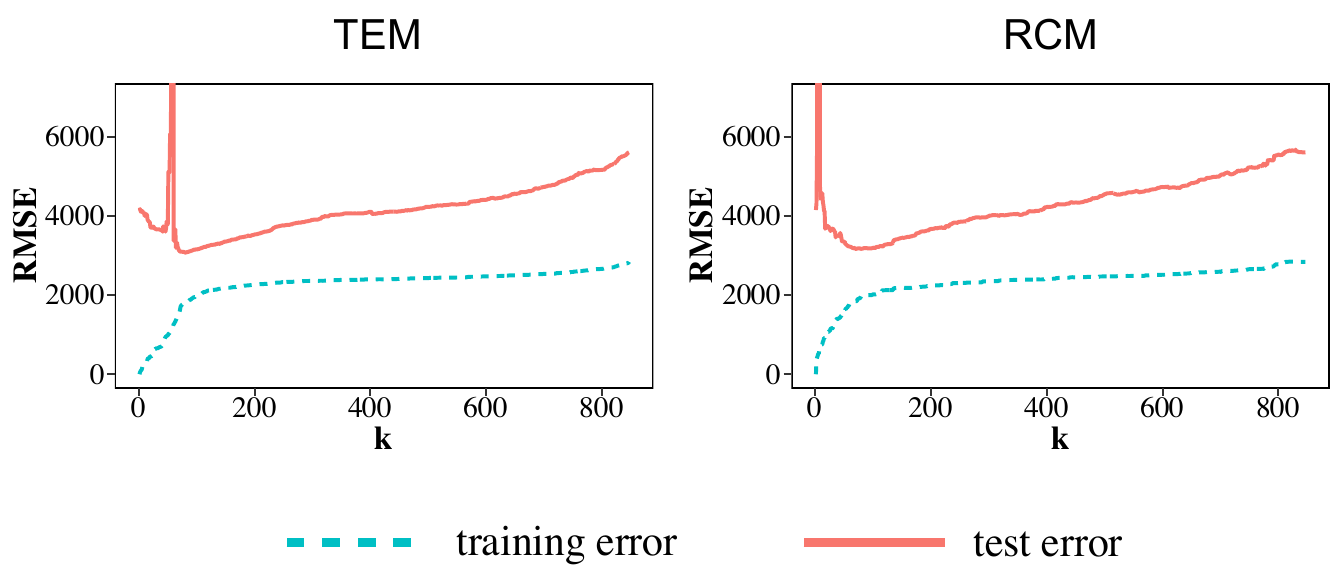}
\caption{RMSE for training and test datasets as a function of the number of clusters $k$ in all areas.}
\label{fig:RMSE ALL}
\end{figure}
For both clustering methods, the training error is almost a monotone-increasing function of $k$, suggesting that the number of clusters is considered as the model complexity. The test error curves for the two methods are similar but exhibit some differences. The RCM shows relatively high instability when $k$ is small, whereas the TEM produces a smoother and clearer double-descent phenomenon, which is similar to the results of Figure \ref{fig:sim_K500} in the simulation study. At the minimum point, the TEM achieves slightly better RMSE than the RCM. Meanwhile, the RCM is much faster than the TEM. Therefore, a choice of the RCM and TEM depends on the user's priority: computational speed or prediction accuracy.

For both TEM and RCM, the descent corresponding to a larger number of clusters achieves better performance than the one corresponding to a smaller number of clusters.
Therefore, our forecasting model in \eqref{eq:forecastmodel} performs well enough to achieve better performance than the interpolation regime.  

\section{Concluding remarks}
We have proposed a novel regression model specifically for aggregate value, called the Aggregate Value Regression with Clustering (AVR-C). AVR-C can be seen as a generalization of Individual Regression (IR) and Aggregate Value Regression (AVR) and exhibits the bias-variance trade-off according to the number of clusters.  Two hierarchical clustering techniques specifically for the AVR-C are proposed: training error minimization (TEM) and residual correlation matrix (RCM). 
Simulation results demonstrated a bias-variance trade-off, which coincides with the theoretical behavior of training and test errors as a function of the number of clusters. The applicability and effectiveness of the proposed methods were validated through the analysis of electricity demand data.

As a future research topic, it would be interesting to focus on the theoretical validation of double descent behavior. The double descent phenomenon has been rapidly developed in recent years in deep learning literature, but a unified theory has not yet been established to our knowledge. 
We would like to carefully investigate the behavior of double descent phenomenon with application to various real data, and hopefully obtain a theoretical result that explains the patterns observed in the empirical analyses.  

Another future research topic would be the extension of AVR-C to spatial data analysis.  In spatial data analysis, we usually take the correlation in the spacial locations into consideration; a similar result is obtained for nearby locations.  When we apply the regression models to each location, the number of regression models, $M$, can be large.  Indeed, with the development of the advancement of censoring technologies, micro-scale data have been frequently obtained. In some cases, the analysis with macro-scale data would result in better performance than that with micro-scale data in terms of aggregate value prediction. Therefore, the scale of the dataset can be selected with AVR-C.  In this case, the clustering can be achieved with both spacial information and correlations of residuals. A theoretical investigation specifically to the AVR-C in spatial data analysis would also be of interest.

\appendix
\section{Appendix}

\subsection{Proof of Lemma \ref{lemma:training_3reg_misspecified}}\label{proof:lemma:training_3reg_misspecified}

First, the term $R$ is written as 
	\begin{align}
			R &= (\bm{y}_3 - \hat{\bm{y}}_3)^T\{(\bm{y} - \hat{\bm{y}}) - (\bm{y} - \hat{\bm{y}}^*)\}\crcr
			 &= (\bm{y}_3 - \hat{\bm{y}}_3)^T(\hat{\bm{y}}^*-\hat{\bm{y}})\crcr
			 &= \bm{y}_3^T(\bm{I}_n - \bm{H}_3)\{(\bm{H}_1 - \bm{H})\bm{y}_1 + (\bm{H}_2 - \bm{H})\bm{y}_2\}\crcr
			 &= (\bm{W}_3\bm{\theta}_3 + \bm{\varepsilon}_3)^T(\bm{I}_n - \bm{H}_3)\{(\bm{H}_1 - \bm{H})(\bm{W}_1\bm{\theta}_1 + \bm{\varepsilon}_1) + (\bm{H}_2 - \bm{H})(\bm{W}_2\bm{\theta}_2 + \bm{\varepsilon}_2)\}\crcr
			 &=\bm{\varepsilon}_3^T(\bm{I}_n - \bm{H}_3)\{(\bm{H}_1 - \bm{H})\bm{\varepsilon}_1 + (\bm{H}_2 - \bm{H})\bm{\varepsilon}_2\}\crcr
			 &\quad+\bm{\varepsilon}_3^T(\bm{I}_n - \bm{H}_3)\{(\bm{H}_1 - \bm{H})\bm{W}_1\bm{\theta}_1 + (\bm{H}_2 - \bm{H})\bm{W}_2\bm{\theta}_2\}\crcr
			 &\quad+(\bm{W}_3\bm{\theta}_3)^T(\bm{I}-\bm{H}_3)(\bm{H}_1 - \bm{H})(\bm{W}_1\bm{\theta}_1)   +(\bm{W}_3\bm{\theta}_3)^T(\bm{I}-\bm{H}_3)(\bm{H}_2-\bm{H})(\bm{W}_2\bm{\theta}_2)\crcr
			 &\quad+(\bm{W}_3\bm{\theta}_3)^T(\bm{I}-\bm{H}_3)(\bm{H}_1 - \bm{H})\bm{\varepsilon}_1   +(\bm{W}_3\bm{\theta}_3)^T(\bm{I}-\bm{H}_3)(\bm{H}_2-\bm{H})(\bm{W}_2\bm{\theta}_2)\bm{\varepsilon}_2.\crcr
			 \label{eq:R_expansion}
\end{align}
	Since $E[\bm{\varepsilon}\bm{\varepsilon}_3^T] = \bm{\Sigma}_{13}+\bm{\Sigma}_{23}$ and $E[\bm{\varepsilon}_m\bm{\varepsilon}_{m'}^T] = \bm{\Sigma}_{mm'}$ $(m \neq m')$, the expectation of first line of \eqref{eq:R_expansion} is expressed as 
\begin{align*}
	&E[\bm{\varepsilon}_3^T(\bm{I}_n - \bm{H}_3)\{(\bm{H}_1 - \bm{H})\bm{\varepsilon}_1 + (\bm{H}_2 - \bm{H})\bm{\varepsilon}_2\}] \\
	=&  \tr\{(\bm{I}_n - \bm{H}_3)(\bm{H}_1 - \bm{H})\bm{\Sigma}_{13}\}
	+\tr\{(\bm{I}_n - \bm{H}_3)(\bm{H}_2 - \bm{H})\bm{\Sigma}_{23}\}. 
	\end{align*}
The expectations of second and fourth lines of \eqref{eq:R_expansion} are zeros because $E[\bm{\varepsilon}_m]=\bm{0}$ ($m=1,2,3$), and the third line is a constant term with respect to $\bm{\varepsilon}_m$ $(m=1,2,3)$.  As a result, we obtain \eqref{eq:proof_W3}.  Furthermore, taking the expectation with respect to $\bm{W}_1$, $\bm{W}_2$, and $\bm{W}_3$, we have   
	\begin{align*}
			E[R]&= \tr\{(\bm{I}_n - \bm{H}_3)(\bm{H}_1 - \bm{H})\bm{\Sigma}_{13}\} + \tr\{(\bm{I}_n - \bm{H}_3)(\bm{H}_2 - \bm{H})\bm{\Sigma}_{23}\}.		\end{align*}
Recall that $\bm{H}_1$ and $\bm{H}$ are the projection matrices onto the linear spaces spanned by the column vectors of $\bm{X}_1$ and $(\bm{X}_1,\bm{X}_2)$, respectively.  Because $\bm{H}$ projects vectors onto a larger subspace than $\bm{H}_1$, we obtain $\bm{H}_1 < \bm{H}$.  As a result,  we have $\tr\{(\bm{I}_n - \bm{H}_3)(\bm{H}_1 - \bm{H})\} = \tr\{(\bm{I}_n - \bm{H}_3)^{1/2}(\bm{H}_1 - \bm{H})(\bm{I}_n - \bm{H}_3)^{1/2}\} \leq 0$.  Similarly, we obtain $\tr\{(\bm{I}_n - \bm{H}_3)(\bm{H}_2 - \bm{H})\} \leq 0$.  Since $\bm{\Sigma}_{13}\geq 0$ and  $\bm{\Sigma}_{23}\geq 0$, we have $E[R] \leq 0$, which complete the proof.

\subsection{Proof of Theorem \ref{thm:correlation_test_misspecified}} \label{proof:thm:correlation_test_misspecified}
	The test error for AVR is
	 \begin{align*}
	&\|(\bm{z}+\bm{z}_3)-(\hat{\bm{y}}+\hat{\bm{y}}_3)\|_2^2 \\
	&= \|(\bm{z}-\bm{X}\hat{\bm{\beta}})+(\bm{z}_3-\bm{X}_3\hat{\bm{\beta}}_3)\|_2^2 \\
	&= \|\bm{z}-\bm{X}\hat{\bm{\beta}}\|_2^2+\|\bm{z}_3-\bm{X}_3\hat{\bm{\beta}}_3\|_2^2 + 2(\bm{z}-\bm{X}\hat{\bm{\beta}})^T(\bm{z}_3-\bm{X}_3\hat{\bm{\beta}}_3) \\
	&= \|\bm{\eta}+\bm{W}\bm{\theta}+\bm{X}(\bm{\beta}-\hat{\bm{\beta}})\|_2^2 + \|\bm{\eta}_3+\bm{W}_3\bm{\theta}_3+\bm{X}_3(\bm{\beta}_3-\hat{\bm{\beta}}_3)\|_2^2 \\
	&\quad+ 2(\bm{\eta}+\bm{W}\bm{\theta}+\bm{X}(\bm{\beta}-\hat{\bm{\beta}}))^T(\bm{\eta}_3+\bm{W}_3\bm{\theta}_3+\bm{X}_3(\bm{\beta}_3-\hat{\bm{\beta}}_3)).
\end{align*}
Here, the prediction values based on the AVR is expressed as
\begin{align*}
	\hat{\bm{y}}=\bm{X}\hat{\bm{\beta}} &= \bm{H}\bm{y}=\bm{H}(\bm{X}\bm{\beta}+\bm{W}\bm{\theta} + \bm{\varepsilon})=\bm{X}\bm{\beta}+\bm{H}\bm{W}\bm{\theta} +\bm{H}\bm{\varepsilon}.
\end{align*}
Therefore, we have
$$
\bm{\eta}+\bm{W}\bm{\theta}+\bm{X}(\bm{\beta}-\hat{\bm{\beta}}) = (\bm{I}-\bm{H})\bm{W}\bm{\theta}+\bm{\eta}-\bm{H}\bm{\varepsilon}.
$$
The expectation of test error for AVR is then expressed as
\begin{align*}
&E\left[\|(\bm{z}+\bm{z}_3)-(\hat{\bm{y}}+\hat{\bm{y}}_3)\|_2^2 \mid \bm{W}_1,\bm{W}_2,\bm{W}_3\right] \\
&=  \|(\bm{I}-\bm{H})\bm{W}\bm{\theta}\|^2 + E[\|\bm{\eta}\|^2] + E[\|\bm{H}\bm{\varepsilon}\|_2^2] \\
& \quad+ \|(\bm{I}-\bm{H}_3)\bm{W}_3\bm{\theta}_3\|^2 + E[\|\bm{\eta}_3\|^2] + E[\|\bm{H}_3\bm{\varepsilon}_3\|_2^2]\\
& \quad+2 (\bm{W}\bm{\theta})^T(\bm{I}-\bm{H})(\bm{I}-\bm{H}_3)\bm{W}_3\bm{\theta}_3 +2 E[\bm{\eta}^T\bm{\eta}_3]+2 E[(\bm{H}\bm{\varepsilon})^T\bm{H}_3\bm{\varepsilon}_3] .
\end{align*}
Similarly, the prediction values based on the IR is expressed as
\begin{align*}
	\bm{X}\hat{\bm{\beta}}^* &= \bm{X}_1\hat{\bm{\beta}}_1^*+\bm{X}_2\hat{\bm{\beta}}_2^*\\
	&= \bm{H}_1\bm{y}_1+\bm{H}_2\bm{y}_2\\
	&=\bm{X}_1\bm{\beta}_1+\bm{H}_1\bm{W}_1\bm{\theta}_1 +\bm{H}_1\bm{\varepsilon}_1 + \bm{X}_2\bm{\beta}_2+\bm{H}_2\bm{W}_2\bm{\theta}_2+\bm{H}_2\bm{\varepsilon}_2\\
	&=\bm{X}\bm{\beta}+\bm{H}_1\bm{W}_1\bm{\theta}_1+\bm{H}_2\bm{W}_2\bm{\theta}_2+\bm{H}_1\bm{\varepsilon}_1 + \bm{H}_2\bm{\varepsilon}_2.
\end{align*}
Combined with the result of
$$
\bm{\eta}+\bm{W}\bm{\theta}+\bm{X}(\bm{\beta}-\hat{\bm{\beta}}^*) = (\bm{I}-\bm{H}_1)\bm{W}_1\bm{\theta}_1 +  (\bm{I}-\bm{H}_2)\bm{W}_2\bm{\theta}_2+\bm{\eta}-\bm{H}_1\bm{\varepsilon}_1-\bm{H}_2\bm{\varepsilon}_2,
$$
the expectation of the test error for IR is 
\begin{align*}
&E\left[\|(\bm{z}+\bm{z}_3)-(\hat{\bm{y}}^*+\hat{\bm{y}}_3)\|_2^2 \mid \bm{W}_1,\bm{W}_2,\bm{W}_3\right] \\
&=   \|(\bm{I}-\bm{H}_1)\bm{W}_1\bm{\theta}_1\|^2+\|(\bm{I}-\bm{H}_2)\bm{W}_2\bm{\theta}_2\|^2 +2(\bm{W}_1\bm{\theta}_1)^T(\bm{I}-\bm{H}_1)(\bm{I}-\bm{H}_2)(\bm{W}_2\bm{\theta}_2)\\
& \quad + E[\|\bm{\eta}\|^2] + E[\|\bm{H}_1\bm{\varepsilon}_1 + \bm{H}_2\bm{\varepsilon}_2\|_2^2] \\
& \quad+ \|(\bm{I}-\bm{H}_3)\bm{W}_3\bm{\theta}_3\|^2 + E[\|\bm{\eta}_3\|^2] + E[\|\bm{H}_3\bm{\varepsilon}_3\|_2^2]\\
& \quad+ 2 (\bm{W}_1\bm{\theta}_1)^T(\bm{I}-\bm{H}_1)(\bm{I}-\bm{H}_3)\bm{W}_3\bm{\theta}_3 + 2 (\bm{W}_2\bm{\theta}_2)^T(\bm{I}-\bm{H}_2)(\bm{I}-\bm{H}_3)\bm{W}_3\bm{\theta}_3 \\
&\quad+2 E[\bm{\eta}^T\bm{\eta}_3]+2 E[(\bm{H}_1\bm{\varepsilon}_1 + \bm{H}_2\bm{\varepsilon}_2)^T\bm{H}_3\bm{\varepsilon}_3] .
\end{align*}
The expectations related to error variables are expressed as 
\begin{align*}
	E[\|\bm{H}\bm{\varepsilon}\|^2]&
	=\tr(\bm{H}V[\bm{\varepsilon}_1+\bm{\varepsilon}_2])=\tr\{\bm{H}(\bm{\Sigma}_{11}+\bm{\Sigma}_{22}+2\bm{\Sigma}_{12})\}\\
	E[\|\bm{H}_1\bm{\varepsilon}_1+\bm{H}_2\bm{\varepsilon}_2\|^2]&=E[\|\bm{H}_1\bm{\varepsilon}_1\|^2] + E[\|\bm{H}_2\bm{\varepsilon}_2\|^2] + 2E[(\bm{H}_1\bm{\varepsilon}_1)^T\bm{H}_2\bm{\varepsilon}_2]\\
	&=\tr(\bm{H}_1\bm{\Sigma}_{11})+\tr(\bm{H}_2\bm{\Sigma}_{22}) + 2\tr(\bm{H}_1\bm{H}_2\bm{\Sigma}_{12}),\\
	E[(\bm{H}\bm{\varepsilon})^T\bm{H}_3\bm{\varepsilon}_3]&=\tr(\bm{H}\bm{H}_3E[\bm{\varepsilon}_3\bm{\varepsilon}^T])=\tr\{\bm{H}\bm{H}_3(\bm{\Sigma}_{13}+\bm{\Sigma}_{23})\}	,\\
	E[(\bm{H}_1\bm{\varepsilon}_1+\bm{H}_2\bm{\varepsilon}_2)^T\bm{H}_3\bm{\varepsilon}_3]&=\tr(\bm{H}_1\bm{H}_3E[\bm{\varepsilon}_3\bm{\varepsilon}_1^T]) + \tr(\bm{H}_2\bm{H}_3E[\bm{\varepsilon}_3\bm{\varepsilon}_2^T])=\tr(\bm{H}_1\bm{H}_3\bm{\Sigma}_{13}) + \tr(\bm{H}_2\bm{H}_3\bm{\Sigma}_{23}).
\end{align*}
Thus, we obtain
\begin{align*}
&E\left[\|(\bm{z}+\bm{z}_3)-(\hat{\bm{y}}+\hat{\bm{y}}_3)\|_2^2 \mid \bm{W}_1,\bm{W}_2,\bm{W}_3\right] - E\left[\|(\bm{z}+\bm{z}_3)-(\hat{\bm{y}}^*+\hat{\bm{y}}_3)\|_2^2 \mid \bm{W}_1,\bm{W}_2,\bm{W}_3\right]\\
&=
\tr\{(\bm{H}-\bm{H}_1)\bm{\Sigma}_{11}\} 
+ \tr\{(\bm{H}-\bm{H}_2)\bm{\Sigma}_{22}\}
+2\tr\{(\bm{H}-\bm{H}_1\bm{H}_2)\bm{\Sigma}_{12}\}\\
&\quad 
+2\tr\{(\bm{H}-\bm{H}_1)\bm{H}_3\bm{\Sigma}_{13}\}
+ 2\tr\{(\bm{H}-\bm{H}_2)\bm{H}_3\bm{\Sigma}_{23}\}\\
&\quad+ \|(\bm{I}-\bm{H})\bm{W}_1\bm{\theta}_1\|^2-\|(\bm{I}-\bm{H}_1)\bm{W}_1\bm{\theta}_1\|^2
+ \|(\bm{I}-\bm{H})\bm{W}_2\bm{\theta}_2\|^2-\|(\bm{I}-\bm{H}_2)\bm{W}_2\bm{\theta}_2\|^2\\
&\quad+2(\bm{W}_1\bm{\theta}_1)^T(\bm{I}-\bm{H})(\bm{W}_2\bm{\theta}_2)   - 2(\bm{W}_1\bm{\theta}_1)^T(\bm{I}-\bm{H}_1)(\bm{I}-\bm{H}_2)(\bm{W}_2\bm{\theta}_2)\\
			 &\quad+2(\bm{W}_3\bm{\theta}_3)^T(\bm{I}-\bm{H}_3)(\bm{H}_1 - \bm{H})(\bm{W}_1\bm{\theta}_1)   +2(\bm{W}_3\bm{\theta}_3)^T(\bm{I}-\bm{H}_3)(\bm{H}_2-\bm{H})(\bm{W}_2\bm{\theta}_2).  
\end{align*}
Finally, taking the expectation with respect to $\bm{W}_1$, $\bm{W}_2$, and $\bm{W}_3$, the last two lines of the above equation becomes zero.  Thus, to complete the proof, we need to prove that
\begin{equation}
    E\left[\|(\bm{I}-\bm{H})\bm{W}_m\bm{\theta}_m\|^2-\|(\bm{I}-\bm{H}_m)\bm{W}_m\bm{\theta}_m\|^2\right] = -p\bm{\theta}_m^\top\bm{\Sigma}_{\bm{W}_m}\bm{\theta}_m \quad (m=1,2).\label{eq:app:(I-H)Wtheta}
\end{equation}
Define $\bm{G}^*=\bm{I}-\bm{H}_m$. For simplicity, we drop the subscript of ``$m$" from $\bm{W}_m$ and $\bm{\theta}_m$ to express their elements, and let $\bm{\theta}_m=(\theta_1,\dots,\theta_p)$, $\bm{G}^*=(g_{ii'})$, $\bm{W}_m=(w_{ik})$, and $\bm{\Sigma}_{\bm{W}_m}=(\sigma_{kk'})$.  Then we have
\begin{align*}
    E\left[\|(\bm{I}-\bm{H}_m)\bm{W}_m\bm{\theta}_m\|^2\right]
    &=E\left[\bm{\theta}_m^\top \bm{W}_m^\top \bm{G}^*\bm{W}_m\bm{\theta}_m\right]\\ &=\sum_{i=1}^n \sum_{i'=1}^n \sum_{k=1}^p \sum_{k'=1}^p \theta_k\theta_{k'}g_{ii'}E[w_{ik}w_{i'k'}]\\
    &=\sum_{i=1}^n \sum_{k=1}^p \sum_{k'=1}^p \theta_k\theta_{k'}g_{ii}\sigma_{kk'}\\
    &=\sum_{i=1}^n g_{ii}\sum_{k=1}^p \sum_{k'=1}^p \theta_k\theta_{k'}\sigma_{kk'}\\
    &=\tr(\bm{I}-\bm{H}_m) \cdot \bm{\theta}_m^\top\bm{\Sigma}_{\bm{W}_m}\bm{\theta}_m
\end{align*}
Similarly, we obtain $E\left[\|(\bm{I}-\bm{H})\bm{W}_m\bm{\theta}_m\|^2\right] =\tr(\bm{I}-\bm{H}) \cdot \bm{\theta}_m^\top\bm{\Sigma}_{\bm{W}_m}\bm{\theta}_m$.  Since $\tr(\bm{I}-\bm{H}_m)=n-p$ and $\tr(\bm{I}-\bm{H})=n-2p$, we get Eq. \eqref{eq:app:(I-H)Wtheta}. The proof is complete.



\subsection{Simulation results for RCM}\label{sec:app:simulation}
Figure \ref{fig:sim_K50_RCM} shows MSEs of AVR-C using RCM for clustering with $M=50$ in the simulation study given in Section \ref{sec:simulation}.  
The MSEs are similar to those using TEM described in Section \ref{sec:simulation} in terms of values of MSEs and variations with the number of clusters; the training MSEs are monotonically increasing with the number of clusters, whereas the test MSEs exhibit a bias-variance trade-off.

Results for $M=500$ given in Figure \ref{fig:sim_K500_RCM} also show similar tendencies to those using TEM.  
However, the MSEs are unstable for each iteration at the smaller numbers of clusters, in which the MSEs are diverged.  
This result indicates that the clustering based on the correlation between residuals of a pair of regression models would be unstable.
\begin{figure}[!t]
	\begin{center}
		\includegraphics[width=0.3\hsize, bb= 0 0 450 300]{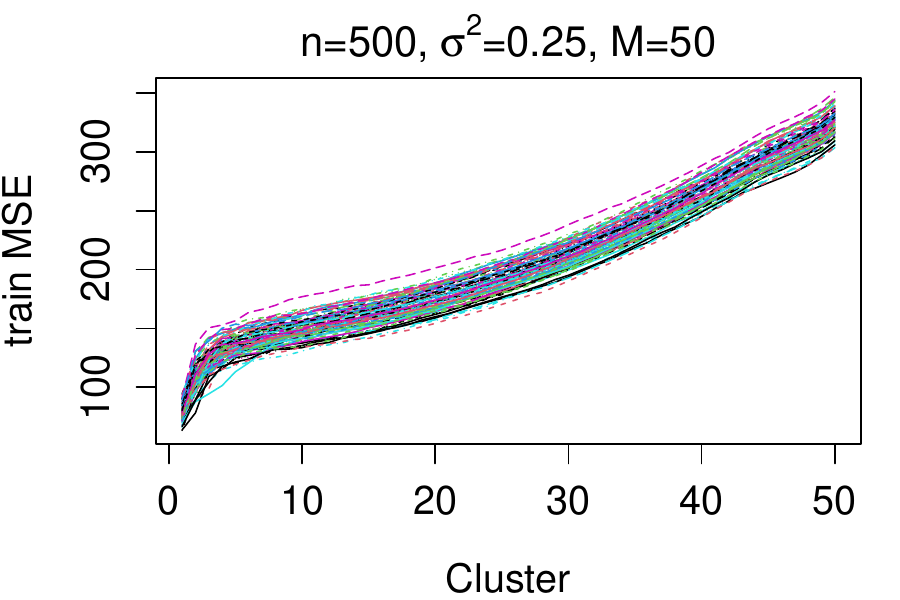}
		\includegraphics[width=0.3\hsize, bb= 0 0 450 300]{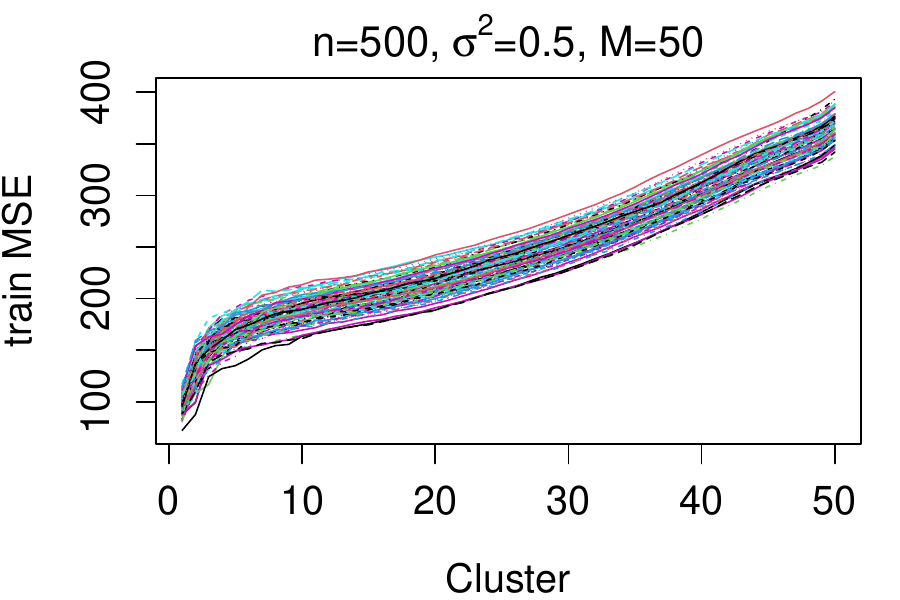}
		\includegraphics[width=0.3\hsize, bb= 0 0 450 300]{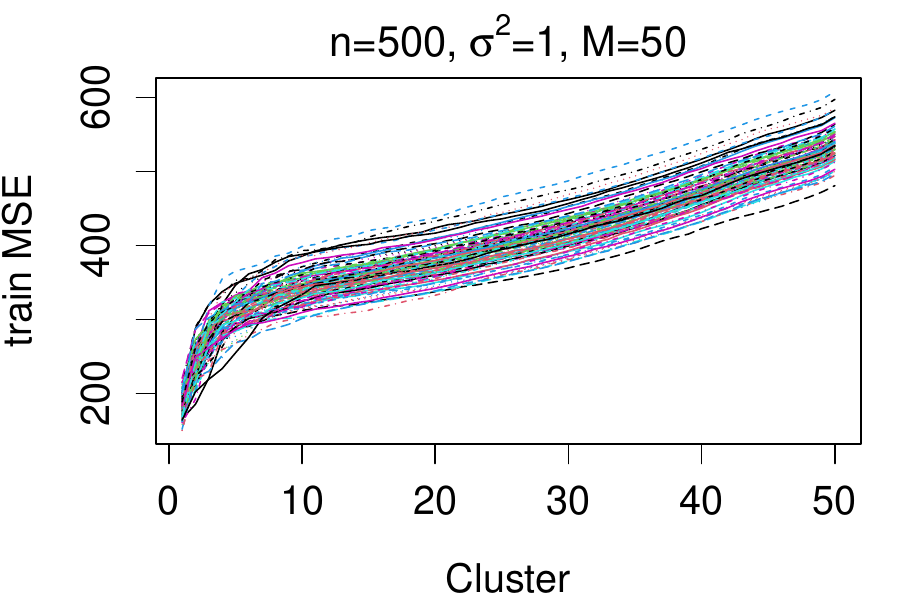} \\
		\includegraphics[width=0.3\hsize, bb= 0 0 450 300]{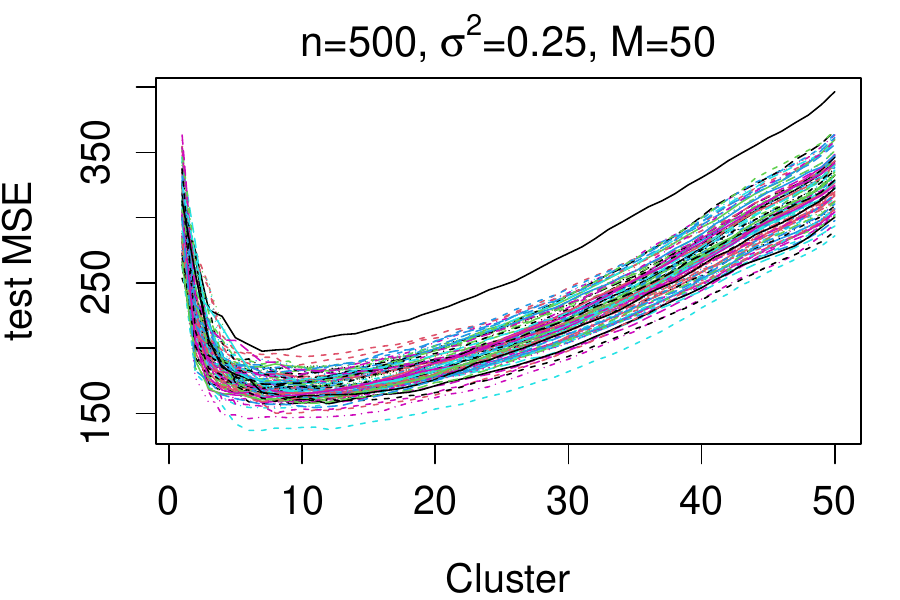}
		\includegraphics[width=0.3\hsize, bb= 0 0 450 300]{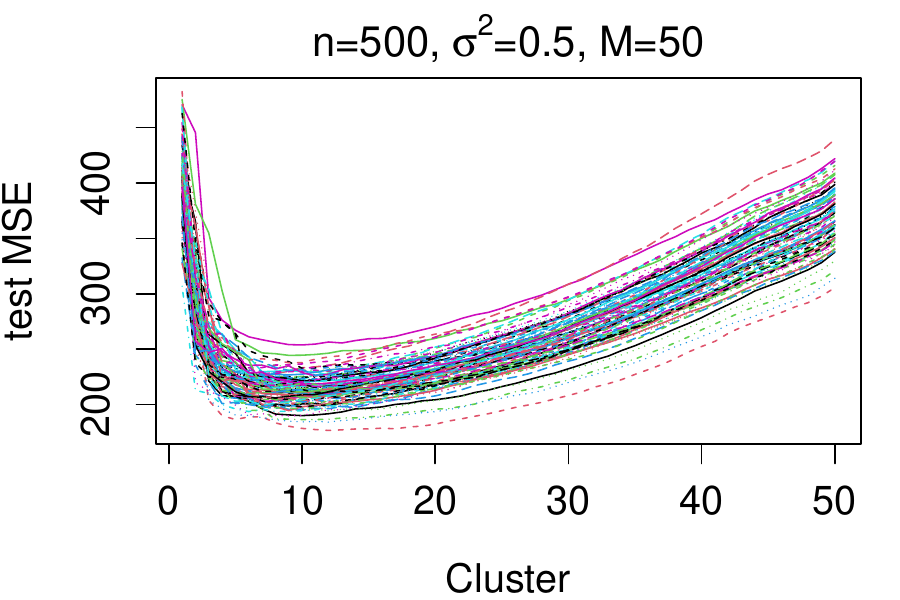}
		\includegraphics[width=0.3\hsize, bb= 0 0 450 300]{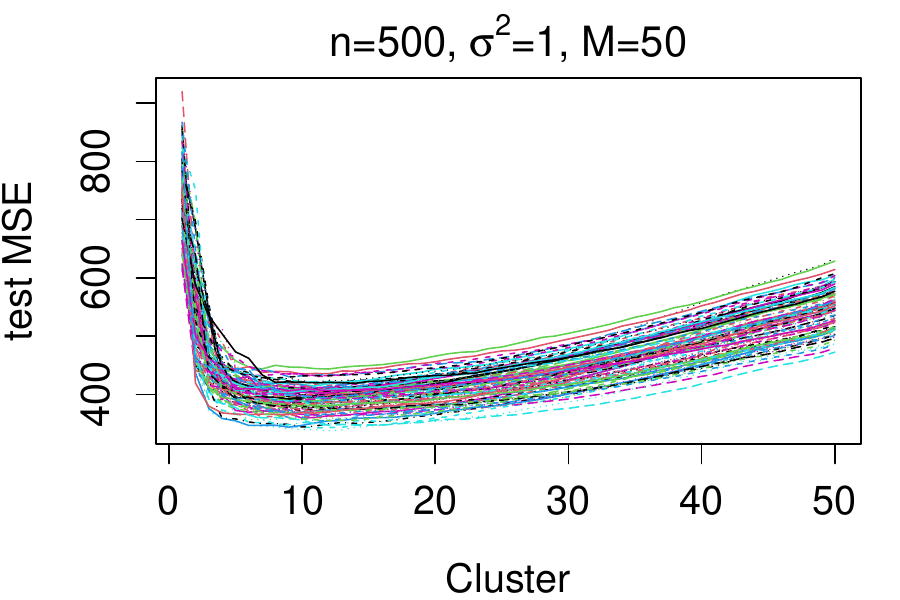}         
	\end{center}
         \vspace{-0.5cm}
	\begin{center}
		\includegraphics[width=0.3\hsize, bb= 0 0 450 300]{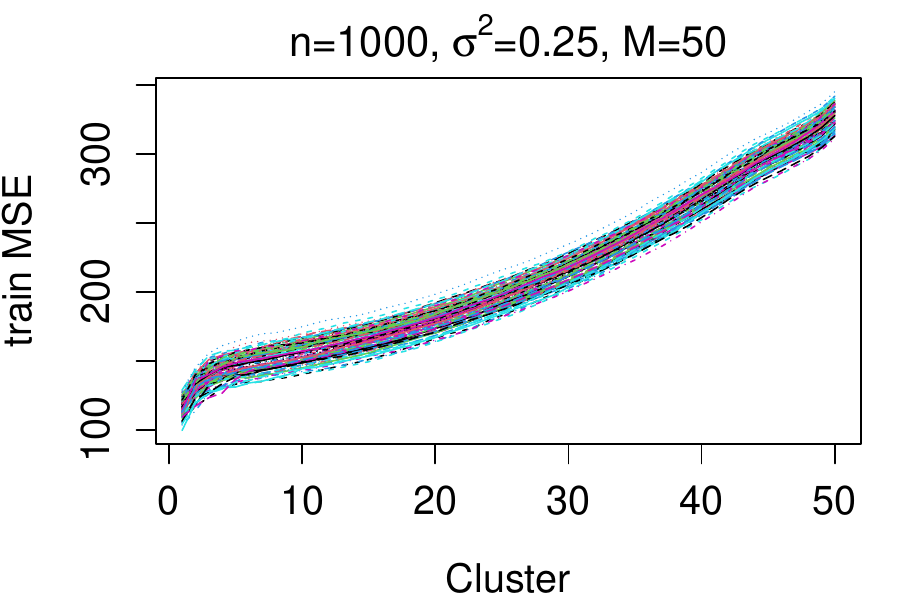}
		\includegraphics[width=0.3\hsize, bb= 0 0 450 300]{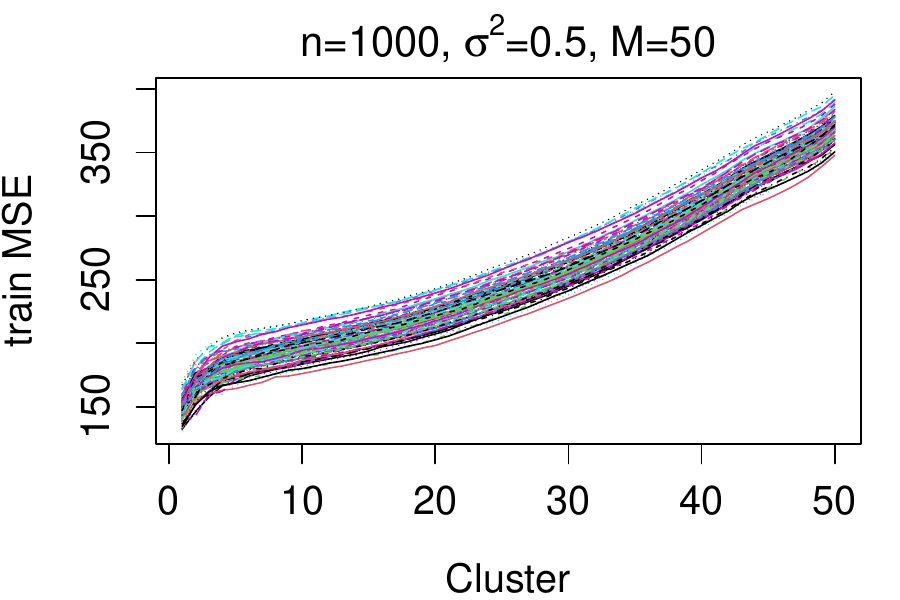}
		\includegraphics[width=0.3\hsize, bb= 0 0 450 300]{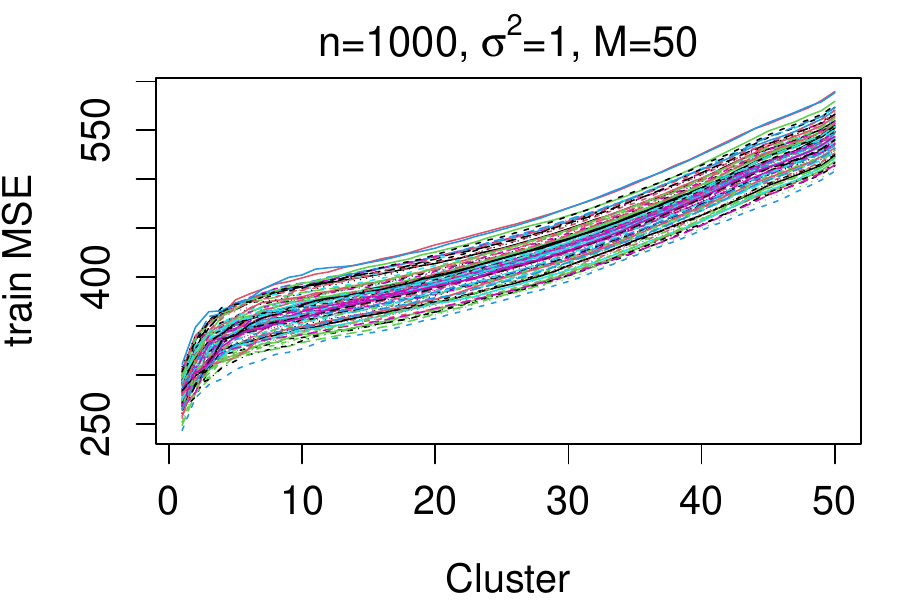} \\
		\includegraphics[width=0.3\hsize, bb= 0 0 450 300]{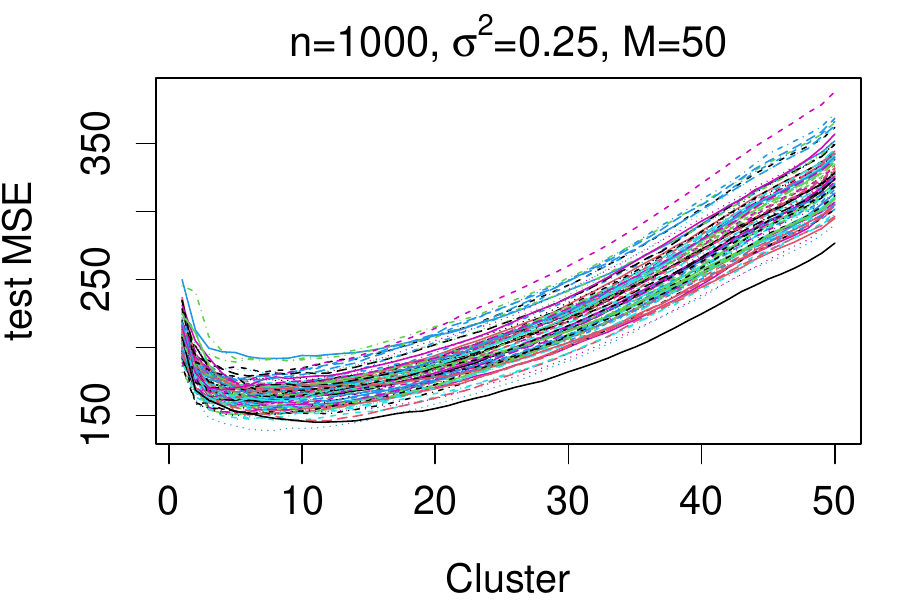}
		\includegraphics[width=0.3\hsize, bb= 0 0 450 300]{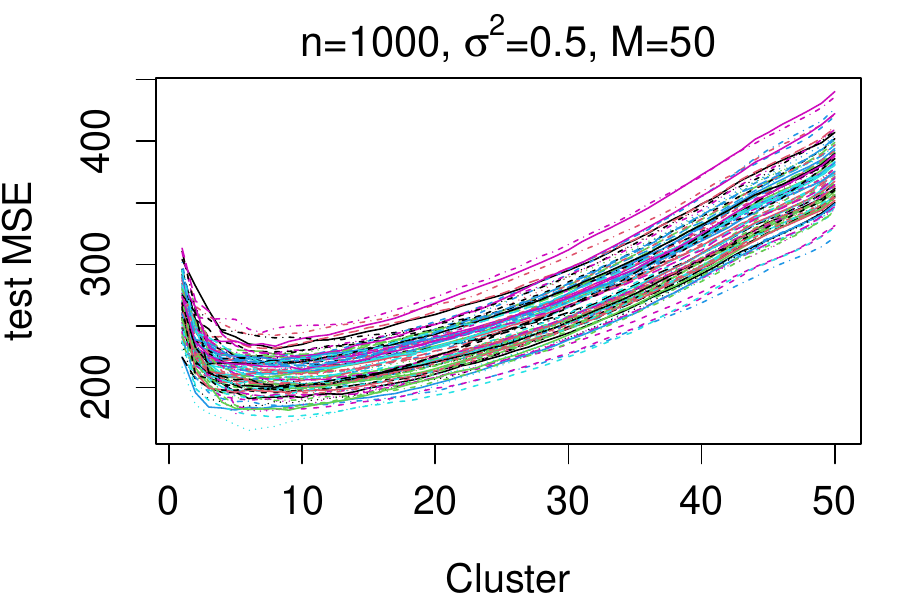}
		\includegraphics[width=0.3\hsize, bb= 0 0 450 300]{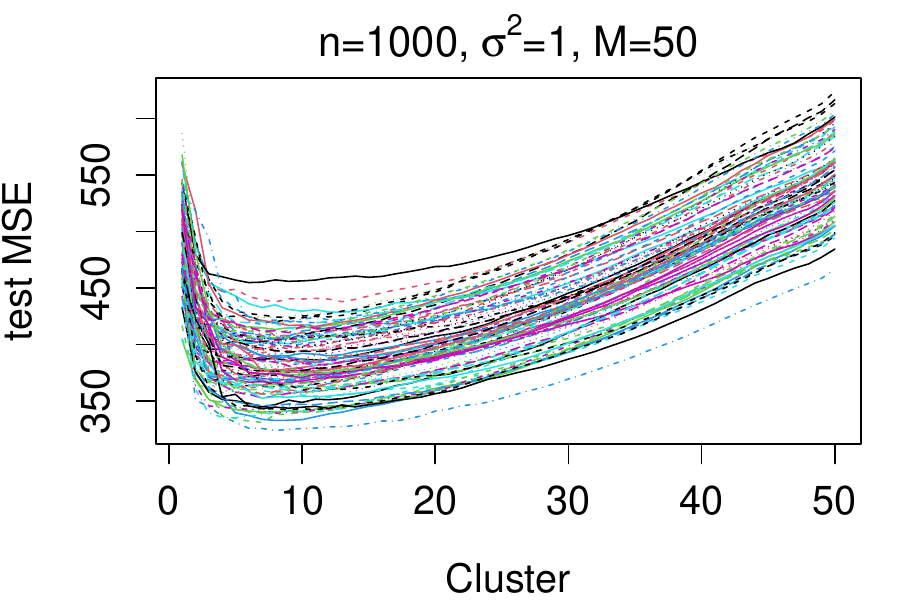}         
	\end{center}
         \vspace{-0.5cm}
 	\begin{center}
		\includegraphics[width=0.3\hsize, bb= 0 0 450 300]{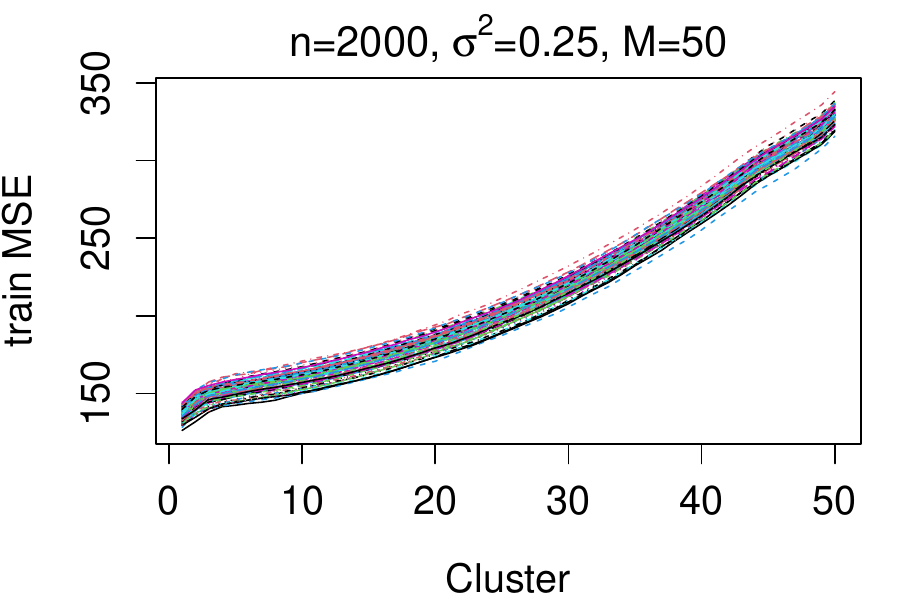}
		\includegraphics[width=0.3\hsize, bb= 0 0 450 300]{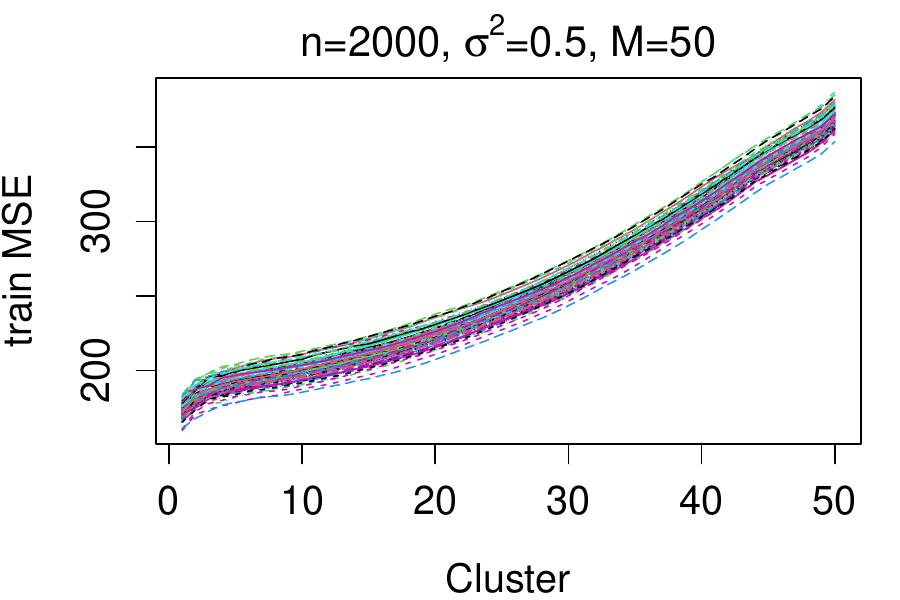}
		\includegraphics[width=0.3\hsize, bb= 0 0 450 300]{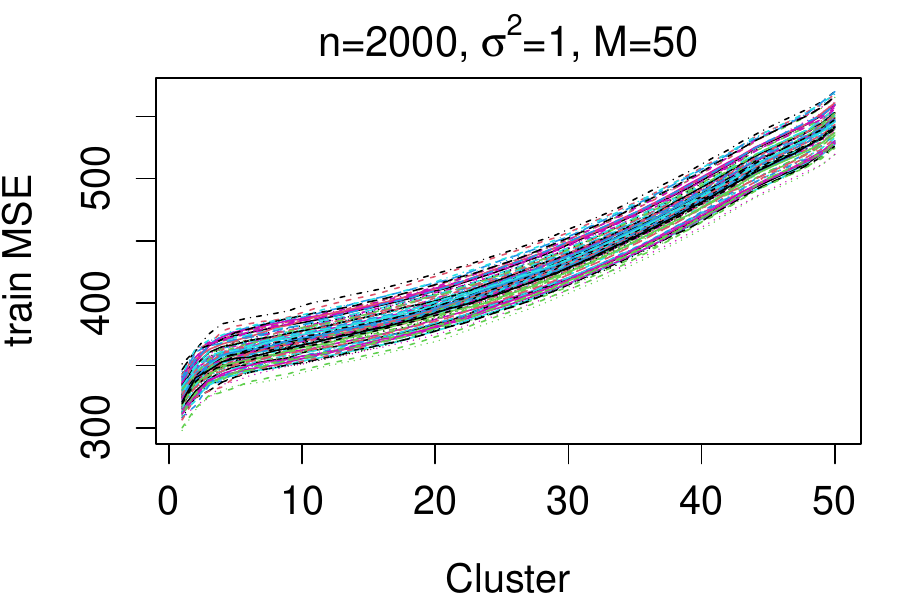} \\
		\includegraphics[width=0.3\hsize, bb= 0 0 450 300]{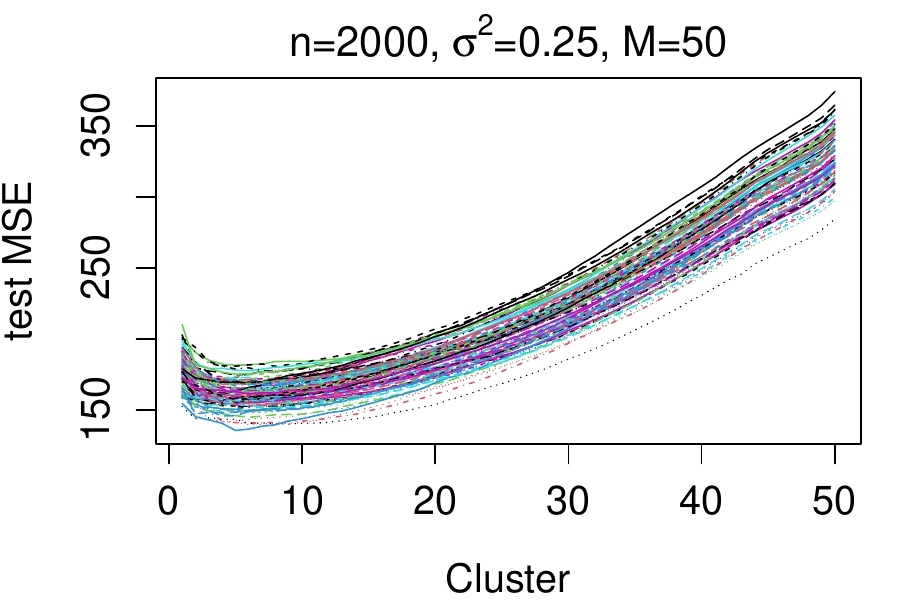}
		\includegraphics[width=0.3\hsize, bb= 0 0 450 300]{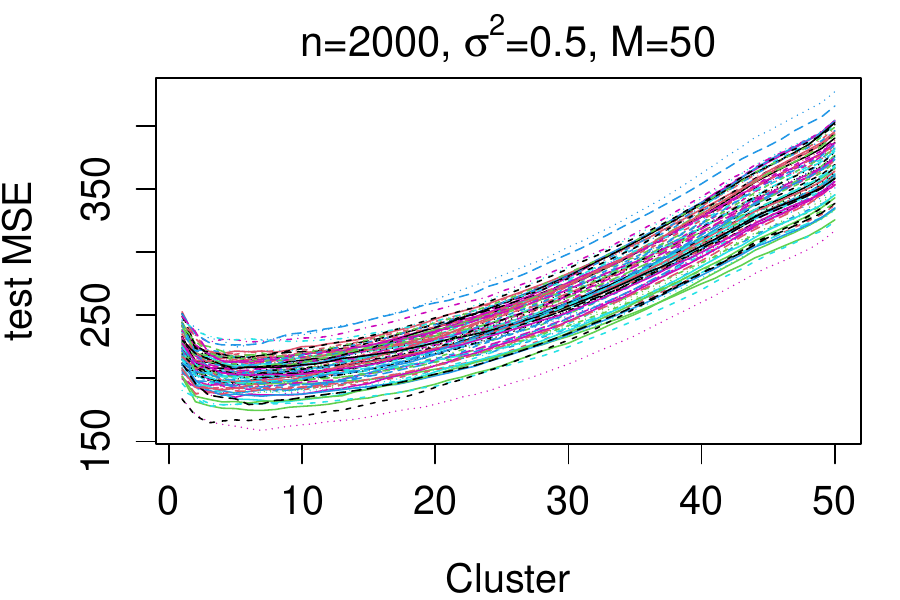}
		\includegraphics[width=0.3\hsize, bb= 0 0 450 300]{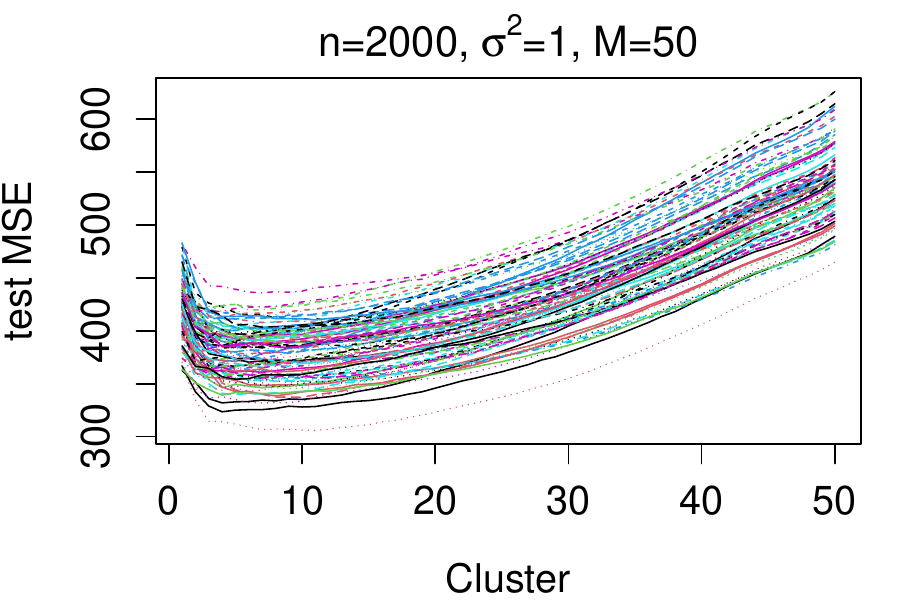}         
	\end{center}
	\caption{Simulation results for $M=50$ using RCM.} 
	\label{fig:sim_K50_RCM}

\end{figure}
 \begin{figure}[!t]
	\begin{center}
		\includegraphics[width=0.3\hsize, bb= 0 0 450 300]{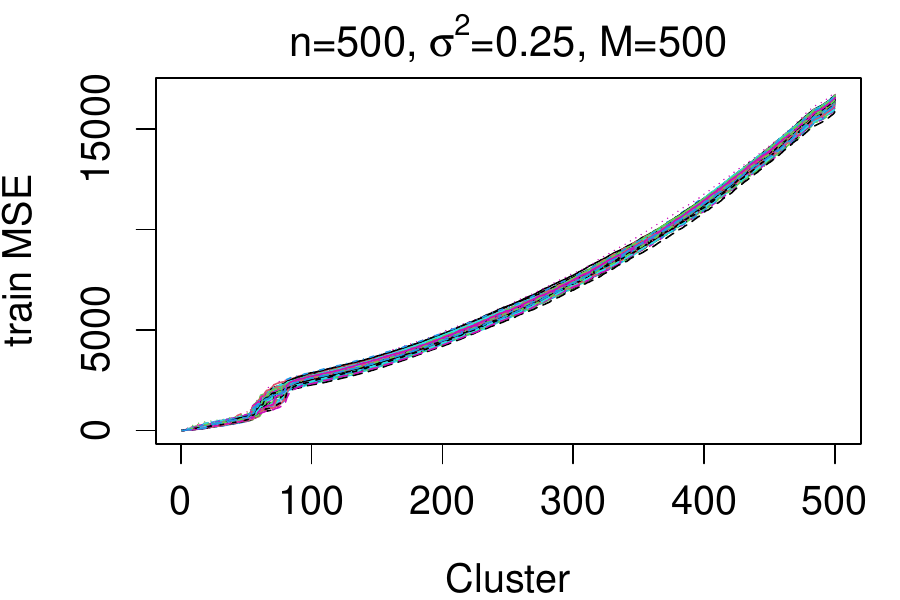}
		\includegraphics[width=0.3\hsize, bb= 0 0 450 300]{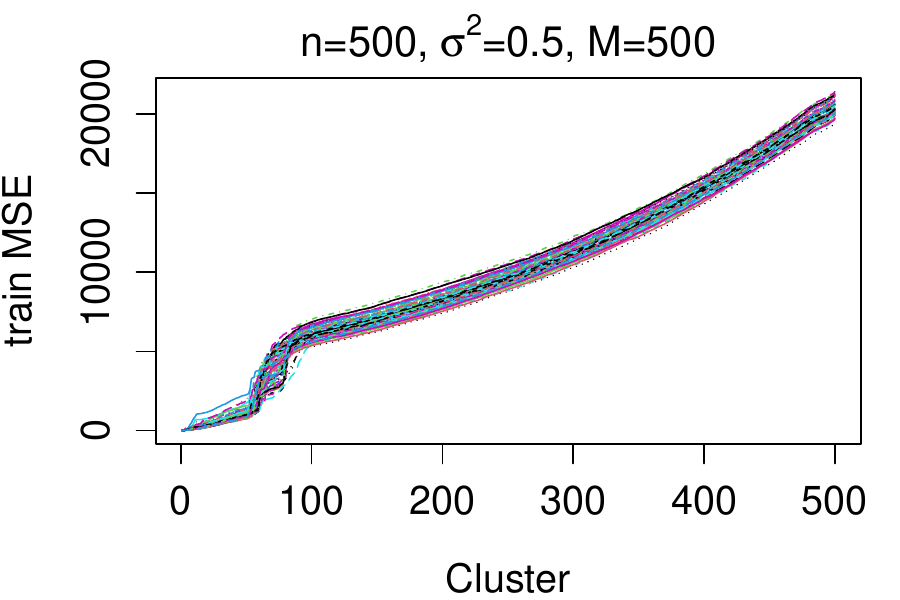}
		\includegraphics[width=0.3\hsize, bb= 0 0 450 300]{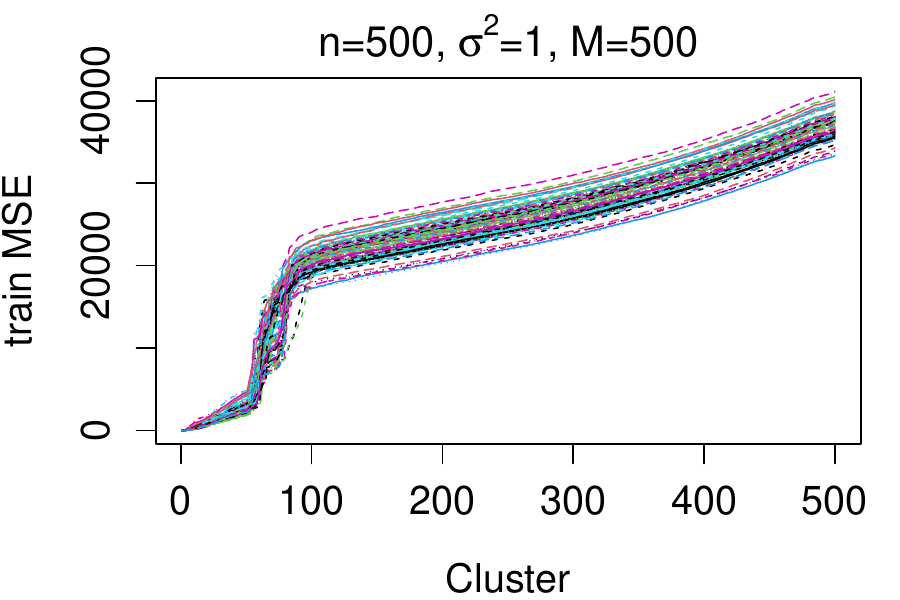} \\
		\includegraphics[width=0.3\hsize, bb= 0 0 450 300]{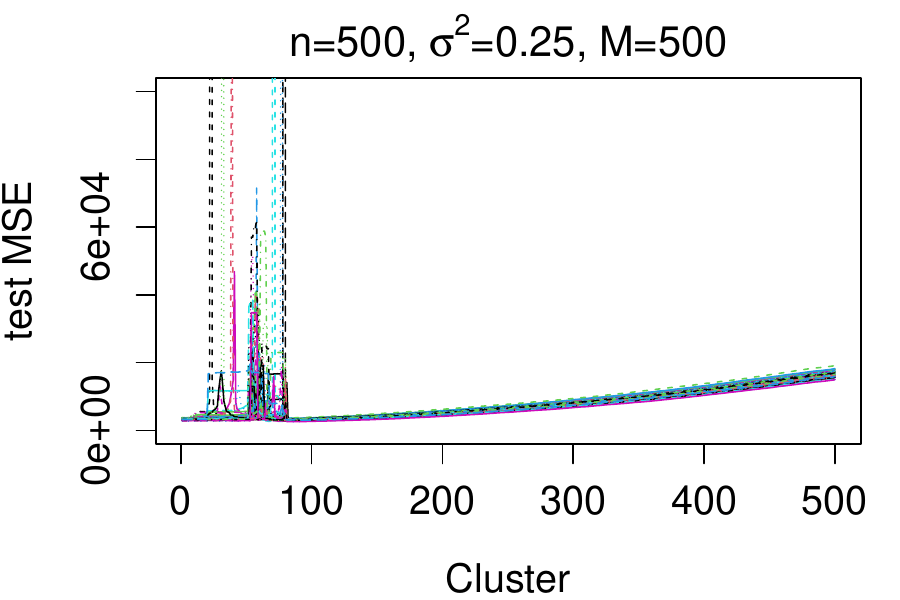}
		\includegraphics[width=0.3\hsize, bb= 0 0 450 300]{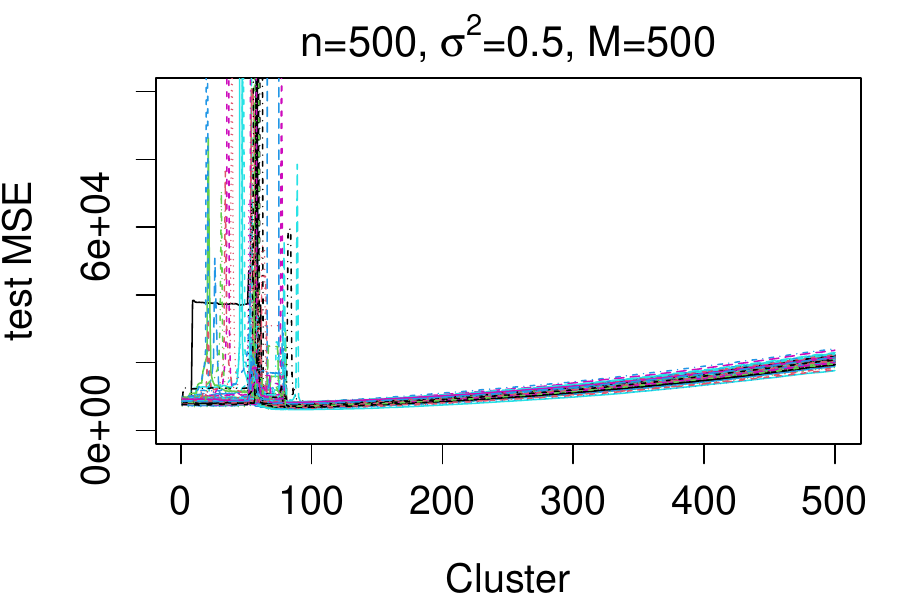}
		\includegraphics[width=0.3\hsize, bb= 0 0 450 300]{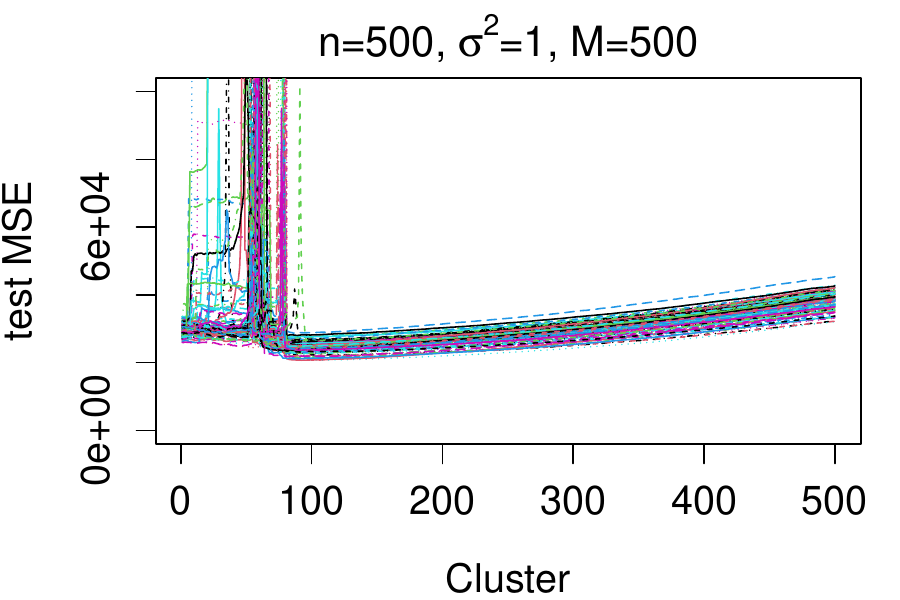}         
	\end{center}
         \vspace{-0.5cm}
	\begin{center}
		\includegraphics[width=0.3\hsize, bb= 0 0 450 300]{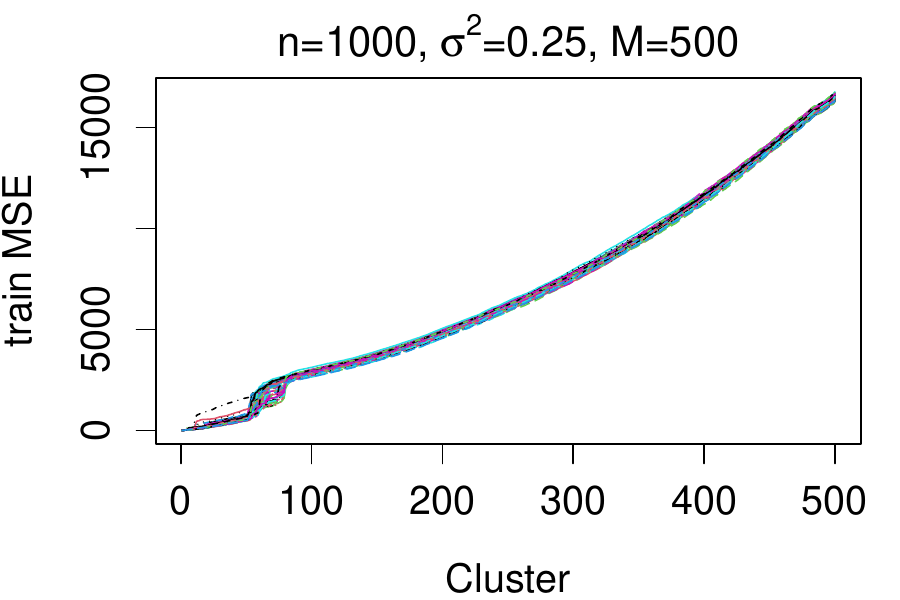}
		\includegraphics[width=0.3\hsize, bb= 0 0 450 300]{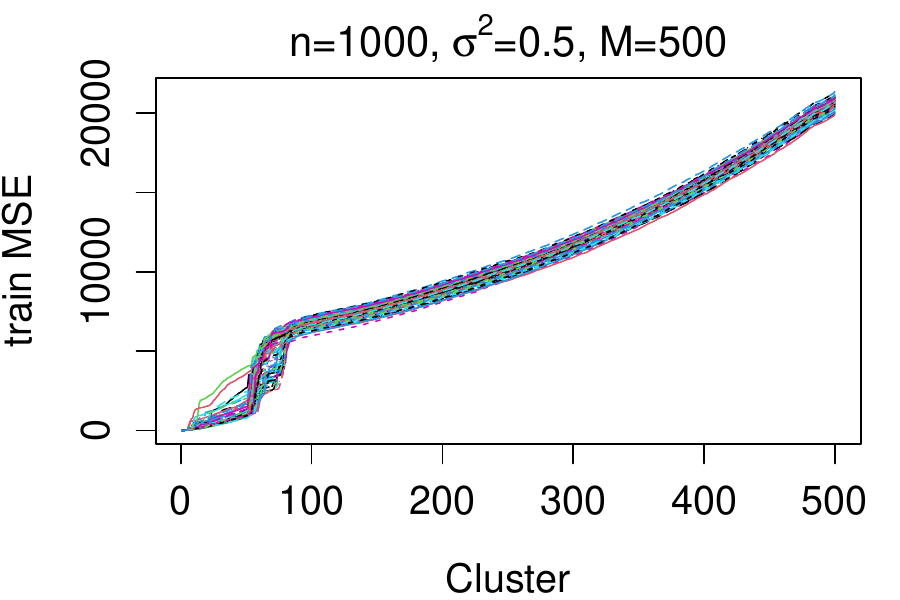}
		\includegraphics[width=0.3\hsize, bb= 0 0 450 300]{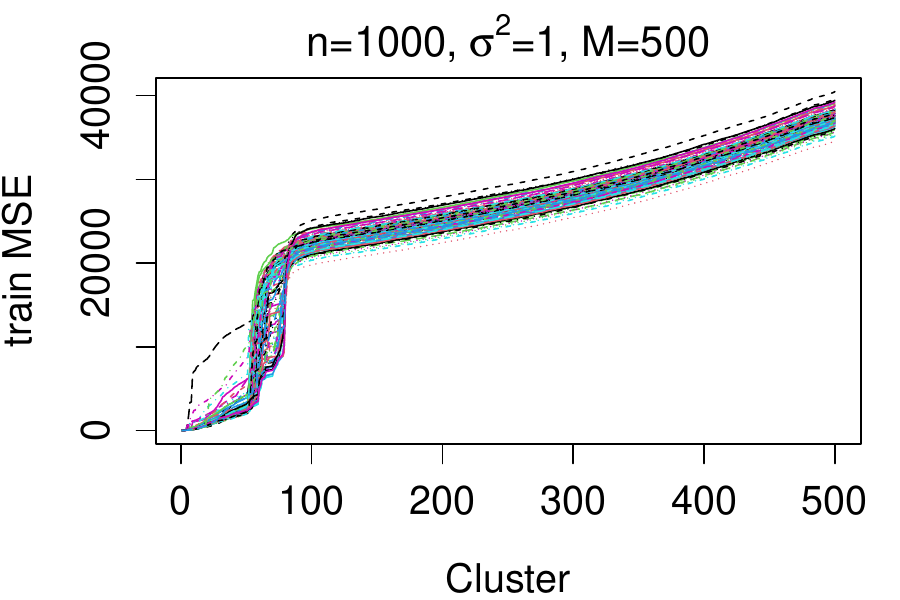} \\
		\includegraphics[width=0.3\hsize, bb= 0 0 450 300]{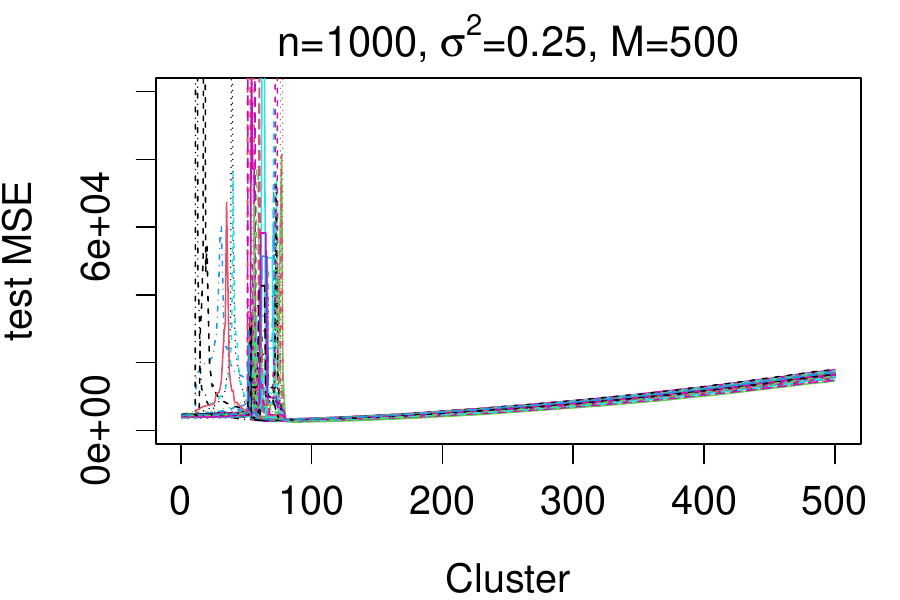}
		\includegraphics[width=0.3\hsize, bb= 0 0 450 300]{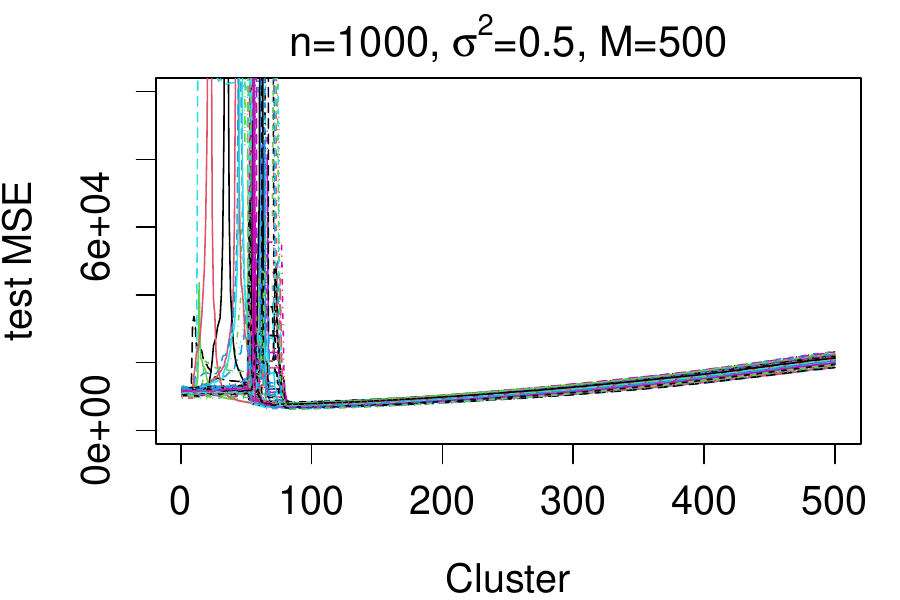}
		\includegraphics[width=0.3\hsize, bb= 0 0 450 300]{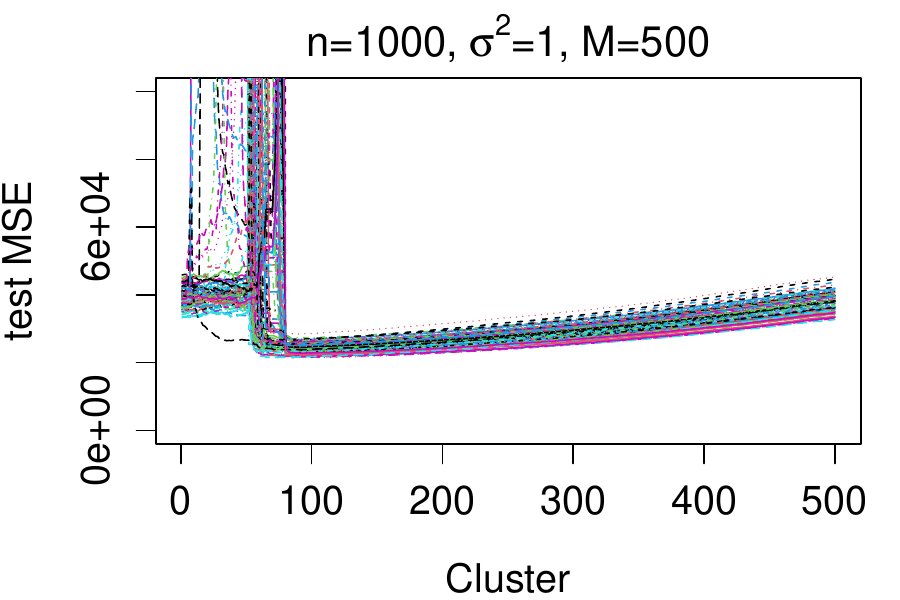}         
	\end{center}
         \vspace{-0.5cm}
 	\begin{center}
		\includegraphics[width=0.3\hsize, bb= 0 0 450 300]{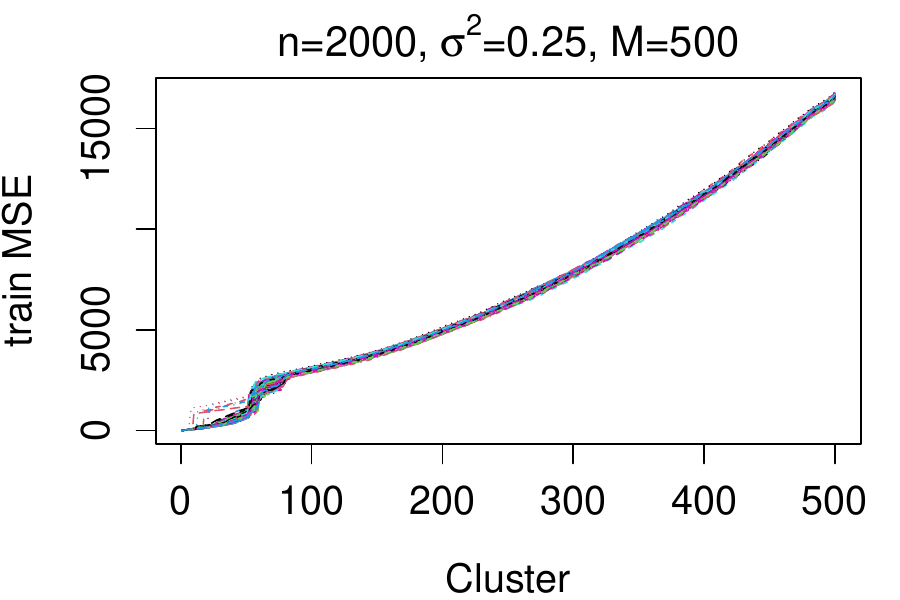}
		\includegraphics[width=0.3\hsize, bb= 0 0 450 300]{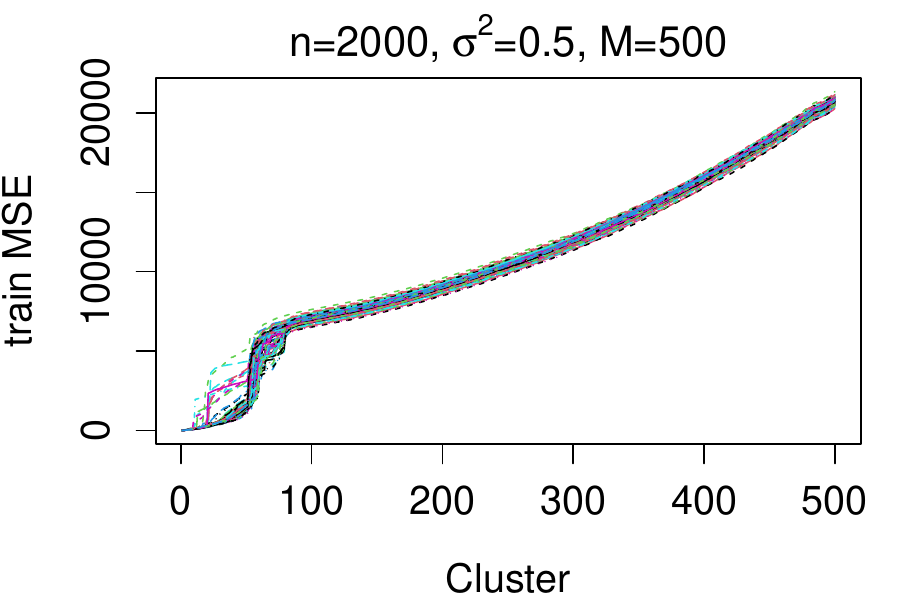}
		\includegraphics[width=0.3\hsize, bb= 0 0 450 300]{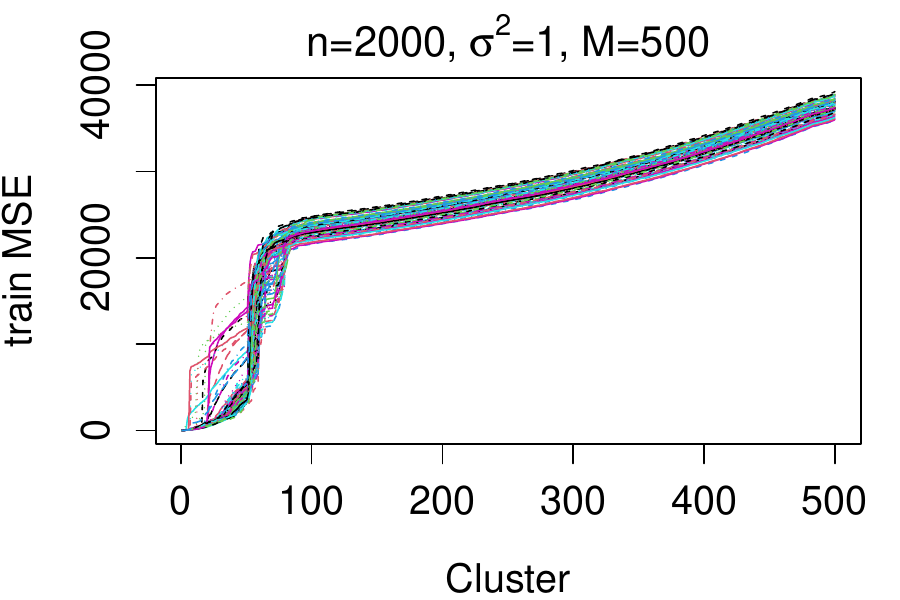} \\
		\includegraphics[width=0.3\hsize, bb= 0 0 450 300]{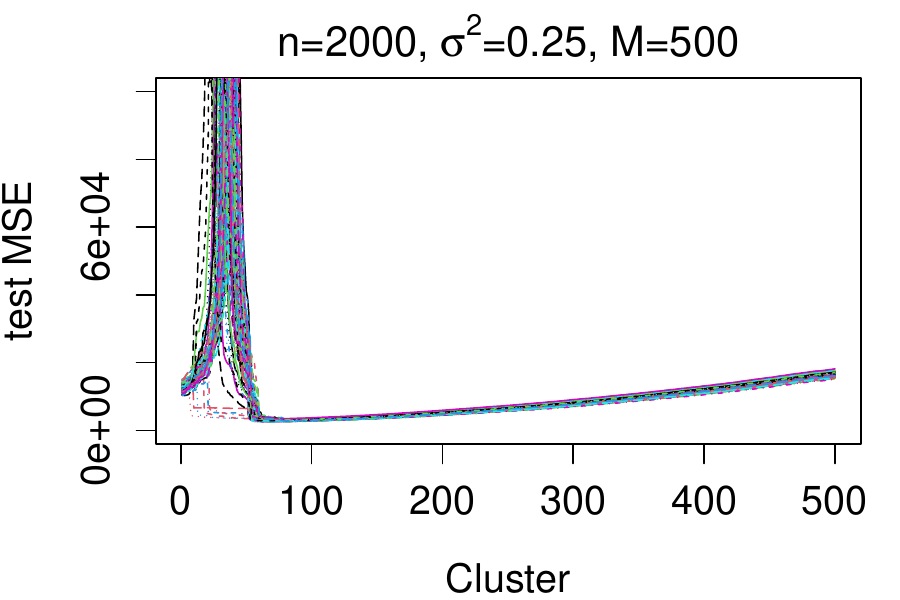}
		\includegraphics[width=0.3\hsize, bb= 0 0 450 300]{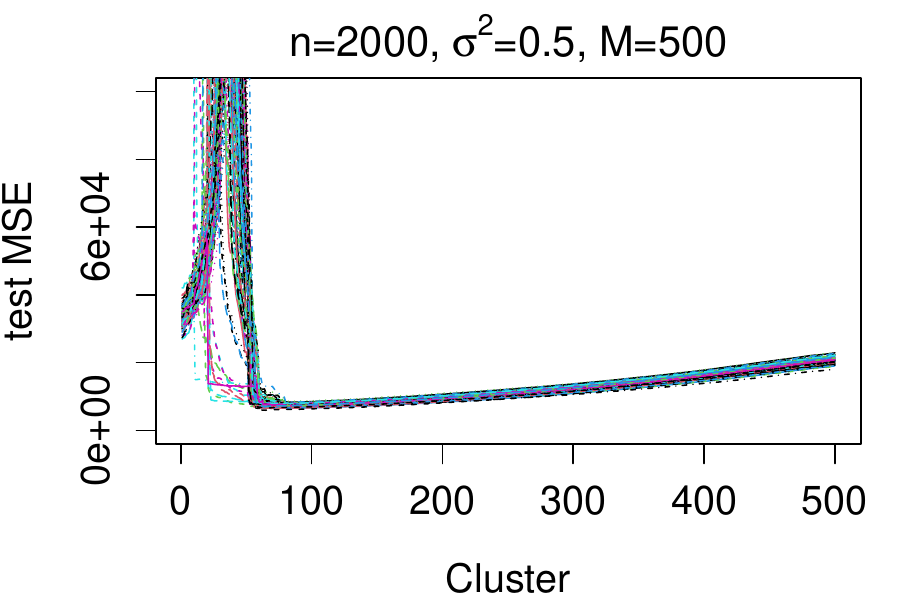}
		\includegraphics[width=0.3\hsize, bb= 0 0 450 300]{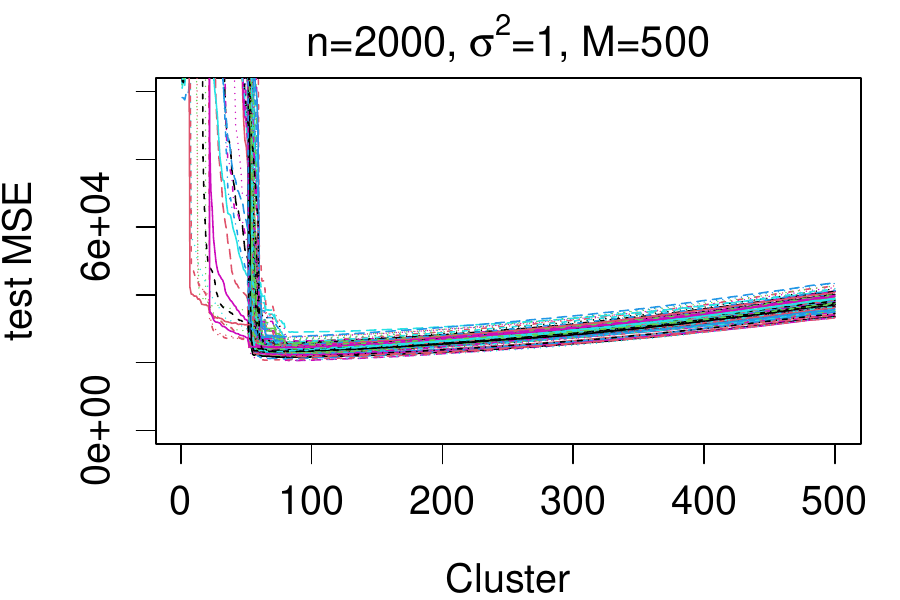}         
	\end{center}
	\caption{Simulation results for $M=500$ using RCM.}
	\label{fig:sim_K500_RCM}

\end{figure}

\bibliographystyle{rss}
\bibliography{paper-ref}

\end{document}